\documentclass[11pt,a4paper]{article}
\usepackage{amsfonts}
\usepackage{epsfig}
\usepackage{graphicx}
\usepackage{amsmath}
\usepackage{amssymb}
\usepackage{color}

\newcommand \cotan{\mathop{\rm cotan}\nolimits}
\newcommand \sn{\mathop{\rm sn}\nolimits}

\newcommand \dn{\mathop{\rm dn}\nolimits}

\begin{document}

\title{Orbits in the problem of two fixed centers on the sphere}

\author{M.A. Gonzalez Leon$^1$\footnote{E-mail: magleon@usal.es}, J. Mateos Guilarte$^2$\footnote{E-mail: guilarte@usal.es} and M. de la Torre Mayado$^2$\footnote{E-mail: marina@usal.es}\\
              \textsl{\small $^1$Departamento de Matem\'atica Aplicada, University of Salamanca}\\ \textsl{\small Avd. Filiberto Villalobos 119,
37007 Salamanca, Spain} \\
              \textsl{\small $^2$Departamento de F\'{\i}sica Fundamental, University of Salamanca}\\ \textsl{\small Facultad de Ciencias, Casas del Parque II, 37008 Salamanca, Spain}}

%\date{\small{\today}}
\date{}

\maketitle

\begin{abstract}

\noindent A trajectory isomorphism between the two Newtonian fixed center problem in the sphere and two associated planar two
fixed center problems is constructed by performing two simultaneous gnomonic projections in $S^2$. This isomorphism converts
the original quadratures into elliptic integrals and allows the bifurcation diagram of the spherical problem to be analyzed in
terms of the corresponding ones of the planar systems. The dynamics along the orbits in the different regimes for the problem in $S^2$
is expressed in terms of Jacobi elliptic functions.
\end{abstract}



%\tableofcontents

\section{Introduction}
\label{intro}

The two fixed center problem on the two-dimensional sphere goes back to Killing \cite{Killing1885}, and in modern times to Kozlov and
Harin \cite{Kozlov1992}, who proved the separability of the problem, thus its integrability, in sphero-conical coordinates. These
coordinates on $S^2$ were introduced by Liouville \cite{Liouville1846}, and independently by Neumann \cite{Neumann} in one of the
first examples of dynamics in spaces of constant curvature, and they are closely related to elliptic coordinates, in fact, the
first system of coordinates plays a role in the dynamics on the sphere completely similar to the second system with respect to the
Euclidian case. Integrability and Hamilton-Jacobi separability in sphero-conical coordinates has been constructed for different
physical systems defined on the sphere, see, for instance, \cite{Bolsinov}. In particular, the authors analyzed in this context
the Neumann problem and the Garnier system on $S^2$ in order to study solitary waves in one-dimensional nonlinear $S^2$-sigma models,
see \cite{nosPRL} and \cite{nosJHEP}. A detailed historical review of several systems defined in spaces of constant curvature,
including open problems, has been recently performed in \cite{Borisov2016}, where a precise bibliography is contained.

The two fixed center problem on the sphere is the superposition of two Kepler problems on $S^2$. An explicit expression for the second constant of motion for this problem and also for some generalizations was given in \cite{Mamaev2003, Borisov2005}. In \cite{Borisov2007} Borisov and Mamaev,
inspired by a previous work of Albouy and Stuchi \cite{Al1, Al2}, established a trajectory isomorphism between the orbits lying in the half-sphere that contains the two attractive centers and the bounded orbits of an associated planar system of two attractive centers.

This isomorphism is constructed using a gnomonic projection from $S^2$ to the tangent plane at the middle point between the centers.
The idea of relating planar and spherical problems in the framework of general force fields by considering the gnomonic projection
goes back to Appell \cite{Appell1890, Appell1891}, who also explained in this context the previous results of Serret \cite{Serret1859}
about the one fixed center problem on the sphere. Almost one century later Higgs \cite{Higgs1979} rediscovered these techniques.
More recently, Albouy has developed these ideas and extended their scope to a general projective dynamics \cite{Al1, Al2, Al3, Albouy2013, Al4}.

In this work we extend the results of \cite{Borisov2007} to the whole sphere, i.e., we establish a trajectory isomorphism between
the complete set of orbits of the original problem and the corresponding one to two associated planar problems. The underlying
idea is to identify each trajectory crossing the equator with the conjunction of two planar unbounded orbits, one of the two
attractive center problem and another one for the system of two repulsive centers.

The extended trajectory isomorphism allows us to understand the bifurcation diagram of the spherical problem, previously analyzed in \cite{Vozmischeva2000, Vozmischeva2002, Vozmischeva}, as the
superposition of the diagrams corresponding to the planar problem of two attractive centers \cite{Waalkens} plus those associated
to two repulsive centers \cite{Seri}. This point of view permits the identification of the domains of allowable motions and the
different cases for orbits, already described in \cite{Vozmischeva}, in terms of their partners for the planar diagrams. Finally,
these identifications lead in a natural way to the determination of the quadratures (elliptic integrals) for the parametric
equations of the orbits in terms of a local time. Using adequately the properties of elliptic integrals, the quadratures are
inverted to obtain explicit formulas in terms of Jacobi elliptic functions for the different types of orbits of the spherical
problem.

The existence of closed orbits in the sphere is guaranteed for the case of commensurability between the involved periods of the
Jacobi functions.

The structure of the paper is as follows: The problem is presented in Section 2 using sphero-conical coordinates on $S^2$. In
Section 3 the extended trajectory isomorphism is defined, and thus the quadratures are converted into elliptic integrals. The
bifurcation diagram for the spherical problem is constructed from the diagrams of the two associated planar problems in Section 4.
Finally, in Section 5, the process of inversion of elliptic integrals in $S^2$ is detailed, and general expressions for the
solutions are shown.

The complete list of analytical expressions for the different types of orbits in $S^2$, in terms of a local time, is included in the
Appendix, together with a gallery of figures for all the significative cases.

\section{The two Newtonian centers problem in $S^2$}
\label{sec:1}

We consider the problem of a unit mass lying on the sphere $S^2$ of radius $R$, viewed as immersed in the Euclidean space
${\mathbb R}^3$ with Cartesian coordinates $(X,Y,Z)$:
\[
X^2+Y^2+Z^2=R^2
\]
under the influence of the superposition of two Kepler potentials on $S^2$, i.e., the potential:
\begin{equation}
{\cal U}(\theta_1,\theta_2)\, =\, -\frac{\gamma_1}{R} \cotan \theta_1 -\frac{\gamma_2}{R} \cotan \theta_2\label{pot1},
\end{equation}

\begin{figure}
\centerline{\includegraphics[width=0.6\textwidth]{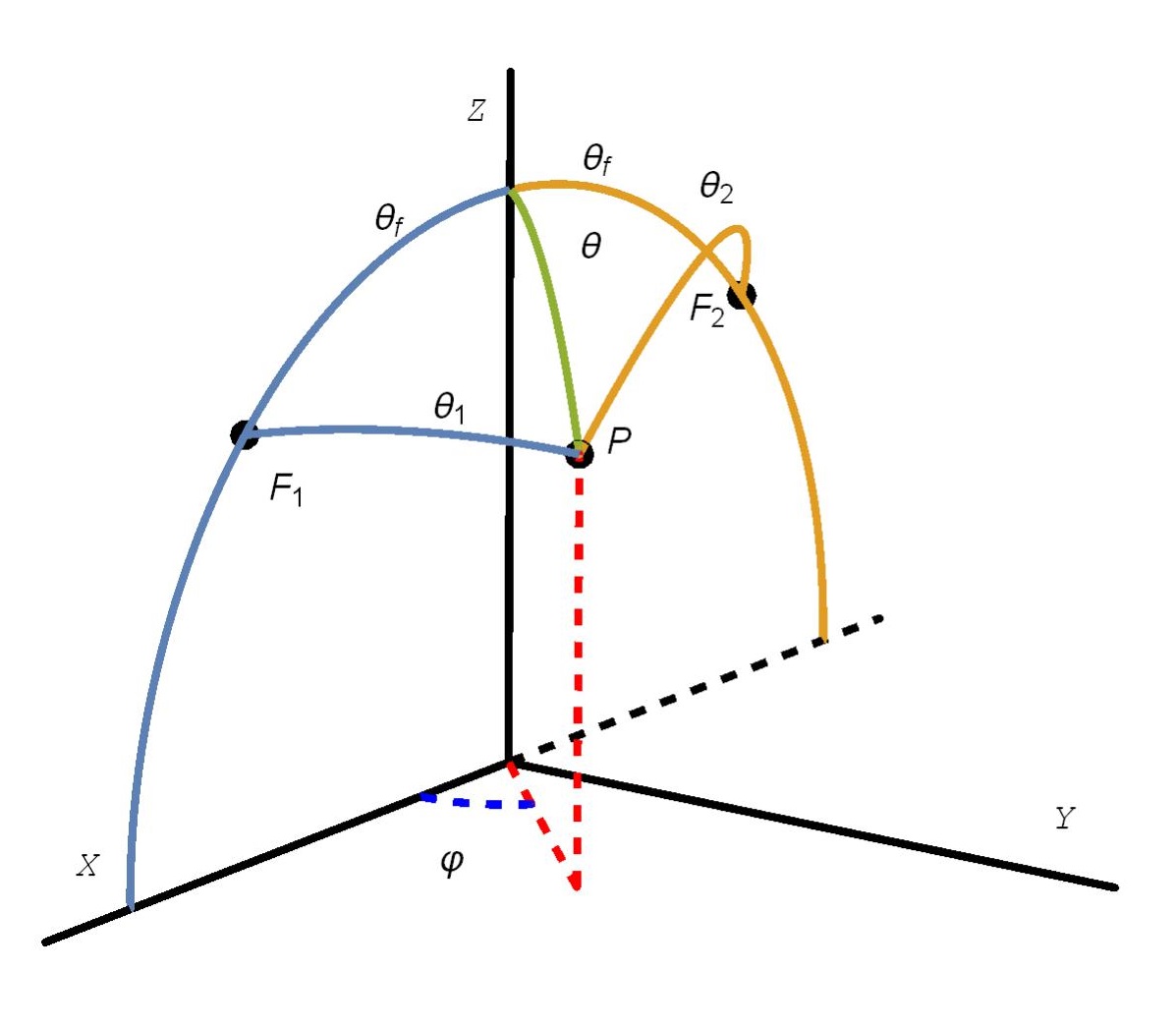}}
\caption{Location of the two Newtonian centers $F_1$ and $F_2$ in $S^2$. The angular separation is $2 \theta_f$, with $0<\theta_f<\frac{\pi}{2}$. $\theta_1$ and $\theta_2$ denote the great circle angles between a given point $P\in S^2$ and $F_1$ and $F_2$, respectively.}
\label{fig:1}       % Give a unique label
\end{figure}

\noindent where $\theta_1$ and $\theta_2$ denote the great circle angles between the location of the centers $F_1$ and $F_2$, see Fig. \ref{fig:1}, and a given point $P$ on $S^2$, in such a way that $R\theta_1$ and $R\theta_2$ are the orthodromic distances from $F_1$ and $F_2$ to $P$, respectively. $\gamma_1$ and $\gamma_2$ are the strengths of the centers, where we have considered $0<\gamma_2\leq \gamma_1$,
i.e., the test mass feels the presence of two attractive centers in $F_1$ and $F_2$, and correspondingly two repulsive centers
at their antipodal points $\bar{F}_1$ and $\bar{F}_2$. Without loss of generality, the chosen points, notation and
orientation are shown in Fig. \ref{fig:1}. Thus, the Cartesian coordinates of $F_1$ and $F_2$ are $(R\sin \theta_f, 0,R\cos \theta_f)=(R\bar{\sigma},0,R\sigma)$ and
$(-R\sin \theta_f, 0,R\cos \theta_f)=(-R\bar{\sigma},0,R\sigma)$, respectively. Parameters $\sigma=\cos \theta_f$ and $\bar{\sigma}=\sin \theta_f$ have been
introduced in order to alleviate the notation.

\noindent  This problem is completely integrable, see, e.g., \cite{Killing1885, Kozlov1992}, there exist two constants of motion,
the Hamiltonian:
\begin{equation}
{\cal H}=\frac{1}{2R^2} \left( L_X^2+L_Y^2+L_Z^2\right)-  \frac{1}{R} \left( \frac{\gamma_1 (\sigma\,  Z+\bar{\sigma}\,  X)}{\sqrt{R^2-(\sigma\, Z+\bar{\sigma}\,
   X)^2}}+\frac{\gamma_2 (\sigma\, Z-\bar{\sigma}\,
   X)}{\sqrt{R^2-(\sigma\, Z-\bar{\sigma}\,  X)^2}}\right)\label{HamCar},
   \end{equation}
where $\vec{L}=\vec{X} \times \vec{P}$, $\vec{P}=(P_X,P_Y,P_Z), \vec{X}=(X,Y,Z)$, and the second invariant:
\begin{equation}
\Omega=
\frac{1}{2R^2}\left( L_X^2 +  \sigma^2 L_Y^2\right) -\frac{\sigma}{R} \left( \frac{\gamma_1 \, Z}{\sqrt{R^2-(\sigma\, Z+\bar{\sigma}\,
   X)^2}}+\frac{\gamma_2 \, Z}{\sqrt{R^2-(\sigma\, Z-\bar{\sigma}\,  X)^2}}\right) \label{SConstCar}.
   \end{equation} % $\Omega_{BM}= \sigma^2 {\cal H}-\Omega$
This constant of motion is slightly different, but equivalent to the invariant obtained by Borisov and Mamaev in \cite{Mamaev2003, Borisov2005}.
Potential (\ref{pot1}) can be rewritten as
\begin{equation}
{\cal U}(\theta_1,\theta_2)= - \, \frac{(\gamma_1+\gamma_2) \sin \frac{\theta_1+\theta_2}{2}\, \cos \frac{\theta_1+\theta_2}{2}+(\gamma_1-\gamma_2) \sin \frac{\theta_2-\theta_2}{2}\,  \cos \frac{\theta_2-\theta_1}{2}}{R\left( \sin^2 \frac{\theta_1+\theta_2}{2}-\sin^2 \frac{\theta_2-\theta_1}{2}\right)} \label{pot0}
\end{equation}
in such a way that it is natural to introduce an \textsl{\'a la Euler} version of sphero-conical coordinates on $S^2$,
i.e.,
\[
U=\sin \frac{\theta_1+\theta_2}{2}\  ,\quad V=\sin \frac{\theta_2-\theta_1}{2}\  ;\quad -\bar{\sigma}<V<\bar{\sigma}\  ,\quad \bar{\sigma}<U<1
\]
Coordinate lines with fixed $U$ or $V$ resemble \lq \lq spherical ellipses" or \lq \lq spherical hyperbolas", respectively, with
foci $F_1$ and $F_2$, with the understanding that ``spherical hyperbolas" are nothing more than ``spherical ellipses" with respect to the pair of
foci $\bar{F}_1$ and $F_2$ or $F_1$ and $\bar{F}_2$.

The change of coordinates
\begin{equation}
X= \frac{R}{\bar{\sigma}} U\, V\  ,\quad Y^2=\frac{R^2}{\sigma^2\bar{\sigma}^2} ( U^2-\bar{\sigma}^2 ) \,  ( \bar{\sigma}^2-V^2 ) \  ,\quad Z^2=\frac{R^2}{\sigma^2}  ( 1-U^2 ) \, (1-V^2)\label{coor1}
\end{equation}
is a four-to-one map because of the ambiguities in the signs of $Y$ and $Z$. Obviously, coordinates $U$ and $V$ are dimensionless.

\noindent Potential (\ref{pot0}) is written in these sphero-conical coordinates with two different expressions depending
on the hemisphere that it is considered. For $S^2_+=\{ (X,Y,Z) \in S^2, \  Z\geq 0\}$, we have
\begin{equation}
{\cal U}_+(U,V)= - \frac{1}{R(U^2-V^2)}\left( (\gamma_1+\gamma_2) U\sqrt{1-U^2} +(\gamma_1-\gamma_2) V\sqrt{1-V^2}\right)\label{potUV+},
\end{equation}
whereas in $S^2_-=\{ (X,Y,Z) \in S^2, \  Z\leq 0 \}$ the potential reads:
\begin{equation}
{\cal U}_-(U,V)= - \frac{1}{R(U^2-V^2)}\left( -(\gamma_1+\gamma_2) U\sqrt{1-U^2} +(\gamma_1-\gamma_2) V\sqrt{1-V^2}\right)\label{potUV-}.
\end{equation}
Note that both expressions (\ref{potUV+}) and (\ref{potUV-}) coincide on the Equator $Z=0$, or $U=1$.
Thus, Hamiltonian (\ref{HamCar}) has also to be split into two different expressions:
\begin{equation}
{\cal H}_{\pm}= \frac{1}{2R^2(U^2-V^2)} \left( (U^2-\bar{\sigma}^2) (1-U^2) p_U^2+(\bar{\sigma}^2-V^2) (1-V^2) p_V^2\right) + {\cal U}_{\pm} (U,V)\label{HamSC}.
\end{equation}

\medskip

\noindent The Hamilton-Jacobi equations coming from (\ref{HamSC})
\begin{equation}
{\cal H}_{\pm}\left( \frac{\partial S}{\partial U},\frac{\partial S}{\partial V},U,V\right)+\frac{\partial S}{\partial t}=0\label{HJeq}
\end{equation}
are separable into two ordinary differential equations if we look for solutions of the form:
$S_{\pm}(t;U,V)=S_t(t)+S_{U\pm}(U)+S_V(V)$. Introducing nondimensional variables
\[
{\cal H}_{\pm} \to \frac{\gamma_1+\gamma_2}{R} {\cal H}_{\pm}\  ,\quad t \to \frac{\sqrt{R^3}}{\sqrt{\gamma_1+\gamma_2}}\, t\  ,\quad
p_{U,V} \to \sqrt{R(\gamma_1+\gamma_2)} p_{U,V}
\]
and defining the parameter
\[
\gamma=\frac{\gamma_2}{\gamma_1+\gamma_2},
\]
the complete solution of (\ref{HJeq}) is
\begin{eqnarray*}
S_{\pm}(t; U,V)&=&-H t+{\rm sg}(p_U) \sqrt{2} \int_{\bar{\sigma}}^U\frac{\sqrt{H  U^2\pm  U \sqrt{1-U^2}-G}}{\sqrt{(1-U^2)(U^2-\bar{\sigma}^2)}}dU\nonumber \\ &&+{\rm sg}(p_V) \sqrt{2} \int_{-{\bar{\sigma}}}^V\frac{\sqrt{- H V^2+ (1-2\gamma) V \sqrt{1-V^2}+G }}{\sqrt{(\bar{\sigma}^2-V^2)(1-V^2)}}dV,\label{HJsol}
   \end{eqnarray*}
where $H$ and $G$ are the values of the constants of motion: ${\cal H}=H$, ${\cal G}=G$; ${\cal G}$ is the separation constant
related to $\Omega$ and ${\cal H}$, (\ref{SConstCar}) and (\ref{HamCar}), by the expression
\[
{\cal G}\, = {\cal H}- \Omega.
\]
Given the local time $\varsigma$ by $d\varsigma=\frac{dt}{U^2-V^2}$, the standard separation procedure leads us to the first-order
equations
\begin{eqnarray}
\frac{dU}{d\varsigma}&=&  {\rm sg}(p_U)\, \sqrt{2} \, \sqrt{(1-U^2)(U^2-\bar{\sigma}^2) (HU^2+U\sqrt{1-U^2}-G)}\label{uno}\\
\frac{dV}{d\varsigma}&=&  {\rm sg}(p_V) \, \sqrt{2} \, \sqrt{(1-V^2)(\bar{\sigma}^2-V^2) (-HV^2+(1-2\gamma)V\sqrt{1-V^2}+G)}\label{dos}
\end{eqnarray}
for the problem in the Northern hemisphere $S_+^2$, and
\begin{eqnarray}
\frac{dU}{d\varsigma}&=&  {\rm sg}(p_U)\, \sqrt{2} \, \sqrt{(1-U^2)(U^2-\bar{\sigma}^2) (HU^2-U\sqrt{1-U^2}-G)}\label{uno2}\\
\frac{dV}{d\varsigma}&=&  {\rm sg}(p_V) \, \sqrt{2} \, \sqrt{(1-V^2)(\bar{\sigma}^2-V^2) (-HV^2+(1-2\gamma)V\sqrt{1-V^2}+G)}\label{dos2}
\end{eqnarray}
for the Southern $S_-^2$ one.

\noindent A direct attack to the quadratures involved looks apparently cumbersome and, as far as we know, they are not solved
in the literature. Nevertheless, some of the qualitative and topological properties of these orbits have been analyzed
in \cite{Vozmischeva2000, Vozmischeva2002, Vozmischeva}.

\section{Trajectory isomorphism between the spherical and two different planar problems}

Following Borisov \& Mamaev \cite{Borisov2007}, we go back to Cartesian coordinates $(X,Y,Z)$ where the potential (\ref{pot1})
can be written as
\begin{equation}
{\cal U}(X,Y,Z)\, =\, -\frac{1}{R} \left( \frac{\gamma_1 (\sigma\,  Z+\bar{\sigma}\,  X)}{\sqrt{R^2-(\sigma\, Z+\bar{\sigma}\,
   X)^2}}+\frac{\gamma_2 (\sigma\, Z-\bar{\sigma}\,
   X)}{\sqrt{R^2-(\sigma\, Z-\bar{\sigma}\,  X)^2}}\right).\label{pot2}
   \end{equation}
The corresponding Newton equations for this problem are
\begin{equation}
\ddot{X}\,=\,  -\frac{\partial {\cal U}}{\partial X}\, +\, \lambda X\  ,\quad
\ddot{Y}\,=\, -\frac{\partial {\cal U}}{\partial Y}\, +\, \lambda Y\  ,\quad \ddot{Z}= -\frac{\partial {\cal U}}{\partial Z}\, +\, \lambda Z,\label{newton1}
\end{equation}
where the dots represent derivatives with respect to the physical (dimensional) time $t$ and $\lambda$ is the Lagrange multiplier.
In \cite{Borisov2007} it was proved that the gnomonic projection from $S_+^2$ to the tangent plane $\Pi_+$ at the point $(0,0,R)$, together with a linear transformation in $\Pi_+$, maps Newton equations (\ref{newton1}) to the Newton equations of an associated problem of two attractive centers in ${\mathbb R}^2$.

\noindent Here, we shall also consider simultaneously another gnomonic projection, from $S_-^2$ to the tangent plane $\Pi_-$,
at $(0,0,-R)$. The projected coordinates $(x,y)$ are given in the two planes by
\begin{equation}
\Pi_+:\quad x= \frac{R}{Z}\, X\  ,\quad y=  \frac{R}{Z}\, Y  \quad ; \quad \Pi_-:\quad
x= \frac{R}{-Z}\, X\  ,\quad y=  \frac{R}{-Z}\, Y.   \label{projection}
\end{equation}
for $Z\neq 0$. The Equator is mapped to the infinity in both $\Pi_+$ and $\Pi_-$ planes.
 
We will use throughout the paper the following criteria: uppercase letters describe magnitudes and variables specifically
defined in the sphere, whereas lowercase letters will be associated to the planar cases.
Following \cite{Borisov2007}, we perform in $\Pi_+$ the linear transformation:
\begin{equation}
x_1\equiv x\  ,\quad x_2\equiv \frac{y}{\sigma}\label{LinearT}.
\end{equation}
The Newton equations (\ref{newton1}) for potential (\ref{pot2}) are rewritten in transformed projected coordinates $(x_1,x_2)$
on $\Pi_+$ as
\begin{equation}
x_1''(\tau) = -\frac{\partial {\cal V_{+}}}{\partial x_1}\  ,\quad
x_2''(\tau) = -\frac{\partial {\cal V_{+}}}{\partial x_2}\label{newtrans}
\end{equation}
\begin{equation}
{\cal V}_+(x_1,x_2)\, =\, -\frac{\alpha_1}{\sqrt{\left( x_1-a\right)^2+x_2^2}}-\frac{\alpha_2}{\sqrt{\left( {x_1}+a\right)^2+x_2^2}}\label{potNORTH}
\end{equation}
where the primes denote the derivative with respect to a new time $\tau$ defined by
\begin{equation*}
d\tau \, =\, \frac{R^2}{Z^2}\, dt\label{time1}
\end{equation*}
and we have introduced the parameters: $a=R\frac{\bar{\sigma}}{\sigma}$, $\alpha_1=\frac{\gamma_1}{\sigma^2}$ and $\alpha_2=\frac{\gamma_2}{\sigma^2}$.

\noindent In a similar way, the Newton equations (\ref{newton1}) restricted to $S_-^2$ can be projected into $\Pi_-$
using (\ref{projection}) and, after applying transformation (\ref{LinearT}), the equations
\begin{equation}
x_1''(\tau) = -\frac{\partial {\cal V_{-}}}{\partial x_1}\  ,\quad
x_2''(\tau) = -\frac{\partial {\cal V_{-}}}{\partial x_2}\nonumber
\end{equation}
\begin{equation}
{\cal V}_-(x_1,x_2)\, =\, \frac{\alpha_2}{\sqrt{\left( {x_1}-a\right)^2+x_2^2}}+\frac{\alpha_1}{\sqrt{\left( x_1+a\right)^2+x_2^2}}\label{potSOUTH}
\end{equation}
are obtained.

\noindent Note that ${\cal V}_-(x_1,x_2)$ in $\Pi_-$ is nothing more than the planar potential of two repulsive centers, where
the roles of the points $(\pm a,0)$, and thus the strengths of the centers in modulus, are interchanged with respect to the
attractive potential ${\cal V}_+(x_1,x_2)$ in $\Pi_+$.

\noindent Thus, while the restriction of the Newton equations (\ref{newton1}) to the Northern hemisphere $S_+^2$ is equivalent
to the Newton equations (\ref{newtrans}) for a planar problem of two attractive centers with potential (\ref{potNORTH}),
the restriction to the Southern hemisphere $S_-^2$ is tantamount to a planar problem of two repulsive centers with
potential (\ref{potSOUTH}).

\noindent Bounded orbits of the attractive planar problem are in a one-to-one correspondence with the orbits of the spherical
problem that lie in $S_+^2$. However, trajectories of the spherical problem crossing the equator have to be described in this
projected picture by two pieces: an unbounded orbit of the attractive planar problem (\ref{potNORTH}) in $\Pi_+$ plus an (unbounded)
orbit of the repulsive planar problem (\ref{potSOUTH}) in $\Pi_-$, corresponding to the parts of the orbit belonging to $S_+^2$
and $S_-^2$, respectively.

It is possible to describe in a compact form the two associated planar problems in $\Pi_+$ and $\Pi_-$, respectively,
by the Hamiltonians
\begin{equation}
{\mathrm h}_{\pm}\, =\, \frac{1}{2} \left( p_1^2+p_2^2\right) +{\cal V}_{\pm}(x_1,x_2)\label{planarham}.
\end{equation}
It is adequate again to use nondimensional variables
\[
x_i\to a x_i\,   ,\  p_i \to \frac{\sqrt{\alpha_1+\alpha_2}}{\sqrt{a}} p_i\,  ,\  \tau \to \frac{\sqrt{a^3}}{\sqrt{\alpha_1+\alpha_2}} \tau\, ,\
 {\mathrm h}_{\pm}\, \to \,  \frac{\alpha_1+\alpha_2}{a} {\mathrm h}_{\pm}\  ; \quad \alpha=\frac{\alpha_2}{\alpha_1+\alpha_2}=\gamma
 \]
and to introduce ``radial", $u$, and ``angular", $v$, elliptic (Euler) coordinates in ${\mathbb R}^2$:
\[
u=\frac{ \sqrt{( x_1+1)^2+x_2^2}+\sqrt{( x_1-1)^2+x_2^2}}{2}\,  ,\  v=\frac{\sqrt{( x_1+1)^2+x_2^2}-\sqrt{( x_1-1)^2+x_2^2}}{2}
\]
\[
x_1=uv\  ,\quad x_2=\pm \sqrt{u^2-1}\sqrt{1-v^2}\  , \qquad
v\in (-1,1)\  ,\quad u>1
\]
in such a way that the Hamiltonians (\ref{planarham}) are written in terms of these coordinates as
\[
{\mathrm h}_\pm\, =\, \frac{1}{u^2-v^2} \left( \frac{u^2-1}{2} \, p_u^2 \mp u\, +\, \frac{1-v^2}{2} \, p_v^2 - (1-2\alpha) v \right),
\]
i.e., two standard Liouville separable systems in elliptic coordinates. It is straightforward to construct the associated
first-order equations with respect to the local time $\zeta=\zeta(\tau)$ defined by
\[
d\zeta =\frac{d\tau}{u^2-v^2},
\]
and we finally obtain the following equations in the $\Pi_+$ plane:
\begin{equation}
\left( \frac{du}{d\zeta}\right)^2= 2 (u^2-1)(h u^2+u-g)\  ,\quad \left( \frac{dv}{d\zeta}\right)^2=2(1-v^2)(-h v^2+(1-2\alpha)v+g)\label{FOeq1},
\end{equation}
which solve the original problem in $S^2_+$. Correspondingly, for the Southern case we obtain in the $\Pi_-$ plane:
\begin{equation}
\left( \frac{du}{d\zeta}\right)^2=2(u^2-1)(\tilde{h}u^2-u-\tilde{g})\  ,\quad \left( \frac{dv}{d\zeta}\right)^2=2(1-v^2)(-\tilde{h}v^2+(1-2\alpha)v+\tilde{g})\label{FOeq2},
\end{equation}
where the constants of motion take the values ${\mathrm h}_+ = h$ and ${\mathrm g}_+= g$ for the energy and the separation
constant in $\Pi_+$, respectively, and ${\mathrm h}_-=\tilde{h}$ and ${\mathrm g}_-=\tilde{g}$ in $\Pi_-$. The quadratures involved in equations (\ref{FOeq1}) and (\ref{FOeq2}) are of elliptic type, and thus expressible in terms of the Jacobi elliptic functions.

\noindent It is possible to synthesize the chain of maps leading from the original problem in the sphere to the pair of planar two center problems  (\ref{FOeq1}) and (\ref{FOeq2}) in a unique one-to-one transformation of coordinates in $S^2$, from sphero-conical $(U,V)$ to planar elliptic $(u,v)$, as follows:
\begin{equation}
U= \frac{\bar{\sigma} u}{\sqrt{\bar{\sigma}^2 u^2+\sigma^2}}\  ;\quad
V= \frac{\bar{\sigma} v}{\sqrt{\bar{\sigma}^2 v^2+\sigma^2}}\label{cambiog}
\end{equation}
together with an equivalence, up to a constant factor, between the nondimensional local time $\varsigma$ of the spherical problem and the nondimensional local time $\zeta$ for the associated planar problems:
\begin{equation}
d\varsigma \, =\, \sqrt{\sigma \bar{\sigma}} \, d\zeta\label{time}.
\end{equation}
The equator $Z=0$, or $U=1$, of $S^2$ is mapped by (\ref{cambiog}) into the point of infinity in the coordinate $u$.

\noindent Equation (\ref{time}) establishes that $\zeta$ can be simultaneously seen as the local time for both problems, remembering
its different meaning when regarded from $S^2$, local time associated with the ``physical" time $t$ in the sphere, or
from $\Pi_{\pm}$, local time corresponding to the projected (nonphysical) time $\tau$ in the planes.

\medskip

\noindent Thus, (\ref{cambiog}) and (\ref{time}) map directly the first-order equations (\ref{uno}, \ref{dos}) in $S^2_+$ to
equations (\ref{FOeq1}) in $\Pi_+$, and (\ref{uno2}, \ref{dos2}) in $S^2_-$ to (\ref{FOeq2}) in $\Pi_-$ via the
identifications
\[
h=\frac{\bar{\sigma}}{\sigma} (H-G)=\frac{\bar{\sigma}}{\sigma} \Omega=\  \tan \theta_f\,  \Omega  \  ,\quad g=\frac{\sigma}{\bar{\sigma}} G=\cotan \theta_f\, G\  , \  \  {\rm in}\  S^2_+
\]
\[
\tilde{h}=\frac{\bar{\sigma}}{\sigma} (H-G)=\frac{\bar{\sigma}}{\sigma} \Omega=\  \tan \theta_f\,  \Omega  \  ,\quad \tilde{g}=\frac{\sigma}{\bar{\sigma}}\   G=\cotan \theta_f\, G\ ,\  \  {\rm in}\  S^2_-
 .\]
It is remarkable that in this projected picture the role of the planar energies $h$ and $\tilde{h}$ is played, up to a factor,
by the projection of the second constant of motion $\Omega$ and not by the projection of the spherical Hamiltonian. This fact is
a consequence of the behavior of constants of motion under central projections, as was explained in \cite{Albouy2013}, see also \cite{Al4}.

\medskip

\noindent Consequently, the transformation (\ref{cambiog}) establishes that fixing in $S^2$ a negative value of the constant of
motion $\Omega$, the orbits of the problem lie in the $S^2_+$ hemisphere and are in a one-to-one correspondence with the bounded
orbits, $h=\frac{\bar{\sigma}}{\sigma} \Omega <0$, of the planar attractive system in the $\Pi_+$ plane. Thus, coordinate $u$ is bounded for $\Omega <0$. However, if $\Omega \geq 0$, orbits cross the equator of $S^2$, and thus the portions of the orbits belonging to $S^2_+$ are described by equations (\ref{FOeq1})
with planar energy $h\geq 0$, unbounded planar orbits in the attractive problem in $\Pi_+$, whereas the portions lying in the
Southern hemisphere $S^2_-$ are determined by equations (\ref{FOeq2}) with $\tilde{h}>0$, i.e., unbounded planar orbits of the
repulsive problem in $\Pi_-$. In this case $u$ is unbounded.

\section{The bifurcation diagrams}

The isomorphic transformation (\ref{cambiog}) allows  us to analyze the bifurcation diagram in $S^2$ starting from the
bifurcation diagrams of the two associated planar problems. From this point of view a global bifurcation diagram for the spherical problem will be constructed out of the diagrams of two planar centers, see \cite{Waalkens} and \cite{Seri}, respectively, attractive in $\Pi_+$ and repulsive in $\Pi_-$ and strengths interchanged. Thus, the results explained in \cite{Vozmischeva2002, Vozmischeva} will now be understood from a different perspective.

Both in $\Pi_+$ and $\Pi_-$ planes, i.e., the images of the North $S^2_+$ and South $S^2_-$ hemispheres, we rewrite (\ref{FOeq1})
and (\ref{FOeq2}) in terms of the ramification points:

\begin{eqnarray}
\left( \frac{du}{d\zeta}\right)^2 &=& 2 h (u^2-1)(u-u_1)(u-u_2)\  ,\quad \left( \frac{dv}{d\zeta}\right)^2=-2 h (1-v^2)(v-v_1)(v-v_2)\label{FOeq1b} \\
\Pi_+ \  :\qquad  u_1&=&\frac{-1}{2h}-\sqrt{\frac{g}{h}+\frac{1}{4h^2}}\quad  ,\quad \qquad  \quad  u_2=\frac{-1}{2h}+\sqrt{\frac{g}{h}+\frac{1}{4h^2}} \nonumber\\
v_1&=&\frac{1-2\alpha}{2h}-\sqrt{\frac{g}{h}+\frac{(1-2\alpha)^2}{4h^2}}\  ,\   v_2=\frac{1-2\alpha}{2h}+\sqrt{\frac{g}{h}+\frac{(1-2\alpha)^2}{4h^2}} \nonumber\quad ,
\end{eqnarray}

\begin{eqnarray}
\left( \frac{du}{d\zeta}\right)^2&=&2\tilde{h} (u^2-1)(u-\tilde{u}_1)(u-\tilde{u}_2)\  ,\quad \left( \frac{dv}{d\zeta}\right)^2=-2 \tilde{h} (1-v^2)(v-\tilde{v}_1)(v-\tilde{v}_2)\label{FOeq2b}\\
\Pi_- \  :\qquad  \tilde{u}_1&=&\frac{1}{2\tilde{h}}-\sqrt{\frac{\tilde{g}}{\tilde{h}}+\frac{1}{4\tilde{h}^2}}\quad  ,\quad \qquad  \quad    \tilde{u}_2=\frac{1}{2\tilde{h}}+\sqrt{\frac{\tilde{g}}{\tilde{h}}+\frac{1}{4\tilde{h}^2}}\nonumber \\
\tilde{v}_1&=&\frac{1-2\alpha}{2\tilde{h}}-\sqrt{\frac{\tilde{g}}{\tilde{h}}+\frac{(1-2\alpha)^2}{4\tilde{h}^2}}\  ,\   \tilde{v}_2=\frac{1-2\alpha}{2\tilde{h}}+\sqrt{\frac{\tilde{g}}{\tilde{h}}+\frac{(1-2\alpha)^2}{4\tilde{h}^2}}. \nonumber
\end{eqnarray}

\begin{figure}
\centerline{\includegraphics[width=0.45\textwidth]{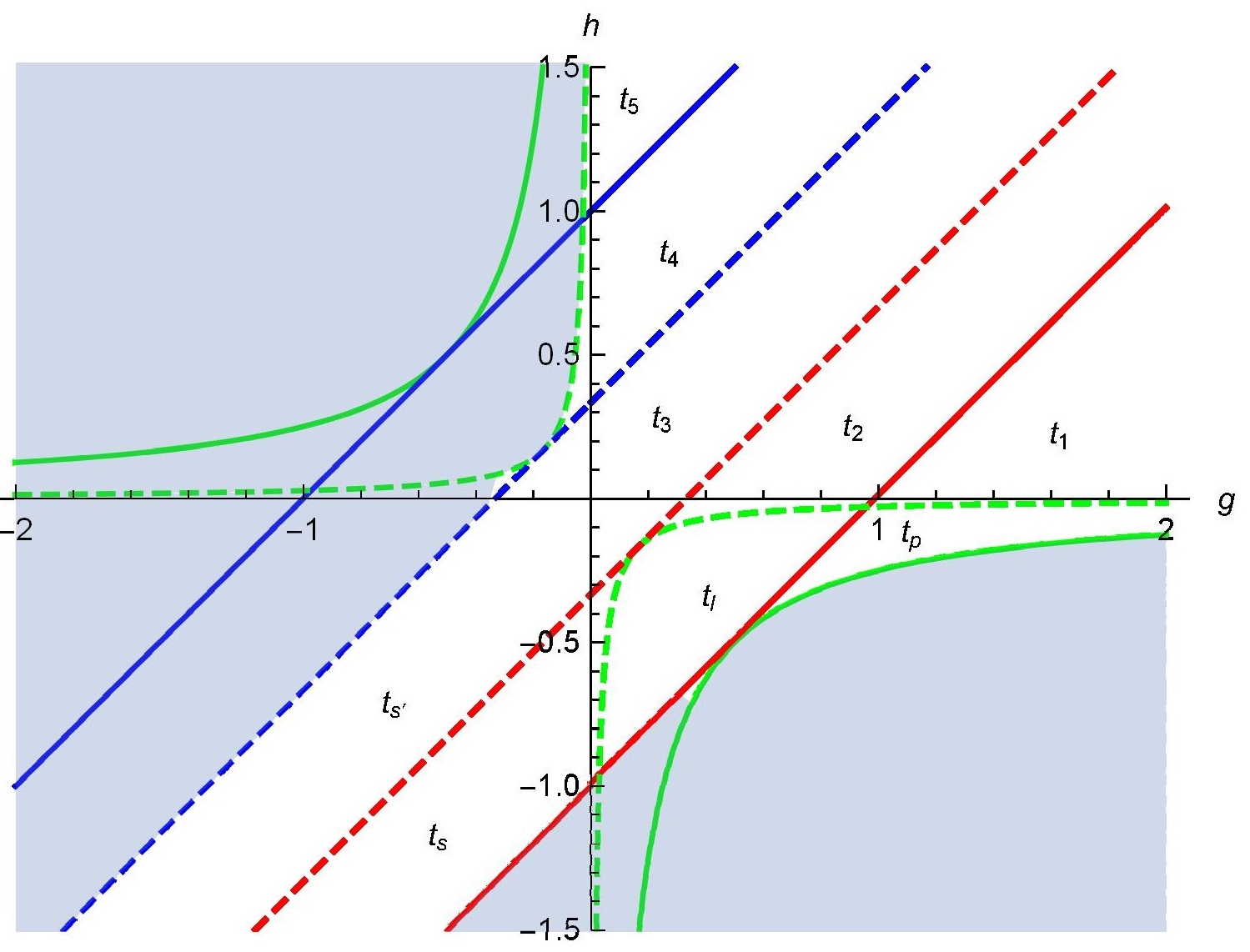}\ \includegraphics[width=0.45\textwidth]{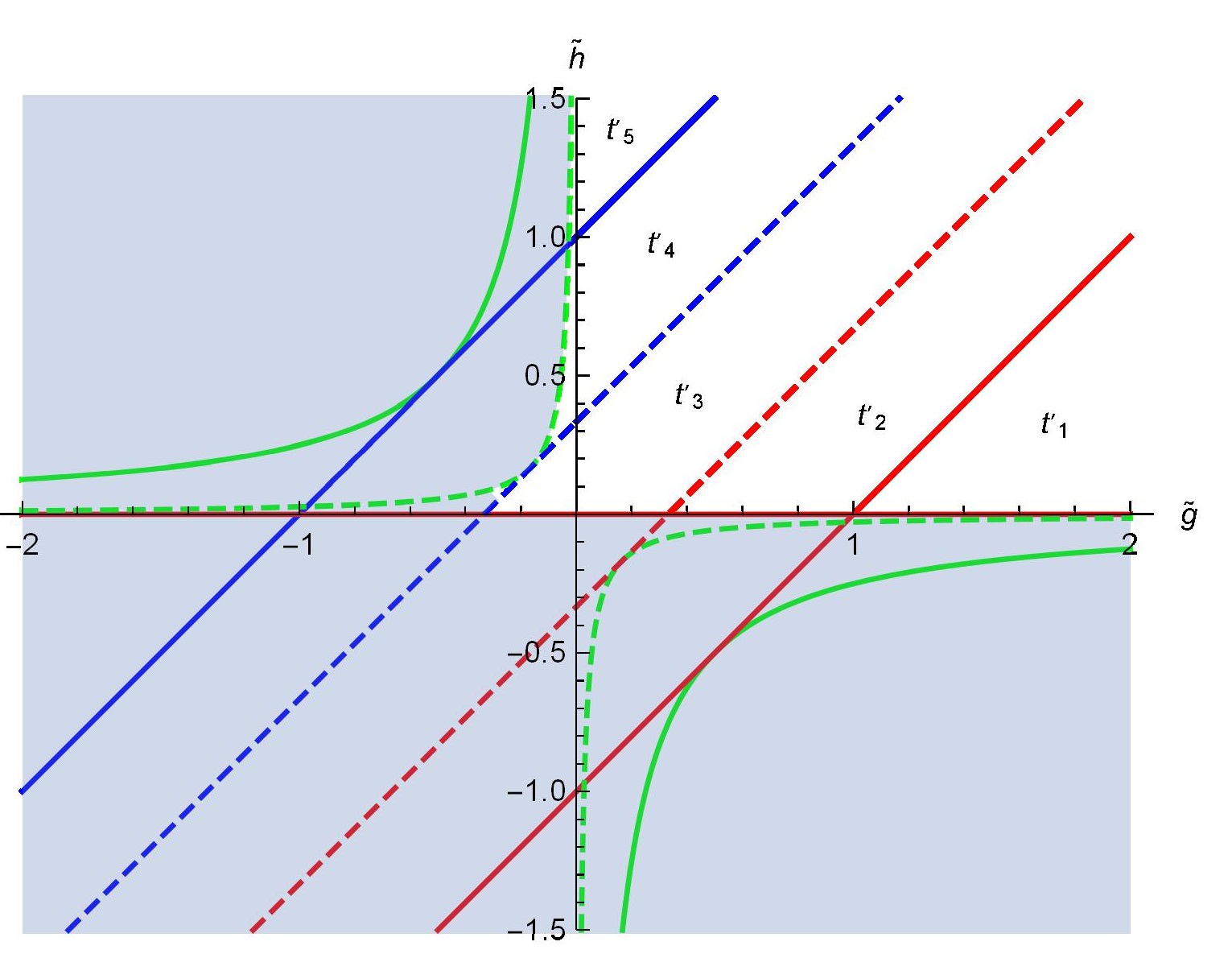}}
\caption{(a) Bifurcation diagram for two attractive centers in the plane. (b) Bifurcation diagram for two repulsive centers in the plane
with the strengths (in modulus) exchanged with respect to the attractive potential. In both cases we choose $\alpha=\frac{1}{3}$.}
\label{bifurplane}
\end{figure}

\noindent In Figure \ref{bifurplane}, plotted for $\alpha=1/3$, we observe the bifurcation diagrams corresponding to the attractive and repulsive planar problems in $\Pi_+$ Fig. \ref{bifurplane}a) and $\Pi_-$ Fig. \ref{bifurplane}b), respectively, with strengths $\alpha_1$, $\alpha_2$, and $\tilde{\alpha}_1=-\alpha_2$, $\tilde{\alpha}_2=-\alpha_1$. Critical curves in both $\{h,g\}$ and $\{\tilde{h},\tilde{g}\}$ planes are determined by the existence of double roots in (\ref{FOeq1b}) and (\ref{FOeq2b}), see \cite{Waalkens, Seri}, and shadowed areas in the diagrams are zones where motion is classically forbidden, i.e., velocities and/or momenta are imaginary.

The allowed motions in the $\Pi_+$-plane are of two types: (1) If $h < 0$, orbits are bounded and are usually labeled as
$\{t_s, t_{s'}, t_l, t_p\}$, for satellitary, lemniscatic and planetary ones, see \cite{Waalkens}. (2) If $h\geq 0$,
see \cite{Seri}, unbounded orbits occur, standardly labeled as $\{t_1, t_2, t_3, t_4, t_5\}$. Separatrices between bounded and
unbounded motions live in the $\{ h = 0 \}$ straight line.

In the $\Pi_-$ plane a similar, but simpler picture is found, see Fig. 2b). On the $\tilde{h}>0$ upper half-plane unbounded
orbits exist in five different classes, labeled as $\{t'_1, t'_2, t'_3, t'_4, t'_5\}$. In this case the line $ \{ \tilde{h} = 0 \}$
does not accommodate separatrices, but rather it corresponds to a limiting behavior of unbounded zero energy orbits reached
from $\tilde{h}>0$.

\medskip

\begin{figure}
\centerline{\includegraphics[width=0.65\textwidth]{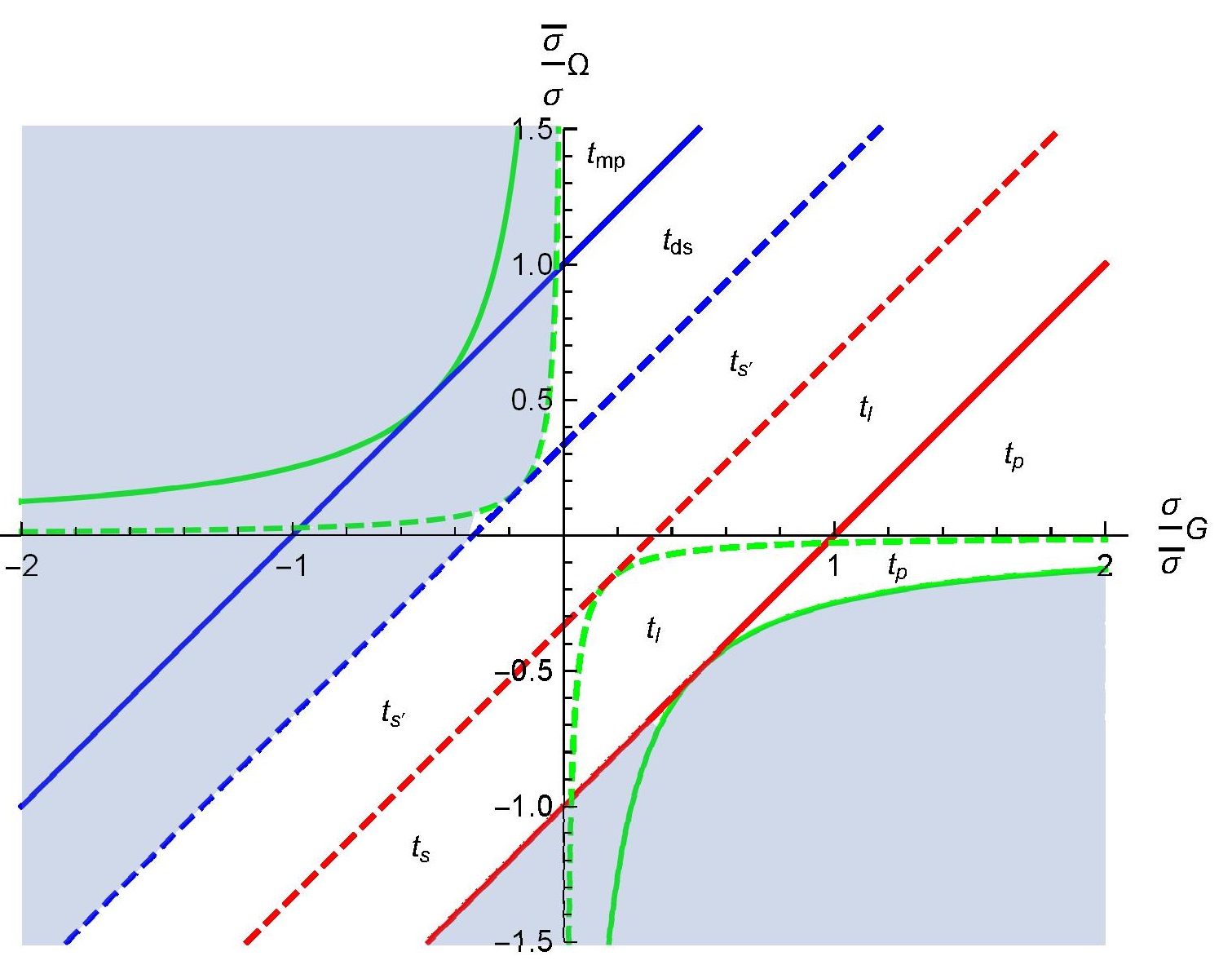}}
\caption{Global bifurcation diagram in $S^2$ with $\gamma=\frac{1}{3}$.}
\label{bifuresfera}
\end{figure}

The bifurcation diagram for the complete problem in $S^2$, Figure \ref{bifuresfera},  can now be constructed from the planar ones
using transformations (\ref{cambiog}) and (\ref{time}). Two morphisms are induced: (1) Orbits in $\Pi_+$ are mapped to
orbits in $S^2_+$ identifying the invariants as follows: $h=\frac{\bar{\sigma}}{\sigma} \Omega$ and $g=\frac{\sigma}{\bar{\sigma}} G$. (2) Orbits in $\Pi_-$ are mapped to orbits in $S^2_-$ if the invariants are translated to: $\tilde{h}=\frac{\bar{\sigma}}{\sigma} \Omega$ and $\tilde{g}=\frac{\sigma}{\bar{\sigma}} G$. The global bifurcation diagram in $S^2$ is thus displayed on the $\{\frac{\bar{\sigma}}{\sigma} \Omega, \frac{\sigma}{\bar{\sigma}} G\}$-plane.

Moreover, the lower half-plane of Figure \ref{bifuresfera}, $\frac{\bar{\sigma}}{\sigma} \Omega <0$, is mapped one-to-one with
the lower half-plane of the problem of two attractive centers in $\Pi_+$, Fig. \ref{bifurplane}a), as it was shown
in \cite{Borisov2007}, orbits lying only in $S_+^2$ are in a bijective correspondence with bounded orbits in $\Pi_+$. However,
with an initial condition fixed, each point $\left(\frac{\bar{\sigma}}{\sigma} \Omega, \frac{\sigma}{\bar{\sigma}} G\right)$
in the upper half-plane of the global diagram represents an orbit that crosses the equator of $S^2$, and thus is mapped
by (\ref{cambiog}) and (\ref{time}) to the union of an unbounded orbit in $\Pi_+$ and another one in $\Pi_-$, with equal planar
energies: $h=\tilde{h}=\frac{\bar{\sigma}}{\sigma} \Omega$.

Critical curves in Figure \ref{bifuresfera} are inherited from the corresponding ones in planar diagrams:

\begin{itemize}

\item Double roots in equations (\ref{FOeq1b}, \ref{FOeq2b}) for the ``radial" variable arise in the:  blue straight line:
${\cal L}_1^2 = \left\{ \frac{\bar{\sigma}}{\sigma} \Omega- \frac{\sigma}{\bar{\sigma}} G-1=0\right\}$, red straight line:
${\cal L}_1^1 = \left\{ \frac{\bar{\sigma}}{\sigma} \Omega- \frac{\sigma}{\bar{\sigma}} G+1=0\right\}$, and green hyperbola:
${\cal L}_1^3 = \left\{ 4 \Omega   G+1=0\right\}$.

\item Analogously, double roots for the ``angular" variable produce the: dashed blue straight line: ${\cal L}_\gamma^1 = \left\{ \frac{\bar{\sigma}}{\sigma} \Omega- \frac{\sigma}{\bar{\sigma}} G-(1-2\gamma)=0\right\}$,
dashed red straight line: ${\cal L}_\gamma^2 = \left\{ \frac{\bar{\sigma}}{\sigma} \Omega- \frac{\sigma}{\bar{\sigma}} G+(1-2\gamma)=0\right\}$, and
dashed green hyperbola: ${\cal L}_\gamma^3 = \left\{ 4 \Omega   G+(1-2\gamma)^2 =0\right\}$.

\end{itemize}

Orbits with $\Omega<0$ are naturally labeled with the inherited standard notation for bounded motion in the planar associated
problem in $\Pi_+$. The branching points $u_1,u_2$ and $v_1, v_2$, understood as functions of $\Omega$ and $G$, allow us to
specify the qualitative features of these orbits in $S_+^2$:

\begin{itemize}

\item Planetary orbits ($t_p$). There are two analytical possibilities that lead to the same type of orbits:
\begin{eqnarray}
(1) &&  -1 < 1 < u_1 < u < u_2 \quad  , \quad  -1 < v < 1 \ , \ v_1,v_2 \in {\mathbb C}, \ {\rm Im} (v_1)=-{\rm Im} (v_2)\neq0 \label{tp1HN} \\
(2) &&  -1 < 1 <  u_1 < u < u_2  \quad , \quad v_1 < v_2 < -1 < v < 1 \label{tp2HN}
\end{eqnarray}
In both cases the bounds $u=u_1$ and $u=u_2$ represent two caustics for these orbits, i.e., two ``spherical ellipses" in
the Northern hemisphere $S_+^2$, see Fig. \ref{graficas1}a, that confine the planetary motion of these ``circumbinary" orbits.

\item Lemniscatic orbits ($t_l$). Analogously, there exist two possibilities:
\begin{eqnarray}
(1) && -1 < u_1 < 1 < u < u_2 \quad , \quad -1 < v < 1 \ , \  v_1,v_2\in {\mathbb C}, \ {\rm Im} (v_1)=-{\rm Im} (v_2)\neq0 \label{tl1HN}\\
(2) && -1 < u_1 < 1 < u < u_2  \quad , \quad v_1 < v_2 < -1 < v  < 1.  \label{tl2HN}
\end{eqnarray}
A unique caustic, $u=u_2$, appears in this case. The orbits describe a lemniscatic motion around the two centers in $S_2^+$. See Fig. \ref{graficas1}b.

\item Satellitary orbits ($t_s$): Each point $\left(\frac{\bar{\sigma}}{\sigma} \Omega, \frac{\sigma}{\bar{\sigma}} G\right)$ of this region in Figure \ref{bifuresfera} represents two possible orbits:
\begin{equation}
(1) \quad -1 < u_1 < 1 < u <  u_2 \quad , \quad -1 <  v_1 < v_2  < v  <  1 \label{ts1HN}
\end{equation}
around the stronger center, limited by the caustics: $u=u_2$ and $v=v_2$, and:
\begin{equation}
(2) \quad -1 < u_1 < 1 < u <  u_2 \quad , \quad -1 <  v <  v_1  < v_2  <1 \label{ts2HN}
\end{equation}
around the weaker center, bounded by:  $u=u_2$ and $v=v_1$. See Figure \ref{graficas1}g.

\item Satellitary orbits ($t_{s'}$) around the stronger center:
\begin{equation}
-1 < u_1 < 1 < u <  u_2 \quad , \quad v_1 < -1  < v_2  < v  <  1. \label{tspHN}
\end{equation}
For this situation the motion is limited by the caustics: $u=u_2$ and $v=v_2$, see Figure \ref{graficas1}c.

\end{itemize}

For $\Omega >0$, it is possible to extend the standard terminology, Planetary ($t_p$), Lemniscatic ($t_l$) and Satellitary ($t_{s'}$), to the orbits that cross the equator, but have a behavior analogous to the corresponding cases
restricted to the Northern hemisphere. However, two completely new types of orbits arise. There are two zones of admissible
motion without partners between the orbits with $\Omega<0$, which we will call Dual Satellitary ($t_{ds}$) and Meridian
Planetary ($t_{mp}$) orbits, taking into account its qualitative features.

Branching points are now identified by $\tilde{u}_1 = - u_2$, $\tilde{u}_2= - u_1$ and $\tilde{v}_1=v_1$, $\tilde{v}_2=v_2$, because
$h=\tilde{h}=\frac{\bar{\sigma}}{\sigma} \Omega$ and $g=\tilde{g}= \frac{\sigma}{\bar{\sigma}} G$ in order to glue continuously the two orbit pieces on the Northern and Southern
hemispheres at the equator.

\begin{itemize}

\item Planetary orbits ($t_p$): The orbits in $S^2$ are composed by two pieces:
\begin{eqnarray}
S_+^2:&& \qquad u_1\, < \, -1\, <\, 1\, <\, u_2\, <\, u\  ,\qquad  v_1\, <\, -1\, <\, v \, <\, 1 \, <\, v_2  \label{tpHS} \\
S_-^2:&& \qquad \tilde{u}_1\, < \, -1\, <\, 1\, <\, \tilde{u}_2\, <\, u\  ,\qquad  \tilde{v}_1\, <\, -1\, <\, v \, <\, 1 \, <\, \tilde{v}_2. \nonumber
\end{eqnarray}
Note that the limit $u\to \infty$ in both cases is nothing more than $U\to 1$, and thus the map (\ref{cambiog}) applies two unbounded curves to a finite one that crosses the
equator of $S^2$. The Northern pieces present the caustic: $u=u_2$, whereas the Southern ones are limited by the ``spherical
ellipse": $u=\tilde{u}_2$. The motion is confined between these curves in a planetary way and can be seen as the natural continuation of the $t_p$ orbits in $S_+^2$ with $\Omega<0$. See Figure \ref{graficas1}d.

\item Lemniscatic orbits ($t_l$): Analogously, there are two parts:
\begin{eqnarray}
S_+^2:&&\qquad u_1\, < \, -1\, <\, u_2\, <\, 1\, <\, u\  ,\qquad  v_1\, <\, -1\, <\, v \, <\, 1 \, <\, v_2 \label{tlHS} \\
S_-^2: &&\qquad -1 \, < \, \tilde{u}_1\, <\, 1\, <\, \tilde{u}_2\, <\, u\  ,\qquad  \tilde{v}_1\, <\, -1\, <\, v \, <\, 1 \, <\, \tilde{v}_2 \nonumber
\end{eqnarray}
in such a way that there are no caustics in $S_+^2$ and one in $S_-^2$: $u=\tilde{u}_2$. We find again a natural resemblance between these orbits and their partners in the $\Omega<0$ case. See Figure \ref{graficas1}e.

\item Satellitary orbits ($t_{s'}$):
\begin{eqnarray}
S_+^2: && \qquad u_1\, < \, -1\, <\, u_2\, <\, 1\, <\, u\  ,\qquad  -1\, <\, v_1\, <\, v \, <\, 1 \, <\, v_2 \label{tsprHS} \\
S_-^2: &&\qquad -1 \, < \, \tilde{u}_1\, <\, 1\, <\, \tilde{u}_2\, <\, u\  ,\qquad  -1\, <\, \tilde{v}_1\, <\, v \, <\, 1 \, <\, \tilde{v}_2. \nonumber
\end{eqnarray}
The caustics are now: $u=\tilde{u}_2$ in $S_-^2$, and $v=v_1=\tilde{v}_1$ in the two hemispheres. See Figure \ref{graficas1}f.

\item Dual Satellitary orbits ($t_{ds}$):
\begin{eqnarray}
S_+^2:&&\qquad u_1\, < \, -1\, <\, u_2\, <\, 1\, <\, u\  ,\qquad  -1\, <\, v_1\, <\, v \, <\, v_2 \, <\, 1 \label{tdsHS}\\
S_-^2: && \qquad -1 \, < \, \tilde{u}_1\, <\, 1\, <\, \tilde{u}_2\, <\, u\  ,\qquad  -1\, <\, \tilde{v}_1\, <\, v \, <\, \tilde{v}_2 \, <\, 1. \nonumber
\end{eqnarray}
The $t_{ds}$ orbits present a behavior delimited by the two caustics: $v=v_1=\tilde{v}_1$ and $v=v_2=\tilde{v}_2$ in $S^2$,
and: $u=\tilde{u}_2$ in the Southern hemisphere. Thus, the orbits pass between the two centers in $S_+^2$, but do not reach the
South Pole. See Figure \ref{graficas1}h.

\item Meridian Planetary orbits ($t_{mp}$):
\begin{eqnarray}
S_+^2:&&\qquad -1\, < \, u_1\, <\, u_2\, <\, 1\, <\, u\  ,\qquad  -1\, <\, v_1\, <\, v \, <\, v_2 \, <\, 1 \label{tmpHS}\\
S_-^2: &&\qquad -1 \, < \, \tilde{u}_1\, <\, \tilde{u}_2\, <\, 1\, <\, u\  ,\qquad  -1\, <\, \tilde{v}_1\, <\, v \, <\, \tilde{v}_2 \, <\, 1. \nonumber
\end{eqnarray}
The situation is similar to the $t_{ds}$ case, but now only the two ``angular" caustics are allowable. Thus, the orbits complete the passing between the centers not only in $S_+^2$ but also in $S_-^2$. The $t_{mp}$ orbits resemble the planetary ones interchanging the surrounded centers. See Figure \ref{graficas1}i.

\end{itemize}

\noindent Finally, the analysis should be completed with the case $\Omega=0$ whose orbits lie in the $S^2_+$ hemisphere.
These can be easily described as the limit $\Omega \rightarrow 0$ in the $\Omega<0$ case. The caustic $u=u_2$ for the $t_p$, $t_l$
and $t_{s'}$ orbits becomes $u_2\rightarrow \infty$, and thus $U(u_2) \rightarrow 1$, i.e., the equator $Z=0$ of $S^2$.
Consequently, the motions are completely similar to the corresponding ones in $S_+^2$, but now bounded by the equator. %See Figures  \ref{graficas2} (a), (b) and (c).

\section{Evaluation of the quadratures, inversion of the elliptic integrals}

The resolution of the problem of two fixed centers has a long history that, apart from the original results of Euler \cite{Euler1, Euler2},
includes the works of Lagrange \cite{Lagrange}, Legendre \cite{Legendre}, Jacobi \cite{Jacobi}, Liouville \cite{Liouville1846} etc.,
see \cite{Mathuna} and references therein. In more recent times Alexeev \cite{Alexeev} has given a detailed qualitative analysis
of the planar problem. Explicit analytical expressions determining the motion along the orbits are obtained by applying standard procedures
that require the inversion of elliptic integrals, see, for instance, \cite{Byrd, Whittaker}. The quadratures solving the two pairs
of uncoupled ODEs (\ref{FOeq1b}) and (\ref{FOeq2b}) have been thoroughly discussed by several authors, see \cite{Mathuna} and
references therein, see also \cite{Demin1961}.

 \medskip

 We shall briefly report here on the processes of quadrature evaluation/elliptic integral inversion in the context of the
spherical problem, keeping in mind that the variables $(u,v)$, which appear in equations (\ref{FOeq1b}) and (\ref{FOeq2b}),
should be regarded as coordinates in $S^2_+$ and $S^2_-$ through the map transformation (\ref{cambiog}), as it has been explained
in the previous sections.

There are two distinctly different situations for the $\Omega<0$ or $\Omega>0$ ranges:

\medskip

\noindent $\bullet$ $\Omega<0$. In this case the inversion of the elliptic integrals appearing in equations (\ref{FOeq1b})
is standard, we will detail only the planetary case as an example.

The range for the $u$-variable in (\ref{FOeq1b}) (left) is: $u_1 < u <  u_2$, and thus the curves: $u=u_1$ and $u=u_2$,
$\forall v\in (-1,1)$, determine the two caustics. The quadrature solving the $u$-equation in (\ref{FOeq1b}) is
\begin{equation}
\pm\sqrt{-\frac{2\bar{\sigma}}{\sigma} \Omega}\, \zeta\, =\, I(u)-I(u_0)\  ;\quad  I(u)=\int_{u_1}^{u}\, \frac{dz}{\sqrt{(z^2-1)(z-u_1)(u_2-z)}},\label{xi}
\end{equation}
where the initial condition $u(0)=u_0$ is assumed. The elliptic integral of the first kind $I(u)$ in (\ref{xi}) can be inverted by
performing the following change of variable $z\to s$, see \cite{Byrd} case 256:
\[
z\, =\, \frac{u_1(1-u_2) +(u_2-u_1) \sn^2 s}{1-u_2 +(u_2-u_1) \sn^2 s }\quad \Rightarrow\quad I(u) =\, g_u \int_{0}^{s_u} \, ds\, =\, g_u\, s_u,
\]
where $\sn s$ denotes the Jacobi sinus function: $\sn s\equiv \sn (s|k_u^2)$, $g_u$ and the elliptic modulus $k_u$ are
defined in terms of the turning points as
\[
k_u^2=\frac{2 (u_2-u_1)}{(u_2-1) (u_1+1)} \quad ,\quad g_{u}=\frac{2}{\sqrt{(u_2-1)(u_1+1)}} \, \, \, .
\]
Formula (\ref{xi}) is thus simplified to become a linear relation between $s_u$ and the local time $\zeta$ which is easily inverted:
\begin{equation}
g_u\, \left( s_u-s_{u_0} \right) \, =\, \pm \sqrt{-\frac{2\bar{\sigma}}{\sigma} \Omega}\, \zeta\quad \Rightarrow \quad s_u(\zeta)=\frac{\pm \sqrt{-\frac{2\bar{\sigma}}{\sigma} \Omega}}{g_u}\,  \zeta+s_{u_0}\label{sols}
\end{equation}
with: $g_u s_{u_0}=I(u_0)$. Finally, recalling the last change of variable, the explicit inversion of (\ref{xi}) is achieved:
\begin{equation*}
u(\zeta) \, =\, \frac{u_1(1-u_2) +(u_2-u_1) \sn^2 s_u}{1-u_2 +(u_2-u_1) \sn^2 s_u },\label{solR1}
\end{equation*}
where $s_u$ is defined as a function of the local time $\zeta$, $s_u(\zeta)$, in equation (\ref{sols}). Alternatively, using
the properties of Jacobi elliptic functions, $u(\zeta)$ can be rewritten in terms of the Jacobi function $\dn$ in the simpler form:
\begin{equation}
u(\zeta) \, =\, \frac{u_1-1 +(u_1+1)  \dn^2 s_u}{1-u_1+(u_1+1) \dn^2 s_u} \quad , \quad -1 < 1 < u_1 < u <  u_2 \label{platc} \, .
\end{equation}
We stress, by writing the inequalities characterizing this type of orbits, that the analytic expression for $u(\zeta)$ appearing in formula
(\ref{platc}) is compelled to live inside the $(u_1,u_2)$ interval.

The companion expression for $v(\zeta)$, for instance, in the planetary case $-1\, <\, v\, <\, 1$, is given, after a completely analogous procedure, by the
expressions
\[
v(\zeta)\, =\, \frac{ 1-v_2+ 2v_2 \sn^2 s_v}{v_2- 1+2  \sn^2 s_v}
\]
with
\[
s_v(\zeta)\, =\, \frac{\pm \sqrt{-\frac{2\bar{\sigma}}{\sigma} \Omega}}{g_v} \, \zeta\, +\, s_{v_0} \  ,\    k_v^2\, =\, \frac{2
   ( v_2-v_1)}{( v_2-1)  ( 1+v_1)} \  ,\   g_v\, =\, \frac{2}{\sqrt{(v_2-1) ( 1+v_1)}}.
\]
Applying transformation (\ref{cambiog}) to these expressions of $u(\zeta)$ and $v(\zeta)$ and replacing the results in (\ref{coor1}), a
complete description in Cartesian coordinates of planetary orbits in the Northern hemisphere is obtained.

Analogously, all the integrals $I(u)$ and $I(v)$ solving equations (\ref{FOeq1b}) in the different ranges of $u$ and $v$
compatible with $\Omega<0$ can be inverted by similar techniques. The ensuing analytic expressions are assembled in the Appendix.
The $u(\zeta)$ and $v(\zeta)$ functions which, respectively, solve the $u$- and $v$-dynamics are smooth, bounded between
turning points, and periodic with periods, respectively, $T_u \propto K(k_u^2)$ and $T_v \propto K(k_v^2)$, where $K(k^2)$ is the complete elliptic function of the first kind. The trajectories  in all these cases are bounded between caustics in $S^2_+$ and dense, except if the $u$- and $v$-periods are commensurable.

\medskip

\noindent $\bullet$ $\Omega>0$. The procedure is more delicate in this case essentially because the trajectories complying
with the ODE pair (\ref{FOeq1b}) reach the equator, whereas there is admissible motion governed by (\ref{FOeq2b}) that also reaches
the equator coming from the Southern hemisphere. Therefore, it is convenient to investigate the inversion of the quadratures
of the $u$-equations of both (\ref{FOeq1b}) and (\ref{FOeq2b}) in a global form. However, in the \lq\lq angular\rq\rq $v$-integrals
there are no differences with respect to the $\Omega<0$ range.

\noindent Let us focus on planetary orbits. An orbit of this type in $S^2$ is described by two pieces: the portion belonging to $S_+^2$ is a solution of
equations (\ref{FOeq1b}) in the ranges
\[
u_1\, < \, -1\, <\, 1\, <\, u_2\, <\, u\  ,\qquad  v_1\, <\, -1\, <\, v \, <\, 1 \, <\, v_2,
\]
whereas for the $S_-^2$ piece we have equations (\ref{FOeq2b}) and the ranges
\[
\tilde{u}_1\, < \, -1\, <\, 1\, <\, \tilde{u}_2\, <\, u\  ,\qquad  \tilde{v}_1\, <\, -1\, <\, v \, <\, 1 \, <\, \tilde{v}_2.
\]
The first quadrature in (\ref{FOeq1b}) for the ``radial" variable
\begin{equation}
\pm\sqrt{\frac{2\bar{\sigma}}{\sigma}\Omega}\, \zeta \, =\, I(u)-I(u_0)\  ; \quad I(u)=\int_{u_2}^{u}\, \frac{dz}{\sqrt{(z^2-1)(z-u_1)(z-u_2)}}\label{integral} \, \, \,
\end{equation}
can be inverted with a change of variable like that explained before in the $\Omega<0$ case. The solution is
\begin{equation}
u(\zeta)\equiv u(s_u)=\frac{u_2-1 +(u_2+1) \dn^2 s_{u}}{1-u_2 +(u_2+1) \dn^2 s_{u}},\label{uPlan}
\end{equation}
where
\[
s_u(\zeta)=\,  \frac{\pm \sqrt{\frac{2\bar{\sigma}}{\sigma}\Omega}}{g_u}\,  \zeta+s_{u_0}\ ,\   k_u^2=\frac{2(u_2-u_1)}{(1-u_1)(1+u_2)} \  ,\  g_{u}=\frac{2}{\sqrt{(1-u_1)(1+u_2)}}.
\] %One should now try to solve equation (\ref{FOeq2b}) in the $\tilde{u}_2<u< \infty$ half-line. But there is a surprise.

\begin{figure}[ht]
\centerline{\includegraphics[height=4.5cm]{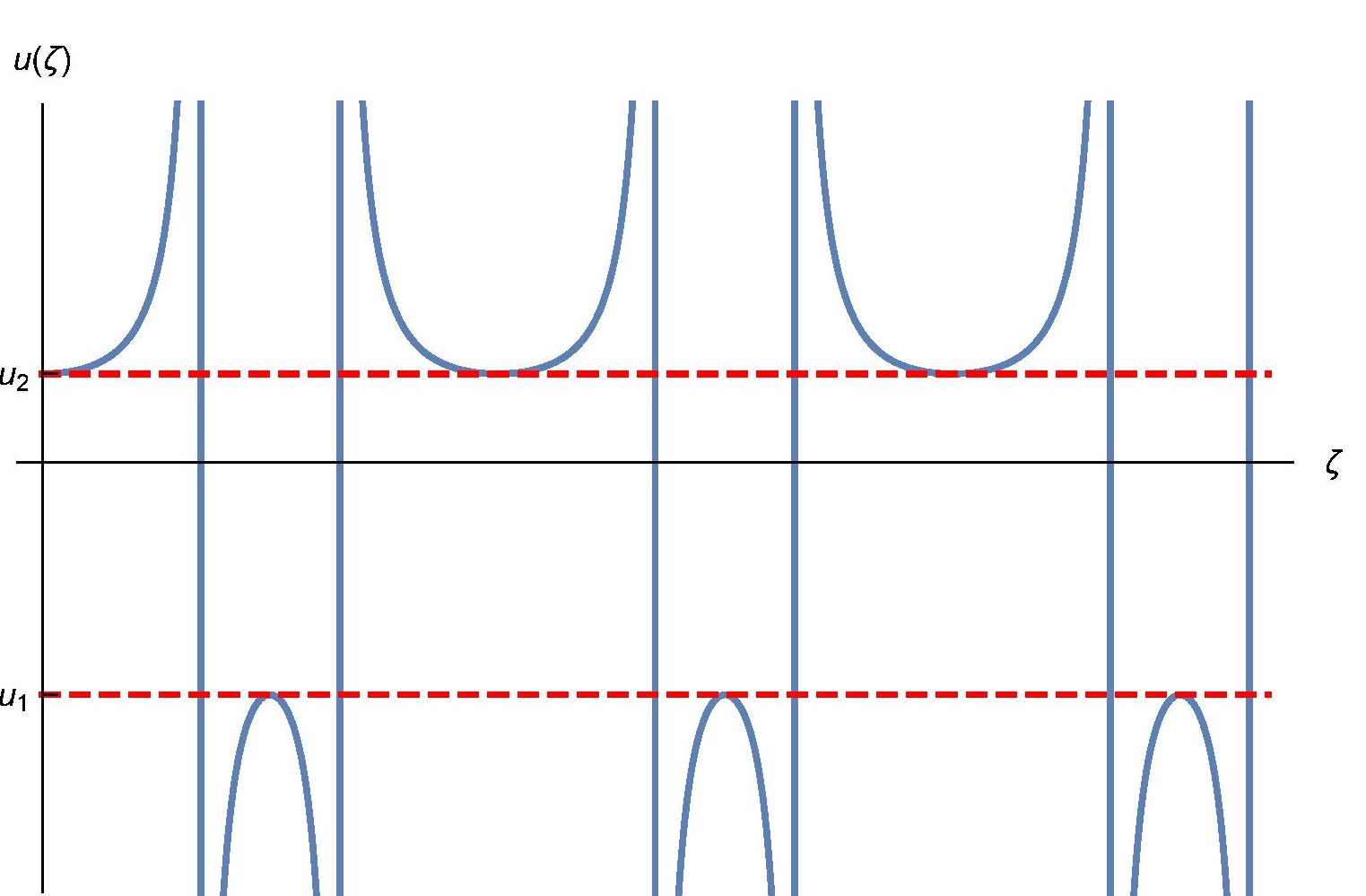}\qquad \includegraphics[height=4.5cm]{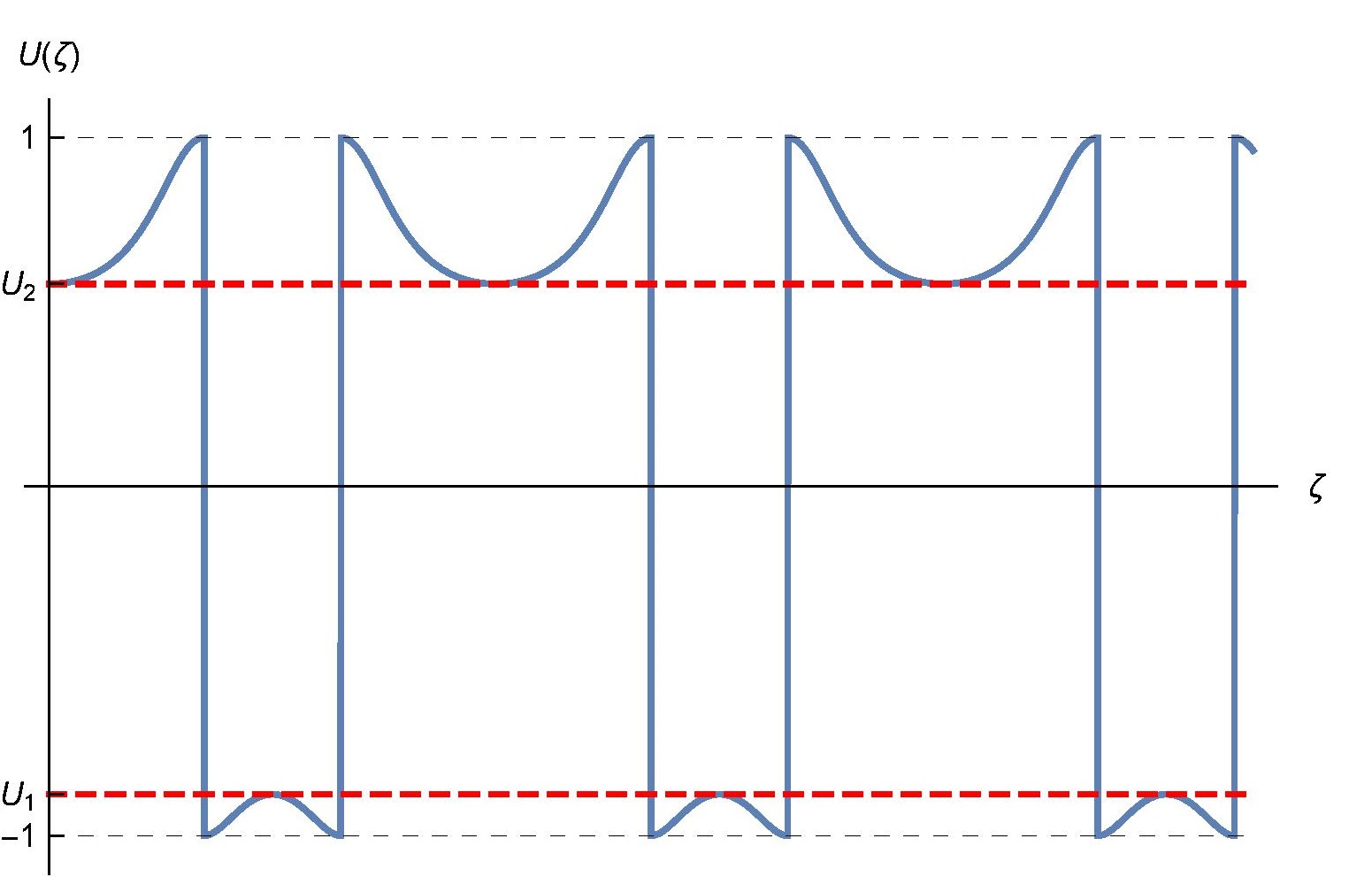}}
\caption{(a) Graphics of the function $u(\zeta)$ defined in (\ref{uPlan}) and (b) its partner $U(\zeta)$ in $S^2$, corresponding
to the values: $\Omega=\sqrt{3}$, $G=\frac{2\sqrt{3}}{3}$, $\sigma=\cos \frac{\pi}{6}$, $s_{u_0}=0$.}
\label{polos}
\end{figure}

\noindent A plot of $u(\zeta)$, see Figure \ref{polos}a, shows several relevant features of $u(\zeta)$. First, the function (\ref{uPlan})
presents infinite poles located at the points where $\dn^2 s_u=\frac{u_2-1}{u_2+1}$. This is an expected result if one
sees $u(\zeta)$ as a solution of the planar problem of two attractive centers with $h>0$ reinterpreting $\zeta$ as the local
time of this planar problem; the trajectory goes to infinity in a finite interval of the local time. However, in the sphere $S^2$
the sphero-conical variable $U(\zeta)$, given by (\ref{cambiog}), is bounded but exhibits finite discontinuities and reaches
its maxima on the equator $U=1$ at the poles of $u(\zeta)$, see Fig. \ref{polos}b. Second, it is remarkable, and a priori unexpected, that
both $u(\zeta)$ and $U(\zeta)$ take negative values. The subtle interpretation of this fact is the understanding that, given
the inversion problem posed by (\ref{integral}), its solution $u(\zeta)\equiv u(s_u)$ solves also the complementary problem:
$y< u_1 <  -1 < 1 < u_2 $, i.e., the inversion problem of the elliptic integral:
\[
I'(y)=\int_y^{u_1} \frac{dz}{\sqrt{(z^2-1)(z-u_1)(z-u_2)}}
\]
in such a way that the inverse function $y(s)$ verifies: $y(s)=u(s_u+K)$, where $K=K(k_u^2)$. Thus, $u(s_u)$ defined in
equation (\ref{uPlan}) represents simultaneously the genuine $u$-\lq\lq radial\rq\rq positive solution, $u\in (u_2,\infty)$,
and the negative $y(s)$-\lq\lq radial\rq\rq  solution with $y\in (-\infty,u_1)$. Note that, according to the plot in
Figure \ref{polos}, these two solutions occur in consecutive intervals of the local time $\zeta$.

\begin{figure}[ht]
\centerline{\includegraphics[height=4.5cm]{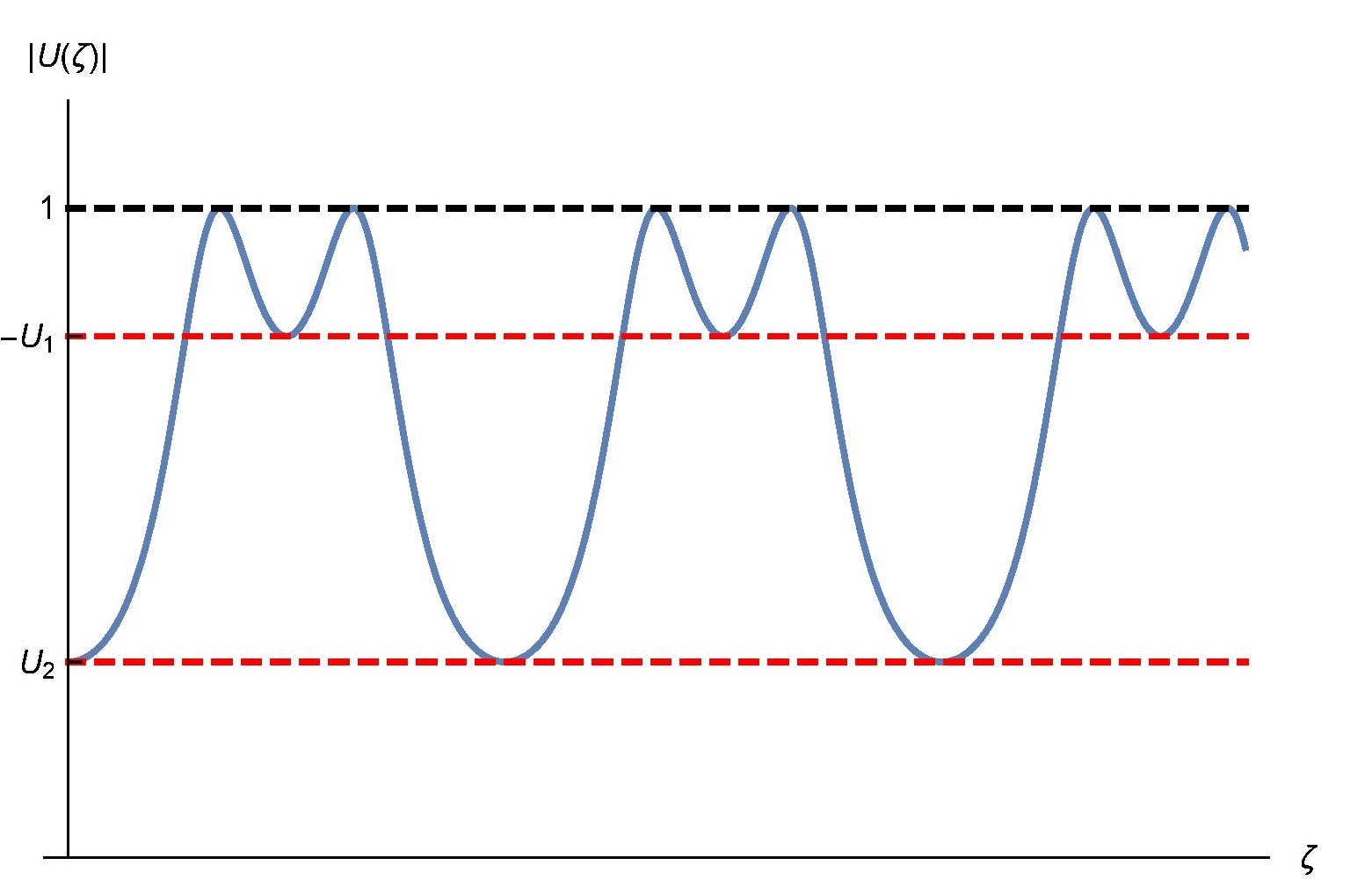}}
\caption{Graphics of the function $|U(\zeta)|$ corresponding to the values: $\Omega=\sqrt{3}$, $G=\frac{2\sqrt{3}}{3}$, $\sigma=\cos \frac{\pi}{6}$, $s_{u_0}=0$.}
\label{polos2}
\end{figure}

\noindent A direct search for the solution of equation (\ref{FOeq2b}) in $S^2_-$, where $\tilde{u}_1 <  -1 < 1 <\tilde{u}_2 < u$,
requires the inversion of the elliptic integral in the following equation:
\[
\pm\sqrt{\frac{2\bar{\sigma}}{\sigma}\Omega}\, \zeta\, =\, \tilde{I}(u)-\tilde{I}(u_0)\  ,\quad \tilde{I}(u)=\int_{\tilde{u}_2}^{u}\, \frac{dz}{\sqrt{(z^2-1)(z-\tilde{u}_1)(z-\tilde{u}_2)}} \, \, \, .
\]
Having in mind that $\tilde{u}_1=-u_2$, $\tilde{u}_2=-u_1$, we can write
\[
\tilde{I}(u) = \int_{-u_1}^{u}\, \frac{dz}{\sqrt{(z^2-1)(z+u_2)(z+u_1)}}= \int_{-u}^{u_1}\, \frac{dw}{\sqrt{(w^2-1)(w-u_2)(w-u_1)}}=I'(-u),
\]
where the change of variable $z=-w$ has been performed. Thus, we conclude that the inversion of $\tilde{I}(u)$, i.e., the \lq\lq
radial" solution in $S^2_-$, is tantamount to the inversion of $I'(-u)$ and consequently to minus the negative part of $u(s_u)$
given in (\ref{uPlan}). Therefore, we represent the \lq\lq radial" solution simultaneously in both $S^2_+$ and $S^2_-$  by simply
taking the absolute value $|u(\zeta)|$ of the solution given in (\ref{uPlan}). Moreover, with this identification the function
$|U(\zeta)|$ is smooth, i.e., the gluing at the equator of the Northern and Southern branches of the orbits is continuous and
differentiable, see Figure \ref{polos2}, with respect to the local time $\zeta$.

\medskip

\noindent This argument is valid also for the ``radial" quadratures of the rest of different types of orbits that cross the
equator. Thus, the general expression for the orbits in Cartesian coordinates over the sphere $S^2$, using (\ref{cambiog})
in (\ref{coor1}), can be written in a compact form valid for all the types of orbits described in the previous section as
\begin{eqnarray}
X(\zeta)&=&\frac{R \bar{\sigma} \, |u(\zeta)|\,  v(\zeta)}{\sqrt{\bar{\sigma}^2 u^2(\zeta) +\sigma^2}\sqrt{\bar{\sigma}^2 v^2(\zeta) +\sigma^2}}\nonumber\\
Y(\zeta)&=&\frac{\pm R \sigma \bar{\sigma} \, \sqrt{u^2(\zeta)-1}\,  \sqrt{1-v^2(\zeta)}}{\sqrt{\bar{\sigma}^2 u^2(\zeta) +\sigma^2}\sqrt{\bar{\sigma}^2 v^2(\zeta) +\sigma^2}}\label{solgeneral}\\
Z(\zeta)&=&\frac{R \sigma \, \mathrm{sg}[u(\zeta)]}{\sqrt{\bar{\sigma}^2 u^2(\zeta) +\sigma^2}\sqrt{\bar{\sigma}^2 v^2(\zeta) +\sigma^2}}\nonumber \, \, \, .
\end{eqnarray}
Here, $ \mathrm{sg}$ denotes the sign function and $(u(\zeta),v(\zeta))$ are the solutions of equations (\ref{FOeq1}) or (\ref{FOeq2}).

\noindent Explicit expressions for (\ref{solgeneral}) in all the different regimes are written in the Appendix.

\medskip

\noindent The quasiperiodicity properties of the functions (\ref{solgeneral}) are inherited from the Jacobi elliptic functions
through the functions $u(\zeta)$ and $v(\zeta)$: solutions (\ref{solgeneral}) are products of periodic functions with
different periods $T_u$ and $T_v$. Consequently, (\ref{solgeneral}) will be periodic, and thus the orbits will be closed,
only if $T_u$ and $T_v$ are commensurable, i.e., there exists $p,q \in \mathbb{N}^*$ such that
 \begin{equation}
 p\, T_{u}\, =\, q\, T_{v},\label{commensurable}
 \end{equation}
otherwise the orbits will be dense inside the allowable region of $S^2$.

The periods $T_u$ and $T_v$ are proportional to $K(k_u^2)$ and $K(k_v^2)$, respectively, with a factor that depends on the
concrete Jacobi functions involved in the respective expressions of $u(\zeta)$ and $v(\zeta)$. The search for a closed orbit,
with the values of $p$, $q$ and $\Omega$ (or $G$) fixed, requires that the transcendental equation (\ref{commensurable}) be solved in the
variable $G$ (alternatively $\Omega$). Explicit expressions for the periods and concrete examples of closed orbits for
different values of $p$ and $q$ are collected in the Appendix.

\medskip

\subsection*{Acknowledgments}

The authors thank the Spanish Ministerio de Econom\'{\i}a y Competitividad (MINECO) for financial support under grant MTM2014-57129-C2-1-P and the Junta de Castilla y Le\'on grant VA057U16.

%\clearpage

\section*{Appendix. Explicit expressions for different types of orbits}

The set of parameters that determines the problem is $R$, $\theta_f$, $\gamma_1$ and $\gamma_2$, but after defining
nondimensional variables the strengths can be measured with only one relative quantity: $\gamma=\frac{\gamma_2}{\gamma_1+\gamma_2}$.

\noindent Our choice of integration constants to characterize the solutions (\ref{solgeneral}) as functions of the nondimensional
local time $\zeta$ introduced in (\ref{time}) is: the two constants of motion $\Omega$ and $G$ and the two initial data $s_{u_0}$
and $s_{v_0}$. The dependence in $\Omega$ and $G$ is given implicitly through the values of the branching points:
\begin{eqnarray*}
u_1&=&\frac{\sigma}{\bar{\sigma}} \left[\frac{-1}{2 \Omega}- \sqrt{\frac{G}{\Omega}+\frac{1}{4\Omega^2}}\right]\qquad \quad  ,\quad  \qquad u_2=\frac{\sigma}{\bar{\sigma}} \left[\frac{-1}{2 \Omega}+ \sqrt{\frac{G}{\Omega}+\frac{1}{4\Omega^2}}\right]\nonumber \\ && \\
v_1&=&\frac{\sigma}{\bar{\sigma}} \left[\frac{(1-2\gamma)}{2\Omega}-\sqrt{\frac{G}{\Omega}+\frac{(1-2\gamma)^2}{4\Omega^2}}\right]\  , \   v_2=\frac{\sigma}{\bar{\sigma}} \left[\frac{(1-2\gamma)}{2\Omega}+\sqrt{\frac{G}{\Omega}+\frac{(1-2\gamma)^2}{4\Omega^2}}\right]\nonumber
\end{eqnarray*}
if $\Omega \neq 0$, and $u_1=\frac{\sigma}{\bar{\sigma}}G$, $v_2=\frac{\sigma}{\bar{\sigma}}\, \frac{-G}{(1-2\gamma)}$ for the $\Omega=0$ case.

\noindent Remember also that the following notation has been introduced throughout the paper:
\[
\sigma=\cos \theta_f\  ,\quad \bar{\sigma}=\sin \theta_f\  ;\quad \sn s_u =\sn (s_u(\zeta)|k_u^2),
\]
and so on for the rest of the Jacobi elliptic functions, where
\begin{eqnarray*}
&& s_u \equiv s_u(\zeta) =\frac{ \pm \sqrt{\frac{2 \bar{\sigma}}{\sigma}|\Omega|}}{g_u} \, \zeta + s_{u_0} \  , \quad
s_v \equiv s_v(\zeta) =\frac{\pm \sqrt{\frac{2 \bar{\sigma}}{\sigma}|\Omega|}}{g_v} \, \zeta + s_{v_0}\qquad {\rm if} \  \Omega\neq 0\\
&& s_u \equiv s_u(\zeta) =\frac{ \pm \sqrt{2}}{g_u} \, \zeta + s_{u_0} \  , \qquad
s_v \equiv s_v(\zeta) =\frac{\pm \sqrt{2}}{g_v} \, \zeta + s_{v_0} \qquad {\rm if} \  \Omega =0
\end{eqnarray*}
in such a way that the initial conditions are $s_{u_0}= s_u(0)$ and $s_{v_0}= s_v(0)$.

\noindent With all these considerations, the orbits for the two fixed centers problem in $S^2$
are:

\medskip

\noindent $\hrulefill$

\centerline{ $\Omega > 0$: Orbits that cross the equator.}
\noindent $\hrulefill$

\noindent $\bullet$   Planetary orbits-$t_p$, see (\ref{tpHS}):
\begin{equation}
\left\{ \begin{array}{l}
X(\zeta) = \frac{R}{\Upsilon_u \Upsilon_v}\, \bar{\sigma}\, (1-u_2 - (u_2+1)\, {\rm dn}^2 s_u)\, (1 - v_1 + 2 v_1\, {\rm sn}^2 s_v) \\ \\
Y(\zeta) = \frac{R}{\Upsilon_u \Upsilon_v}\, 4 \sigma \bar{\sigma}\, \sqrt{u_2^2-1}\, \sqrt{v_1^2-1}\  {\rm dn} s_u\, {\rm sn} s_v \, {\rm cn} s_v\\ \\
Z(\zeta) = \frac{R}{\Upsilon_u \Upsilon_v}\, \sigma \, (u_2-1-(u_2 + 1)\, {\rm dn}^2 s_u)\, (v_1-1 + 2\, {\rm sn}^2 s_v)
\end{array} \right.
\end{equation}
where
\begin{eqnarray*}
\Upsilon_u &=& \sqrt{(u_2-1)^2 - 2 (u_2^2-1) (\sigma^2-\bar{\sigma}^2)\, {\rm dn}^2 s_u + (u_2+1)^2\, {\rm dn}^4 s_u} \\
  \Upsilon_v &=& \sqrt{(v_1-1)^2 + 4 (1-v_1)(\bar{\sigma}^2 v_1- \sigma^2)\, {\rm sn}^2 s_v + 4 (\bar{\sigma}^2 v_1^2+\sigma^2) \, {\rm sn}^4 s_v}
\end{eqnarray*}

\begin{equation*}
k_u^2 = \frac{2(u_2-u_1)}{(1-u_1)(1+u_2)}  \ ,\  g_u = \frac{2}{ \sqrt{(1-u_1)(1+u_2)}}  \ , \
k_v^2 = \frac{2(v_2-v_1)}{(1-v_1)(1+v_2)} \ ,\  g_v = \frac{2}{ \sqrt{(1-v_1)(1+v_2)}}
\end{equation*}

%\begin{eqnarray*}
%k_u^2 &=& \frac{2(u_2-u_1)}{(1-u_1)(1+u_2)}  \quad ,\quad g_u = \frac{2}{ \sqrt{(1-u_1)(1+u_2)}}  \\
%k_v^2 &=& \frac{2(v_2-v_1)}{(1-v_1)(1+v_2)} \quad ,\quad  g_v = \frac{2}{ \sqrt{(1-v_1)(1+v_2)}}
%\end{eqnarray*}

\noindent $\bullet$   Lemniscatic orbits-$t_l$ (\ref{tlHS}):
\begin{equation}
\left\{ \begin{array}{l}
X(\zeta) = \frac{R}{\Upsilon_u \Upsilon_v}\, \bar{\sigma}\, (u_2-1 - 2 u_2\, {\rm dn}^2 s_u)\, (1 - v_1 + 2 v_1\, {\rm sn}^2 s_v) \\ \\
Y(\zeta) = \frac{R}{\Upsilon_u \Upsilon_v}\, 4 \sigma \bar{\sigma}\, k_u\,  \sqrt{1-u_2^2}\, \sqrt{v_1^2-1}\  {\rm dn} s_u\, {\rm sn} s_u\, {\rm sn} s_v \, {\rm cn} s_v\\ \\
Z(\zeta) = \frac{R}{\Upsilon_u \Upsilon_v}\, \sigma \, (1-u_2- 2\, {\rm dn}^2 s_u)\, (v_1-1 + 2\, {\rm sn}^2 s_v)
\end{array} \right.
\end{equation}

\begin{eqnarray*}
\Upsilon_u &=& \sqrt{(u_2-1)^2 + 4(1-u_2)(\bar{\sigma}^2 u_2-\sigma^2)\, {\rm dn}^2 s_u + 4 (\bar{\sigma}^2 u_2^2+\sigma^2)\, {\rm dn}^4 s_u} \\
\Upsilon_v &=& \sqrt{(v_1-1)^2 + 4 (1-v_1)(\bar{\sigma}^2 v_1- \sigma^2)\, {\rm sn}^2 s_v + 4 (\bar{\sigma}^2 v_1^2+\sigma^2) \, {\rm sn}^4 s_v}
\end{eqnarray*}

\begin{equation*}
k_u^2 = \frac{(1-u_1)(1+u_2)}{2(u_2-u_1)} \ ,\  g_u = \frac{\sqrt{2}}{ \sqrt{u_2-u_1}} \ , \
k_v^2 = \frac{2(v_2-v_1)}{(1-v_1)(1+v_2)}  \ , \  g_v = \frac{2}{ \sqrt{(1-v_1)(1+v_2)}}
\end{equation*}

%\begin{eqnarray*}
%k_u^2 &=& \frac{(1-u_1)(1+u_2)}{2(u_2-u_1)} \quad  , \quad  g_u = \frac{\sqrt{2}}{ \sqrt{(u_2-u_1)}} \\
%k_v^2 &=& \frac{2(v_2-v_1)}{(1-v_1)(1+v_2)}  \quad ,  \quad  g_v = \frac{2}{ \sqrt{(1-v_1)(1+v_2)}}
%\end{eqnarray*}

\noindent $\bullet$   Satellitary orbits-$t_{s'}$ (\ref{tsprHS}):
\begin{equation}
\left\{ \begin{array}{l}
X(\zeta) = \frac{R}{\Upsilon_u \Upsilon_v}\, \bar{\sigma}\, (1- u_2 + 2 u_2\, {\rm dn}^2 s_u)\, ( 2 v_1 +(1-v_1)\, {\rm sn}^2 s_v) \\ \\
Y(\zeta) = \frac{R}{\Upsilon_u \Upsilon_v}\, 4 \sigma \bar{\sigma}\, k_u\,  \sqrt{1-u_2^2}\, \sqrt{1-v_1^2}\  {\rm dn} s_u\, {\rm sn} s_u\, {\rm cn} s_v\\ \\
Z(\zeta) = \frac{R}{\Upsilon_u \Upsilon_v}\, \sigma \, (u_2-1 + 2\, {\rm dn}^2 s_u)\, (2 -(1-v_1)\, {\rm sn}^2 s_v)
\end{array} \right.
\end{equation}

\begin{eqnarray*}
\Upsilon_u &=& \sqrt{(u_2-1)^2 + 4(1-u_2)(\bar{\sigma}^2 u_2-\sigma^2)\, {\rm dn}^2 s_u + 4 (\bar{\sigma}^2 u_2^2+\sigma^2)\, {\rm dn}^4 s_u} \\
\Upsilon_v &=& \sqrt{ 4 (\bar{\sigma}^2 v_1^2+\sigma^2) + 4 (1-v_1)(\bar{\sigma}^2 v_1- \sigma^2)\, {\rm sn}^2 s_v +  (v_1-1)^2 \, {\rm sn}^4 s_v}
\end{eqnarray*}

\begin{equation*}
k_u^2 = \frac{(1-u_1)(1+u_2)}{2(u_2-u_1)}  \ , \  g_u = \frac{\sqrt{2}}{ \sqrt{u_2-u_1}} \ , \
k_v^2 = \frac{(1-v_1)(1+v_2)}{2(v_2-v_1)}    \ , \  g_v = \frac{\sqrt{2}}{ \sqrt{v_2-v_1}}
\end{equation*}

%\begin{eqnarray*}
%k_u^2 &=& \frac{(1-u_1)(1+u_2)}{2(u_2-u_1)}  \quad , \quad  g_u = \frac{\sqrt{2}}{ \sqrt{(u_2-u_1)}} \\
%k_v^2 &=& \frac{(1-v_1)(1+v_2)}{2(v_2-v_1)}    \quad ,  \quad  g_v = \frac{\sqrt{2}}{ \sqrt{(v_2-v_1)}}
%\end{eqnarray*}

\noindent $\bullet$   Dual Satellitary orbits-$t_{ds}$ (\ref{tdsHS}):
\begin{equation}
\left\{ \begin{array}{l}
X(\zeta) = \frac{R}{\Upsilon_u \Upsilon_v}\, \bar{\sigma}\, (1-u_2+ 2 u_2\, {\rm dn}^2 s_u)\, (1+ v_1-(1-v_1)\, {\rm dn}^2 s_v) \\ \\
Y(\zeta) = \frac{R}{\Upsilon_u \Upsilon_v}\, 4 \sigma \bar{\sigma}\, k_u\,  \sqrt{1-u_2^2}\, \sqrt{1-v_1^2}\  {\rm dn} s_u\, {\rm sn} s_u\,  {\rm dn} s_v\\ \\
Z(\zeta) = \frac{R}{\Upsilon_u \Upsilon_v}\, \sigma \, (u_2-1 + 2\, {\rm dn}^2 s_u)\, (1+v_1 +(1-v_1)\, {\rm dn}^2 s_v)
\end{array} \right.
\end{equation}

\begin{eqnarray*}
\Upsilon_u &=& \sqrt{(u_2-1)^2 + 4 (1-u_2)(\bar{\sigma}^2 u_2-\sigma^2)\, {\rm dn}^2 s_u + 4 (\bar{\sigma}^2 u_2^2+\sigma^2)\, {\rm dn}^4 s_u} \\
\Upsilon_v &=& \sqrt{(1+v_1)^2+ 2 (1-v_1^2)(\sigma^2- \bar{\sigma}^2 )\, {\rm dn}^2 s_v +  (1-v_1)^2 \, {\rm dn}^4 s_v}
\end{eqnarray*}

\begin{equation*}
 k_u^2 = \frac{(1-u_1)(1+u_2)}{2(u_2-u_1)} \ , \  g_u = \frac{\sqrt{2}}{ \sqrt{u_2-u_1}}  \ , \
 k_v^2 = \frac{2(v_2-v_1)}{(1-v_1)(1+v_2)}  \ , \ g_v = \frac{2}{ \sqrt{(1-v_1)(1+v_2)}}
\end{equation*}

%\begin{eqnarray*}
% k_u^2 &=& \frac{(1-u_1)(1+u_2)}{2(u_2-u_1)} \quad , \quad g_u = \frac{\sqrt{2}}{ \sqrt{(u_2-u_1)}}  \\
% k_v^2 &=& \frac{2(v_2-v_1)}{(1-v_1)(1+v_2)}  \quad , \quad g_v = \frac{2}{ \sqrt{((1-v_1)(1+v_2)}}
%\end{eqnarray*}

\noindent $\bullet$   Meridian Planetary orbits-$t_{mp}$ (\ref{tmpHS}):
\begin{equation}
\left\{ \begin{array}{l}
X(\zeta) = \frac{R}{\Upsilon_u \Upsilon_v}\, \bar{\sigma}\, (u_2+ 1- 2 u_2\, {\rm sn}^2 s_u)\, ( 1+v_1-(1-v_1)\, {\rm dn}^2 s_v) \\ \\
Y(\zeta) = \frac{R}{\Upsilon_u \Upsilon_v}\, 4 \sigma \bar{\sigma}\,  \sqrt{1-u_2^2}\, \sqrt{1-v_1^2}\     {\rm cn} s_u\, {\rm sn} s_u\,  {\rm dn} s_v\\ \\
Z(\zeta) = \frac{R}{\Upsilon_u \Upsilon_v}\, \sigma \, (1+u_2- 2\, {\rm sn}^2 s_u)\, (1+v_1 +(1-v_1)\, {\rm dn}^2 s_v)
\end{array} \right.
\end{equation}

\begin{eqnarray*}
\Upsilon_u &=& \sqrt{(1+u_2)^2 - 4 (1+u_2)(\bar{\sigma}^2u_2+\sigma^2)\, {\rm sn}^2 s_u + 4 (\bar{\sigma}^2u_2^2+\sigma^2)\, {\rm sn}^4 s_u} \\
\Upsilon_v &=& \sqrt{(1+v_1)^2+ 2 (1-v_1^2)(\sigma^2- \bar{\sigma}^2 )\, {\rm dn}^2 s_v +  (1-v_1)^2 \, {\rm dn}^4 s_v}
\end{eqnarray*}

\begin{equation*}
k_u^2 = \frac{2(u_2-u_1)}{(1-u_1)(1+u_2)}  \ , \  g_u = \frac{2}{ \sqrt{(1-u_1)(1+u_2)}} \ , \
k_v^2 = \frac{2(v_2-v_1)}{(1-v_1)(1+v_2)}  \ , \   g_v = \frac{2}{ \sqrt{(1-v_1)(1+v_2)}}
\end{equation*}

%\begin{eqnarray*}
%k_u^2 &=& \frac{2(u_2-u_1)}{(1-u_1)(1+u_2)}  \quad , \quad  g_u = \frac{2}{ \sqrt{(1-u_1)(1+u_2)}} \\
%k_v^2 &=& \frac{2(v_2-v_1)}{(1-v_1)(1+v_2)}  \quad , \quad  g_v = \frac{2}{ \sqrt{((1-v_1)(1+v_2)}}
%\end{eqnarray*}

\noindent $\bullet$ Taking into account the Jacobi functions involved in each type of solutions, the $u$- and $v$-periods
for the different orbits with $\Omega>0$ are
\begin{equation*}
 t_p\, {\rm and}\, t_{mp}\   {\rm orbits}: \quad  T_u= \frac{g_u}{\sqrt{\frac{2\bar{\sigma}}{\sigma} \Omega}}\  2 K(k_u^2) \quad , \quad  T_v= \frac{g_v}{\sqrt{\frac{2\bar{\sigma}}{\sigma} \Omega}}\ 2 K(k_v^2)
\end{equation*}
\begin{equation*}
 t_l \, {\rm and}\, t_{ds}\ {\rm orbits}: \quad  T_u= \frac{g_u}{\sqrt{\frac{2\bar{\sigma}}{\sigma} \Omega}}\ 4 K(k_u^2) \quad , \quad  T_v= \frac{g_v}{ \sqrt{\frac{2\bar{\sigma}}{\sigma} \Omega}}\ 2 K(k_v^2)
\end{equation*}
\begin{equation*}
 t_{s'}\ {\rm orbits}: \quad  T_u= \frac{g_u}{ \sqrt{\frac{2\bar{\sigma}}{\sigma} \Omega}}\ 4 K(k_u^2) \quad , \quad  T_v= \frac{g_v}{ \sqrt{\frac{2\bar{\sigma}}{\sigma} \Omega}}\ 4 K(k_v^2)
\end{equation*}

\medskip

\noindent $\hrulefill$

\centerline{$\Omega < 0$: Orbits that lie only in the Northern hemisphere.}

\noindent $\hrulefill$

\noindent $\bullet$   Planetary orbits-$t_p$ of type $1$, (\ref{tp1HN}):
\begin{equation}
\left\{ \begin{array}{l}
X(\zeta) = \frac{R}{\Upsilon_u \Upsilon_v}\, \bar{\sigma}\, (u_1 - 1 + (u_1 + 1)\, {\rm dn}^2 s_u)\, ( |1 + v_1| \, (1-{\rm cn} s_v) - |1 - v_1|\, (1 +  {\rm cn} s_v)) \\ \\
Y(\zeta) = \frac{R}{\Upsilon_u \Upsilon_v}\, 4 \sigma \bar{\sigma}\, \sqrt{u_1^2-1}\, \sqrt{|1-v_1||1+v_1|}\  {\rm dn} s_u\, {\rm sn} s_v \\ \\
Z(\zeta) = \frac{R}{\Upsilon_u \Upsilon_v}\, \sigma \, (1-u_1 + (u_1 + 1)\, {\rm dn}^2 s_u)\, ( |1 + v_1| \, (1-{\rm cn} s_v) + |1 - v_1|\, (1 +  {\rm cn} s_v))
\end{array} \right.
\end{equation}

\begin{eqnarray*}
\Upsilon_u &=& \sqrt{(u_1-1)^2 - 2 (u_1^2-1) (\sigma^2-\bar{\sigma}^2)\, {\rm dn}^2 s_u + (u_1+1)^2\, {\rm dn}^4 s_u} \\
\Upsilon_v &=& \sqrt{|1-v_1|^2 (1+ {\rm cn} s_v)^2 + 2 |1-v_1||1+v_1| (\sigma^2- \bar{\sigma}^2)\, {\rm sn}^2 s_v + |1+v_1|^2 (1 -{\rm cn} s_v)^2}
\end{eqnarray*}

\begin{equation*}
k_u^2 = \frac{2(u_2-u_1)}{(u_1+1)(u_2-1)} \, , \,  g_u = \frac{2}{ \sqrt{(u_1+1)(u_2-1)}}  \, , \,
k_v^2 = \frac{4-(|1-v_1|-|1+v_1|)^2}{4|1-v_1||1+v_1|}  \, , \,  g_v = \frac{1}{ \sqrt{|1-v_1||1+v_1|}}
\end{equation*}

%\begin{eqnarray*}
%k_u^2 &=& \frac{2(u_2-u_1)}{(u_1+1)(u_2-1)} \quad , \quad  g_u = \frac{2}{ \sqrt{(u_1+1)(u_2-1)}}  \\
%k_v^2 &=& \frac{4-(|1-v_1|-|1+v_1|)^2}{4|1-v_1||1+v_1|}  \quad , \quad  g_v = \frac{1}{ \sqrt{|1-v_1||1+v_1|}}
%\end{eqnarray*}

\noindent $\bullet$   Planetary orbits-$t_p$ of type $2$, (\ref{tp2HN}):
\begin{equation}
\left\{ \begin{array}{l}
X(\zeta) = \frac{R}{\Upsilon_u \Upsilon_v}\, \bar{\sigma}\, (1-u_1 - (u_1 + 1)\, {\rm dn}^2 s_u)\, (1-v_2+2 v_2\, {\rm sn}^2 s_v) \\ \\
Y(\zeta) = \frac{R}{\Upsilon_u \Upsilon_v}\, 4 \sigma \bar{\sigma}\, \sqrt{u_1^2-1}\, \sqrt{v_2^2-1}\  {\rm dn} s_u\, {\rm sn} s_v \, {\rm cn} s_v \\ \\
Z(\zeta) = \frac{R}{\Upsilon_u \Upsilon_v}\, \sigma \, (-1+u_1 - (u_1 + 1)\, {\rm dn}^2 s_u)\, ( v_2-1+2\, {\rm sn}^2 s_v )
\end{array} \right.
\end{equation}

\begin{eqnarray*}
\Upsilon_u &=& \sqrt{(u_1-1)^2 - 2 (u_1^2-1) (\sigma^2-\bar{\sigma}^2)\, {\rm dn}^2 s_u + (u_1+1)^2\, {\rm dn}^4 s_u} \\
\Upsilon_v &=& \sqrt{(v_2-1)^2 + 4 (1-v_2)(\bar{\sigma}^2 v_2- \sigma^2)\, {\rm sn}^2 s_v + 4 (\bar{\sigma}^2 v_2^2+\sigma^2) \, {\rm sn}^4 s_v}
\end{eqnarray*}

\begin{equation*}
 k_u^2 = \frac{2(u_2-u_1)}{(u_1+1)(u_2-1)} \ , \  g_u = \frac{2}{ \sqrt{(u_1+1)(u_2-1)}}  \ , \
k_v^2 = \frac{2(v_2-v_1)}{(v_1+1)(v_2-1)}  \ , \  g_v = \frac{2}{ \sqrt{(v_1+1)(v_2-1)}}
\end{equation*}

%\begin{eqnarray*}
% k_u^2 &=& \frac{2(u_2-u_1)}{(u_1+1)(u_2-1)} \quad , \quad  g_u = \frac{2}{ \sqrt{(u_1+1)(u_2-1)}}  \\
%k_v^2 &=& \frac{2(v_2-v_1)}{(v_1+1)(v_2-1)}  \quad , \quad  g_v = \frac{2}{ \sqrt{(v_1+1)(v_2-1)}}
%\end{eqnarray*}

\noindent $\bullet$   Lemniscatic orbits-$t_l$ of type $1$, (\ref{tl1HN}):
\begin{equation}
\left\{ \begin{array}{l}
X(\zeta) = \frac{R}{\Upsilon_u \Upsilon_v}\, \bar{\sigma}\, (1-u_1 + 2 u_1 \, {\rm dn}^2 s_u)\, ( |1 + v_1| \, (1-{\rm cn} s_v) - |1 - v_1|\, (1 +  {\rm cn} s_v)) \\ \\
Y(\zeta) = \frac{R}{\Upsilon_u \Upsilon_v}\, 4 \sigma \bar{\sigma}\, k_u\,  \sqrt{1-u_1^2}\, \sqrt{|1-v_1||1+v_1|}\  {\rm dn} s_u\,  {\rm sn} s_u \,{\rm sn} s_v \\ \\
Z(\zeta) = \frac{R}{\Upsilon_u \Upsilon_v}\, \sigma \, (u_1-1 + 2\, {\rm dn}^2 s_u)\, ( |1 + v_1| \, (1-{\rm cn} s_v) + |1 - v_1|\, (1 +  {\rm cn} s_v))
\end{array} \right.
\end{equation}

\begin{eqnarray*}
\Upsilon_u &=& \sqrt{(u_1-1)^2 + 4 (1-u_1) (\bar{\sigma}^2 u_1-\sigma^2)\, {\rm dn}^2 s_u + 4 (\bar{\sigma}^2 u_1^2+\sigma^2)\, {\rm dn}^4 s_u} \\
\Upsilon_v &=& \sqrt{|1-v_1|^2 (1+ {\rm cn} s_v)^2 + 2 |1-v_1||1+v_1| (\sigma^2- \bar{\sigma}^2)\, {\rm sn}^2 s_v + |1+v_1|^2 (1 -{\rm cn} s_v)^2}
\end{eqnarray*}

\begin{equation*}
 k_u^2 = \frac{(u_1+1)(u_2-1)}{2(u_2-u_1)}  \, , \,  g_u = \frac{\sqrt{2}}{ \sqrt{u_2-u_1}} \, , \,
 k_v^2 = \frac{4-(|1-v_1|-|1+v_1|)^2}{4|1-v_1||1+v_1|} \, , \,  g_v = \frac{1}{ \sqrt{|1-v_1||1+v_1|}}
\end{equation*}

%\begin{eqnarray*}
% k_u^2 &=& \frac{(u_1+1)(u_2-1)}{2(u_2-u_1)}  \quad  , \quad  g_u = \frac{\sqrt{2}}{ \sqrt{(u_2-u_1)}} \\
% k_v^2 &=& \frac{4-(|1-v_1|-|1+v_1|)^2}{4|1-v_1||1+v_1|} \quad , \quad  g_v = \frac{1}{ \sqrt{|1-v_1||1+v_1|}}
%\end{eqnarray*}

\noindent $\bullet$   Lemniscatic orbits-$t_l$ of type $2$, (\ref{tl2HN}):
\begin{equation}
\left\{ \begin{array}{l}
X(\zeta) = \frac{R}{\Upsilon_u \Upsilon_v}\, \bar{\sigma}\, (u_1-1 - 2 u_1 \, {\rm dn}^2 s_u)\, (1-v_2+2 v_2\, {\rm sn}^2 s_v) \\ \\
Y(\zeta) = \frac{R}{\Upsilon_u \Upsilon_v}\, 4 \sigma \bar{\sigma}\, k_u \, \sqrt{1-u_1^2}\, \sqrt{v_2^2-1}\  {\rm dn} s_u\, {\rm sn} s_u\, {\rm sn} s_v \, {\rm cn} s_v \\ \\
Z(\zeta) = \frac{R}{\Upsilon_u \Upsilon_v}\, \sigma \, (1-u_1 - 2\, {\rm dn}^2 s_u)\, ( v_2-1+2\, {\rm sn}^2 s_v )
\end{array} \right.
\end{equation}

\begin{eqnarray*}
\Upsilon_u &=& \sqrt{(u_1-1)^2 + 4 (1-u_1)(\bar{\sigma}^2 u_1- \sigma^2)\, {\rm dn}^2 s_u + 4 (\bar{\sigma}^2 u_1^2+\sigma^2) \, {\rm dn}^4 s_u} \\
\Upsilon_v &=& \sqrt{(v_2-1)^2 + 4 (1-v_2)(\bar{\sigma}^2 v_2- \sigma^2)\, {\rm sn}^2 s_v + 4 (\bar{\sigma}^2 v_2^2+\sigma^2) \, {\rm sn}^4 s_v}
\end{eqnarray*}

\begin{equation*}
k_u^2 = \frac{(u_1+1)(u_2-1)}{2(u_2-u_1)}  \ , \  g_u = \frac{\sqrt{2}}{ \sqrt{u_2-u_1}} \ , \
k_v^2 = \frac{2(v_2-v_1)}{(v_1+1)(v_2-1)} \ , \  g_v = \frac{2}{ \sqrt{(v_1+1)(v_2-1)}}
\end{equation*}

%\begin{eqnarray*}
%k_u^2 &=& \frac{(u_1+1)(u_2-1)}{2(u_2-u_1)}  \quad , \quad  g_u = \frac{\sqrt{2}}{ \sqrt{(u_2-u_1)}} \\
%k_v^2 &=& \frac{2(v_2-v_1)}{(v_1+1)(v_2-1)} \quad , \quad  g_v = \frac{2}{ \sqrt{(v_1+1)(v_2-1)}}
%\end{eqnarray*}

\noindent $\bullet$   Satellitary orbits-$t_{s}$ in zone $1$, (\ref{ts1HN}):
\begin{equation}
\left\{ \begin{array}{l}
X(\zeta) = \frac{R}{\Upsilon_u \Upsilon_v}\, \bar{\sigma}\, (1-u_1 + 2 u_1 \, {\rm dn}^2 s_u)\, ( v_2 (1-v_1) + v_1 (v_2-1) \, {\rm sn}^2 s_v)) \\ \\
Y(\zeta) = \frac{R}{\Upsilon_u \Upsilon_v}\, 2 \sigma \bar{\sigma}\, k_u\,  \sqrt{1-u_1^2}\, \sqrt{1-v_2^2}\, (1-v_1)\,  {\rm dn} s_u\,  {\rm sn} s_u\, {\rm dn} s_v \,{\rm cn} s_v \\ \\
Z(\zeta) = \frac{R}{\Upsilon_u \Upsilon_v}\, \sigma \, (u_1-1 + 2\, {\rm dn}^2 s_u)\, ( 1-v_1-(1-v_2) \, {\rm sn}^2 s_v))
\end{array} \right.
\end{equation}

\begin{eqnarray*}
\Upsilon_u &=& \sqrt{(u_1-1)^2 + 4 (1-u_1) (\bar{\sigma}^2 u_1-\sigma^2)\, {\rm dn}^2 s_u + 4 (\bar{\sigma}^2 u_1^2+\sigma^2)\, {\rm dn}^4 s_u} \\
\Upsilon_v &=& \sqrt{(v_1-1)^2 (\bar{\sigma}^2 v_2^2 +\sigma^2) - 2 (1-v_1)(1-v_2)(\bar{\sigma}^2 v_1v_2 +\sigma^2) \, {\rm sn}^2 s_v + (v_2-1)^2  (\bar{\sigma}^2 v_1^2 +\sigma^2)\, {\rm sn}^4 s_v}
\end{eqnarray*}

\begin{equation*}
k_u^2 = \frac{(u_1+1)(u_2-1)}{2(u_2-u_1)}  \ , \  g_u = \frac{\sqrt{2}}{ \sqrt{u_2-u_1}}  \ , \
k_v^2 = \frac{(1+v_1)(1-v_2)}{(1-v_1)(1+v_2)} \ , \  g_v = \frac{2}{ \sqrt{(1-v_1)(1+v_2)}}
\end{equation*}

%\begin{eqnarray*}
%k_u^2 &=& \frac{(u_1+1)(u_2-1)}{2(u_2-u_1)}  \quad , \quad  g_u = \frac{\sqrt{2}}{ \sqrt{(u_2-u_1)}}  \\
%k_v^2 &=& \frac{(1+v_1)(1-v_2)}{(1-v_1)(1+v_2)} \quad ,\quad  g_v = \frac{2}{ \sqrt{(1-v_1)(1+v_2)}}
%\end{eqnarray*}

\noindent $\bullet$   Satellitary orbits-$t_{s}$ in zone $2$, (\ref{ts2HN}):
\begin{equation}
\left\{ \begin{array}{l}
X(\zeta) = \frac{R}{\Upsilon_u \Upsilon_v}\, \bar{\sigma}\, (1-u_1 + 2 u_1 \, {\rm dn}^2 s_u)\, ( 2 v_2 - (1+v_2) \, {\rm dn}^2 s_v)) \\ \\
Y(\zeta) = \frac{R}{\Upsilon_u \Upsilon_v}\, 4 \sigma \bar{\sigma}\, k_u\, k_v\, \sqrt{1-u_1^2}\, \sqrt{1-v_2^2}\, {\rm dn} s_u\,  {\rm sn} s_u\, {\rm sn} s_v \\ \\
Z(\zeta) = \frac{R}{\Upsilon_u \Upsilon_v}\, \sigma \, (u_1-1 + 2\, {\rm dn}^2 s_u)\, ( 2-(1+v_2) \, {\rm dn}^2 s_v))
\end{array} \right.
\end{equation}

\begin{eqnarray*}
\Upsilon_u &=& \sqrt{(u_1-1)^2 + 4 (1-u_1) (\bar{\sigma}^2 u_1-\sigma^2)\, {\rm dn}^2 s_u + 4 (\bar{\sigma}^2 u_1^2+\sigma^2)\, {\rm dn}^4 s_u} \\
\Upsilon_v &=& \sqrt{4 (\bar{\sigma}^2 v_2^2 +\sigma^2) - 4(1+v_2)(\bar{\sigma}^2 v_2 +\sigma^2) \, {\rm dn}^2 s_v + (1+v_2)^2  \, {\rm dn}^4 s_v}
\end{eqnarray*}

\begin{equation*}
k_u^2 = \frac{(u_1+1)(u_2-1)}{2(u_2-u_1)}  \ , \  g_u = \frac{\sqrt{2}}{ \sqrt{u_2-u_1}}  \ , \
k_v^2 = \frac{(1+v_1)(1-v_2)}{(1-v_1)(1+v_2)}  \ , \  g_v = \frac{2}{ \sqrt{(1-v_1)(1+v_2)}}
\end{equation*}

%\begin{eqnarray*}
%k_u^2 &=& \frac{(u_1+1)(u_2-1)}{2(u_2-u_1)}  \quad , \quad  g_u = \frac{\sqrt{2}}{ \sqrt{(u_2-u_1)}}  \\
%k_v^2 &=& \frac{(1+v_1)(1-v_2)}{(1-v_1)(1+v_2)}  \quad , \quad  g_v = \frac{2}{ \sqrt{(1-v_1)(1+v_2)}}
%\end{eqnarray*}

\noindent $\bullet$   Satellitary orbits-$t_{s'}$ (\ref{tspHN}):
\begin{equation}
\left\{ \begin{array}{l}
X(\zeta) = \frac{R}{\Upsilon_u \Upsilon_v}\, \bar{\sigma}\, (1-u_1 + 2 u_1 \, {\rm dn}^2 s_u)\, ( 2 v_2 + (1-v_2) \, {\rm sn}^2 s_v)) \\ \\
Y(\zeta) = \frac{R}{\Upsilon_u \Upsilon_v}\, 4 \sigma \bar{\sigma}\, k_u\,  \sqrt{1-u_1^2}\, \sqrt{1-v_2^2}\  {\rm dn} s_u\,  {\rm sn} s_u \,{\rm cn} s_v \\ \\
Z(\zeta) = \frac{R}{\Upsilon_u \Upsilon_v}\, \sigma \, (u_1-1 + 2\, {\rm dn}^2 s_u)\, ( 2-(1-v_2) \, {\rm sn}^2 s_v))
\end{array} \right.
\end{equation}

\begin{eqnarray*}
\Upsilon_u &=& \sqrt{(u_1-1)^2 + 4 (1-u_1) (\bar{\sigma}^2 u_1-\sigma^2)\, {\rm dn}^2 s_u + 4 (\bar{\sigma}^2 u_1^2+\sigma^2)\, {\rm dn}^4 s_u} \\
\Upsilon_v &=& \sqrt{4 (\bar{\sigma}^2 v_2^2 +\sigma^2) + 4 (1-v_2)(\bar{\sigma}^2 v_2 -\sigma^2) \, {\rm sn}^2 s_v + (v_2-1)^2 \, {\rm sn}^4 s_v}
\end{eqnarray*}

\begin{equation*}
k_u^2 = \frac{(u_1+1)(u_2-1)}{2(u_2-u_1)}  \ , \  g_u = \frac{\sqrt{2}}{ \sqrt{u_2-u_1}} \ , \
k_v^2 = \frac{(v_1+1)(v_2-1)}{2(v_2-v_1)}    \ , \   g_v =  \frac{\sqrt{2}}{ \sqrt{v_2-v_1}}
\end{equation*}

%\begin{eqnarray*}
%k_u^2 &=& \frac{(u_1+1)(u_2-1)}{2(u_2-u_1)}  \quad , \quad  g_u = \frac{\sqrt{2}}{ \sqrt{(u_2-u_1)}} \\
%k_v^2 &=& \frac{(v_1+1)(v_2-1)}{2(v_2-v_1)}   \quad ,\quad  g_v =  \frac{\sqrt{2}}{ \sqrt{(v_2-v_1)}}
%\end{eqnarray*}

\noindent $\bullet$ The $u$- and $v$- periods in the case $\Omega<0$ are

\begin{equation*}
 t_p\, (2) \  {\rm orbits}: \qquad  T_u= \frac{g_u}{\sqrt{-\frac{2\bar{\sigma}}{\sigma} \Omega}}\  2 K(k_u^2) \quad , \quad  T_v= \frac{g_v}{ \sqrt{-\frac{2\bar{\sigma}}{\sigma} \Omega}}\ 2 K(k_v^2)
\end{equation*}
\begin{equation*}
 t_p\, (1) \  {\rm orbits}: \qquad  T_u= \frac{g_u}{\sqrt{-\frac{2\bar{\sigma}}{\sigma} \Omega}}\  2 K(k_u^2) \quad , \quad  T_v= \frac{g_v}{ \sqrt{-\frac{2\bar{\sigma}}{\sigma} \Omega}}\ 4 K(k_v^2)
\end{equation*}
\begin{equation*}
t_l\, (2)\ {\rm orbits}:  T_u= \frac{g_u}{\sqrt{-\frac{2\bar{\sigma}}{\sigma} \Omega}}\ 4 K(k_u^2) \quad , \quad  T_v= \frac{g_v}{ \sqrt{-\frac{2\bar{\sigma}}{\sigma} \Omega}}\ 2 K(k_v^2)
\end{equation*}
\begin{equation*}
 t_l\, (1), t_s\, (1), t_s\, (2)\  {\rm and}\, t_{s'}\ {\rm orbits}:  T_u= \frac{g_u}{\sqrt{-\frac{2\bar{\sigma}}{\sigma} \Omega}}\ 4 K(k_u^2) \quad , \quad  T_v= \frac{g_v}{ \sqrt{-\frac{2\bar{\sigma}}{\sigma} \Omega}}\ 4 K(k_v^2)
\end{equation*}

\medskip

\noindent $\hrulefill$

\centerline{ $\Omega = 0$: Orbits that lie in the Northern hemisphere bounded by the equator.}

\noindent $\hrulefill$

\noindent $\bullet$   Planetary orbits-$t_p$ in this case are \qquad  $-1 < 1 < u_1 < u \quad , \quad v_2 < -1 < v < 1$.
\begin{equation}
\left\{ \begin{array}{l}
X(\zeta) = \frac{R}{\Upsilon_u \Upsilon_v}\, \bar{\sigma}\, (u_1 - {\rm sn}^2 s_u)\, (- 1- v_2 + v_2\, {\rm dn}^2 s_v) \\ \\
Y(\zeta) = \frac{R}{\Upsilon_u \Upsilon_v}\, 2 \sigma \bar{\sigma}\,  \sqrt{u_1^2-1}\, \sqrt{\frac{1+v_2}{v_2-1}}\  {\rm dn} s_u\,  {\rm sn} s_v \,{\rm cn} s_v \\ \\
Z(\zeta) = \frac{R}{\Upsilon_u \Upsilon_v}\, \sigma \,  {\rm cn}^2 s_u\,  {\rm dn}^2 s_v
\end{array} \right.
\end{equation}

\begin{eqnarray*}
\Upsilon_u &=& \sqrt{(\bar{\sigma}^2 u_1^2 +\sigma^2) - 2 (\bar{\sigma}^2 u_1+\sigma^2)\, {\rm sn}^2 s_u + {\rm sn}^4 s_u} \\
\Upsilon_v &=& \sqrt{ \bar{\sigma}^2 (1+v_2)^2  - 2 \bar{\sigma}^2 v_2 (1+v_2) \, {\rm dn}^2 s_v +  (\bar{\sigma}^2 v_2^2 +\sigma^2)\, {\rm dn}^4 s_v}
\end{eqnarray*}

\begin{equation*}
k_u^2 = \frac{2}{(u_1+1)}  \quad , \quad  g_u = \frac{2}{\sqrt{1+u_1}} \quad , \quad
k_v^2 = \frac{2}{(1-v_2)}   \quad ,\quad  g_v =  \frac{2}{\sqrt{1-v_2}}
\end{equation*}

\noindent $\bullet$   Lemniscatic orbits-$t_l$: \qquad $-1 <  u_1 < 1 < u \quad , \quad  v_2 < -1 < v < 1$.
\begin{equation}
\left\{ \begin{array}{l}
X(\zeta) = \frac{R}{\Upsilon_u \Upsilon_v}\, \bar{\sigma}\, (1- u_1 \, {\rm sn}^2 s_u)\, (- 1- v_2 + v_2\, {\rm dn}^2 s_v) \\ \\
Y(\zeta) = \frac{R}{\Upsilon_u \Upsilon_v}\, 2 \sqrt{2}\, \sigma \bar{\sigma}\,  \sqrt{1-u_1}\, \sqrt{\frac{1+v_2}{v_2-1}}\  {\rm dn} s_u \,  {\rm sn} s_u \,  {\rm sn} s_v \,{\rm cn} s_v \\ \\
Z(\zeta) = \frac{R}{\Upsilon_u \Upsilon_v}\, \sigma \,  {\rm cn}^2 s_u\,  {\rm dn}^2 s_v
\end{array} \right.
\end{equation}

\begin{eqnarray*}
\Upsilon_u &=& \sqrt{1 - 2 (\bar{\sigma}^2 u_1+\sigma^2)\, {\rm sn}^2 s_u + (\bar{\sigma}^2 u_1^2+\sigma^2)\, {\rm sn}^4 s_u}  \\
\Upsilon_v &=& \sqrt{ \bar{\sigma}^2 (1+v_2)^2  - 2 \bar{\sigma}^2 v_2 (1+v_2) \, {\rm dn}^2 s_v +  (\bar{\sigma}^2 v_2^2 +\sigma^2)\, {\rm dn}^4 s_v}
\end{eqnarray*}

\begin{equation*}
k_u^2 = \frac{(u_1+1)}{2}  \quad , \quad  g_u = \sqrt{2}  \quad , \quad
k_v^2 = \frac{2}{(1-v_2)}   \quad ,\quad  g_v =  \frac{2}{\sqrt{1-v_2}}
\end{equation*}

\noindent $\bullet$   Satellitary orbits-$t_{s'}$: \qquad  $-1 <  u_1 < 1 < u \quad , \quad  -1 < v_2 < v < 1$.
\begin{equation}
\left\{ \begin{array}{l}
X(\zeta) = \frac{R}{\Upsilon_u \Upsilon_v}\, \bar{\sigma}\, (1-u_1 \, {\rm sn}^2 s_u)\, ( 1 + v_2 - {\rm dn}^2 s_v) \\ \\
Y(\zeta) = \frac{R}{\Upsilon_u \Upsilon_v}\, \sqrt{2} \sigma \bar{\sigma}\,  \sqrt{1-u_1}\, \sqrt{1-v_2^2}\  {\rm dn} s_u\,  {\rm sn} s_u \,{\rm cn} s_v \\ \\
Z(\zeta) = \frac{R}{\Upsilon_u \Upsilon_v}\, \sigma \,  {\rm cn}^2 s_u\,  {\rm dn}^2 s_v
\end{array} \right.
\end{equation}

\begin{eqnarray*}
\Upsilon_u &=& \sqrt{1 - 2 (\bar{\sigma}^2 u_1+\sigma^2)\, {\rm sn}^2 s_u + (\bar{\sigma}^2 u_1^2+\sigma^2)\, {\rm sn}^4 s_u} \\
\Upsilon_v &=& \sqrt{ \bar{\sigma}^2 (1+v_2)^2  - 2 \bar{\sigma}^2 (1+v_2) \, {\rm dn}^2 s_v +  {\rm dn}^4 s_v}
\end{eqnarray*}

\begin{equation*}
k_u^2 = \frac{(u_1+1)}{2}  \quad , \quad  g_u = \sqrt{2} \quad , \quad
k_v^2 = \frac{(1-v_2)}{2}   \quad ,\quad  g_v =  \sqrt{2}
\end{equation*}

\noindent $\bullet$ Finally, the $u$- and $v$-periods for these orbits with $\Omega=0$ are

\begin{equation*}
 t_p \ {\rm orbits}: \qquad \quad   T_u= \frac{g_u}{ \sqrt{2}}\  2 K(k_u^2) \quad , \quad  T_v= \frac{g_v}{\sqrt{2}}\ 2 K(k_v^2)
\end{equation*}
\begin{equation*}
 t_l \ {\rm orbits}: \qquad \quad   T_u= \frac{g_u}{ \sqrt{2}}\  4 K(k_u^2) \quad , \quad  T_v= \frac{g_v}{\sqrt{2}}\ 2 K(k_v^2)
\end{equation*}
\begin{equation*}
t_{s'}\ {\rm orbits}:\quad   T_u= \frac{g_u}{\sqrt{2}}\ 4 K(k_u^2) \quad , \quad  T_v= \frac{g_v}{\sqrt{2}}\ 4 K(k_v^2).
\end{equation*}

\medskip

\noindent Several orbits with $R=1$, $\gamma=\frac{1}{3}$ and $\theta_f=\frac{\pi}{6}$, one for each different situation, are represented in Figure \ref{graficas1}. These orbits are dense in all the cases and they are depicted in the interval $\zeta \in [0,70]$.

\medskip

\noindent We can see in Figure \ref{graficas2} several closed orbits of different types, with specification of the values of $p$, $q$ and initial data $s_{u_0}$ and $s_{v_0}$. In all the cases $p$, $q$ and the constant of motion $\Omega$ have been fixed, thus $G$ has been calculated by solving numerically equation (\ref{commensurable}).

\clearpage

\begin{center}
\begin{tabular}{ccc}
\includegraphics[height=4.3cm]{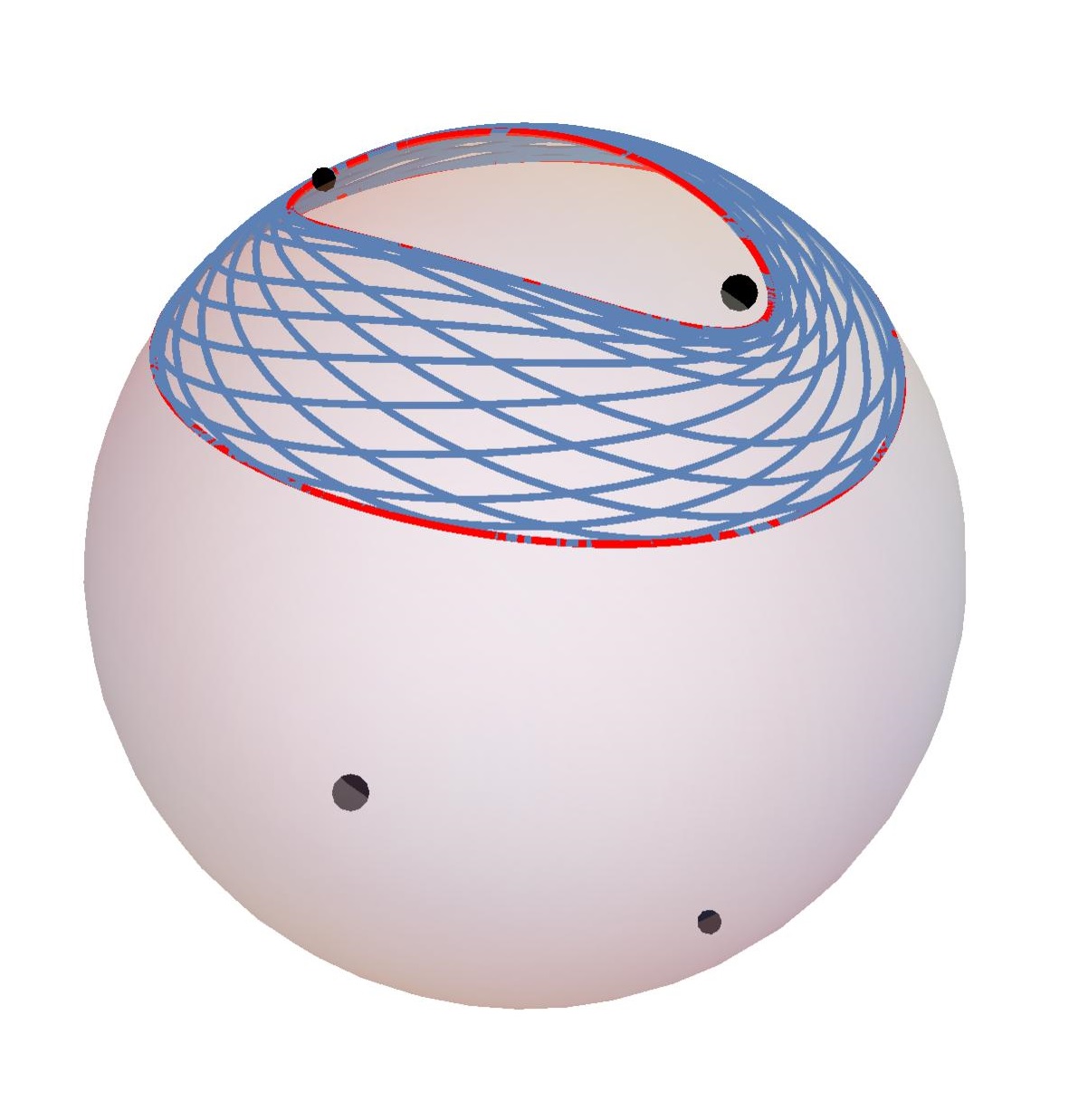}  &  \includegraphics[height=4.1cm]{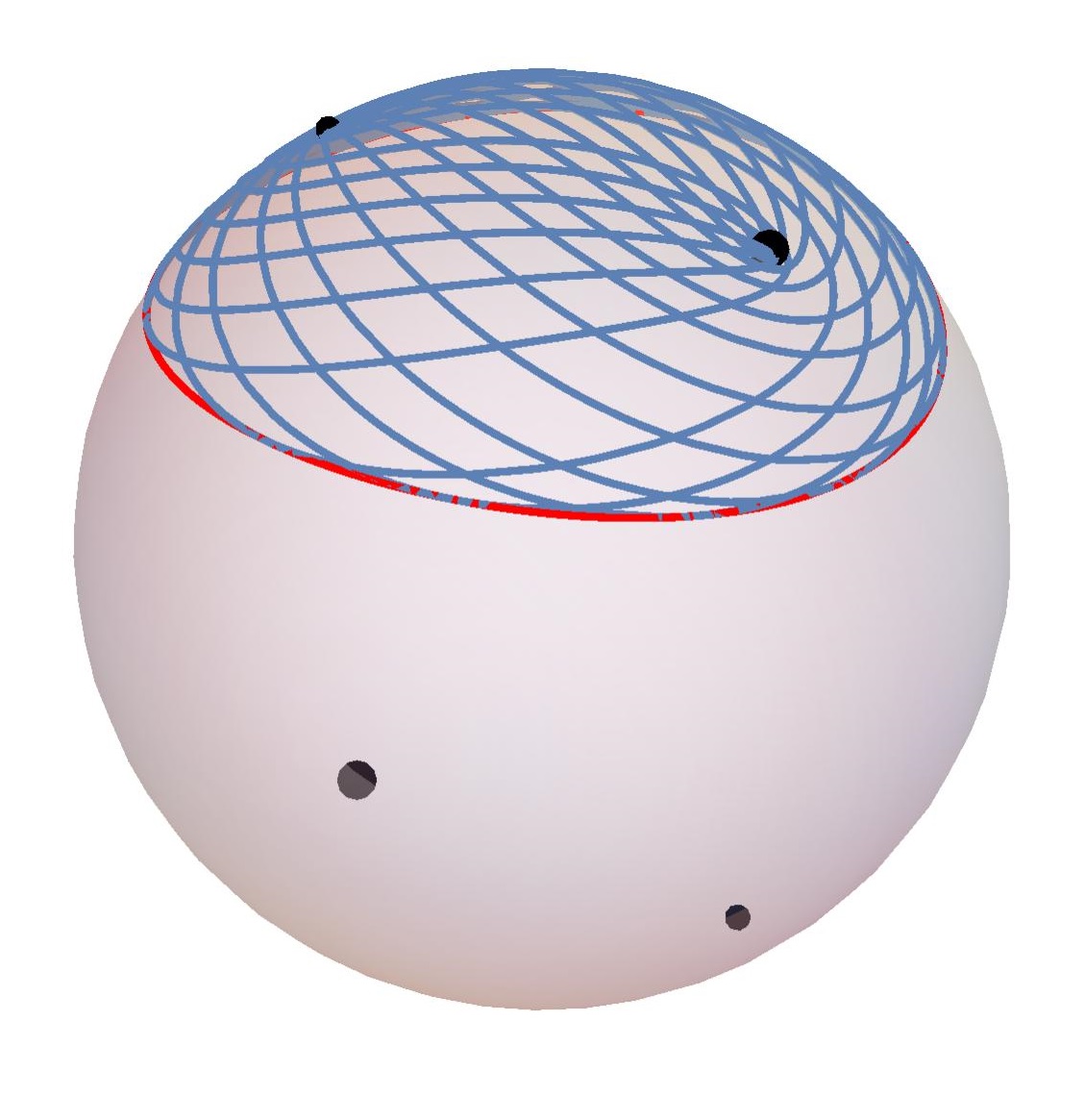}  & \includegraphics[height=3.8cm]{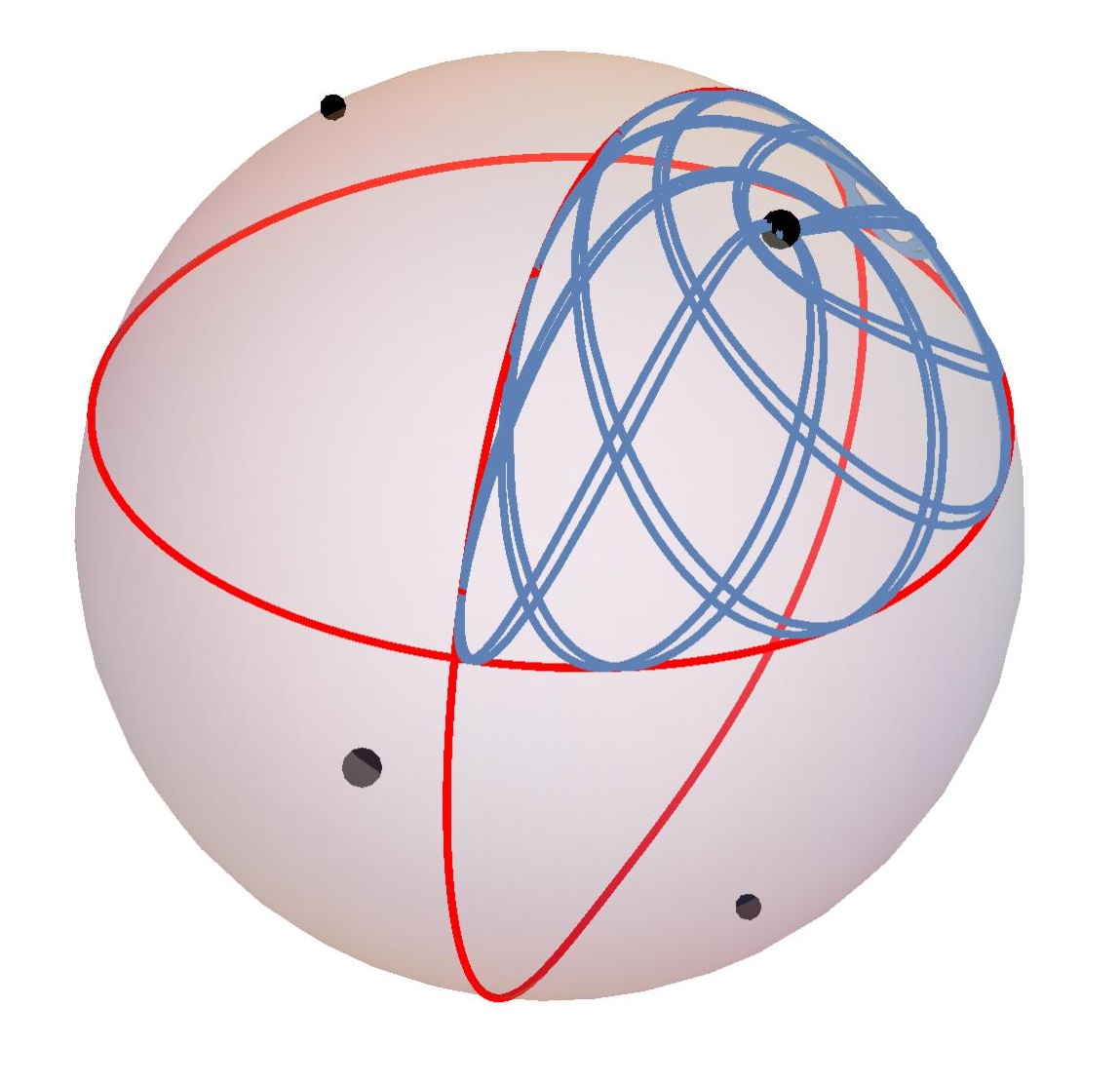} \\
 \\ (a) {\small $t_p: \frac{\bar{\sigma}}{\sigma}\Omega = -0.27, \ \frac{\sigma}{\bar{\sigma}}G = 0.8$} & (b) {\small $t_l: \frac{\bar{\sigma}}{\sigma}\Omega = -0.3, \ \frac{\sigma}{\bar{\sigma}}G = 0.6$} & (c) {\small $t_{s'}: \frac{\bar{\sigma}}{\sigma}\Omega = -0.2, \ \frac{\sigma}{\bar{\sigma}}G =-0.1$} \\ {\small $s_{u_0}=0$, $\, s_{v_0}=0$} & {\small $\, s_{u_0}=1$, $\, s_{v_0}=0$} & {\small $s_{u_0}=1$, $\, s_{v_0}=0$}\\
\\ \includegraphics[height=3.8cm]{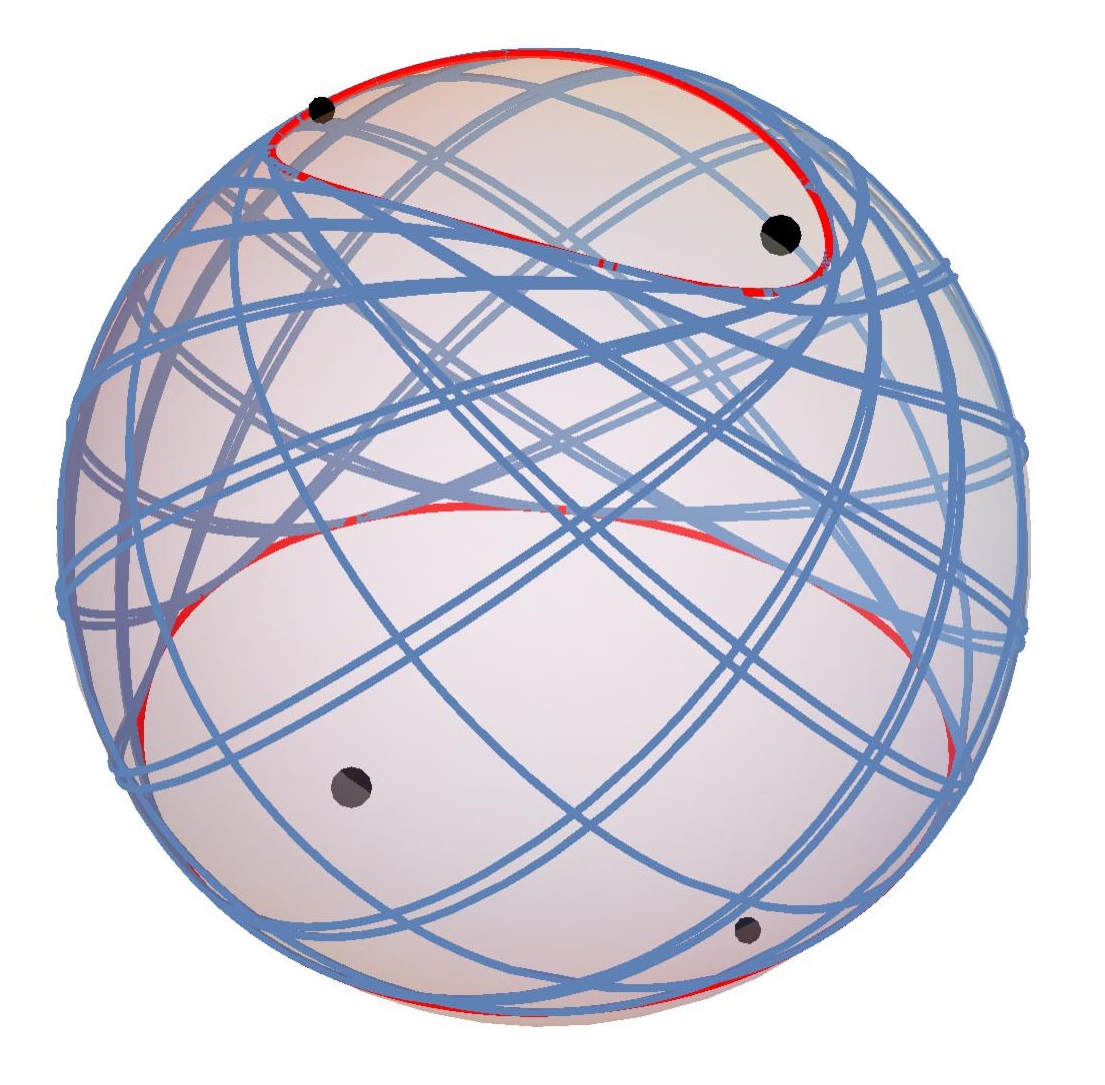}  & \includegraphics[height=3.8cm]{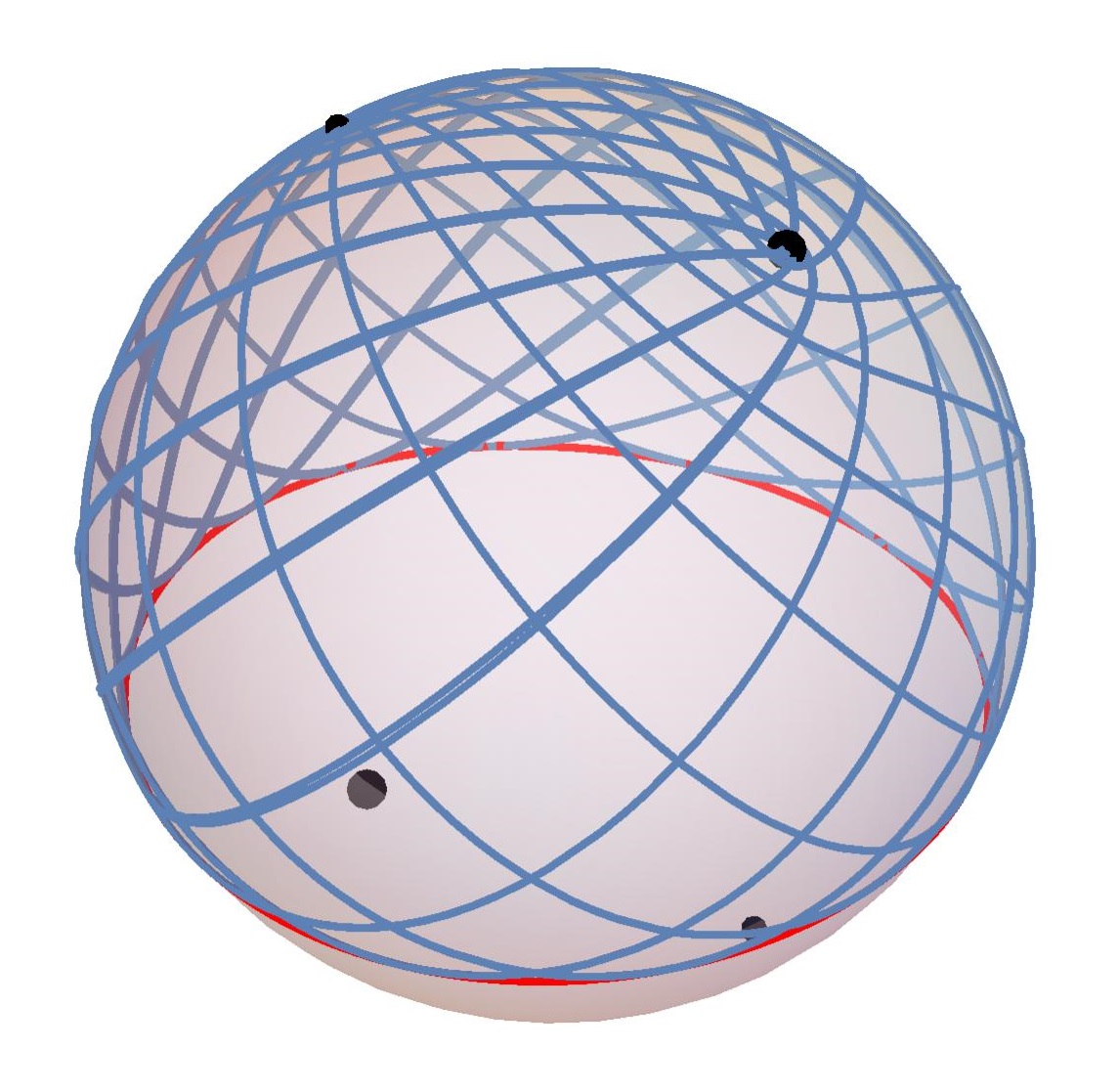}  & \includegraphics[height=3.8cm]{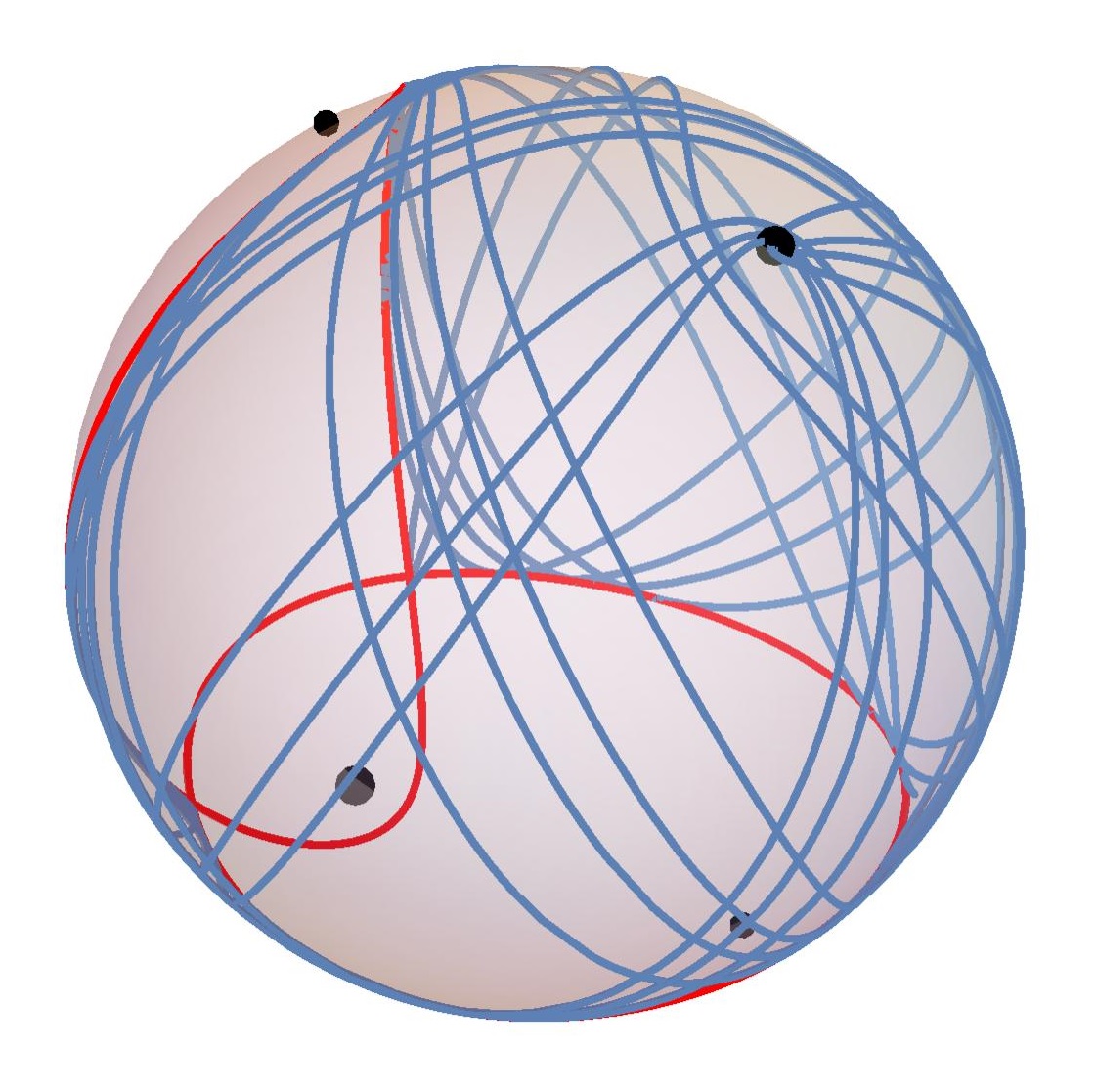}  \\
\\ (d) {\small $t_p: \frac{\bar{\sigma}}{\sigma}\Omega = 0.5, \ \frac{\sigma}{\bar{\sigma}}G = 2$} & (e) {\small $t_l: \frac{\bar{\sigma}}{\sigma}\Omega = 0.25, \ \frac{\sigma}{\bar{\sigma}}G = 1$} & (f) {\small $t_{s'}: \frac{\bar{\sigma}}{\sigma}\Omega = 0.5, \ \frac{\sigma}{\bar{\sigma}}G =0.5$} \\ {\small $s_{u_0}=1$, $\, s_{v_0}=2$} & {\small $\, s_{u_0}=0$, $\, s_{v_0}=0$} & {\small $s_{u_0}=1$, $\, s_{v_0}=2$}\\ \\
\includegraphics[height=3.8cm]{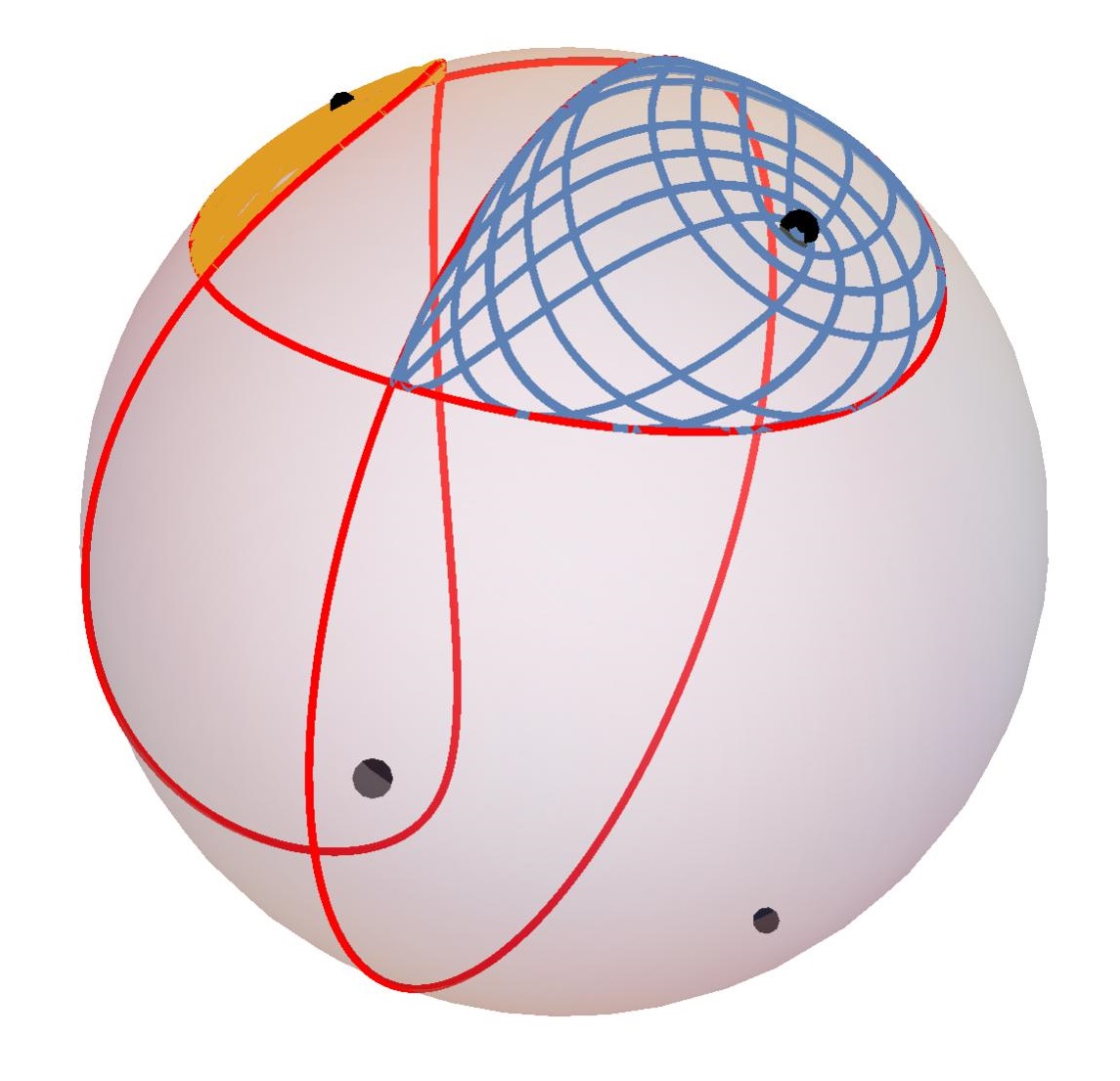}  &  \includegraphics[height=3.8cm]{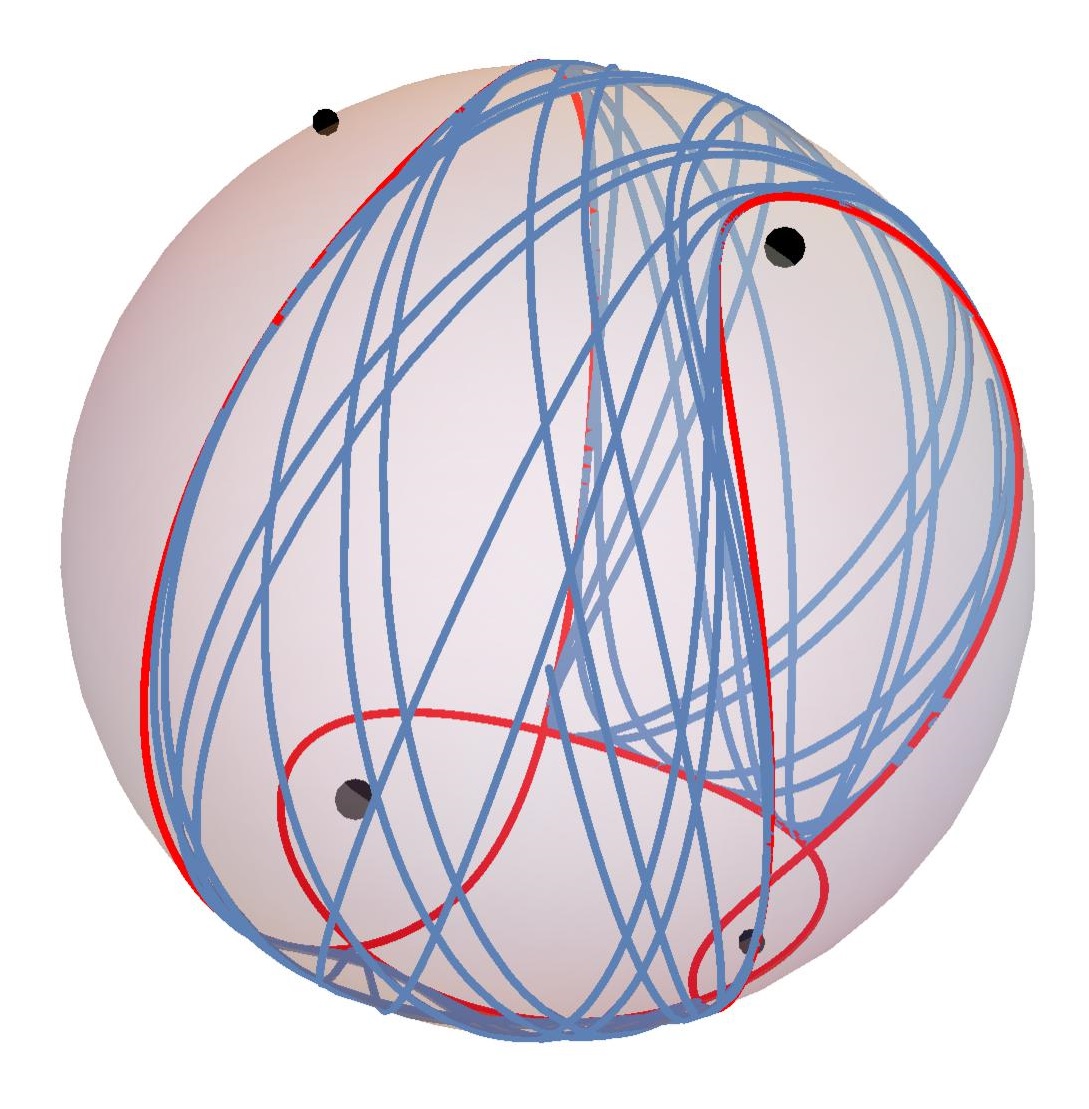}  & \includegraphics[height=3.8cm]{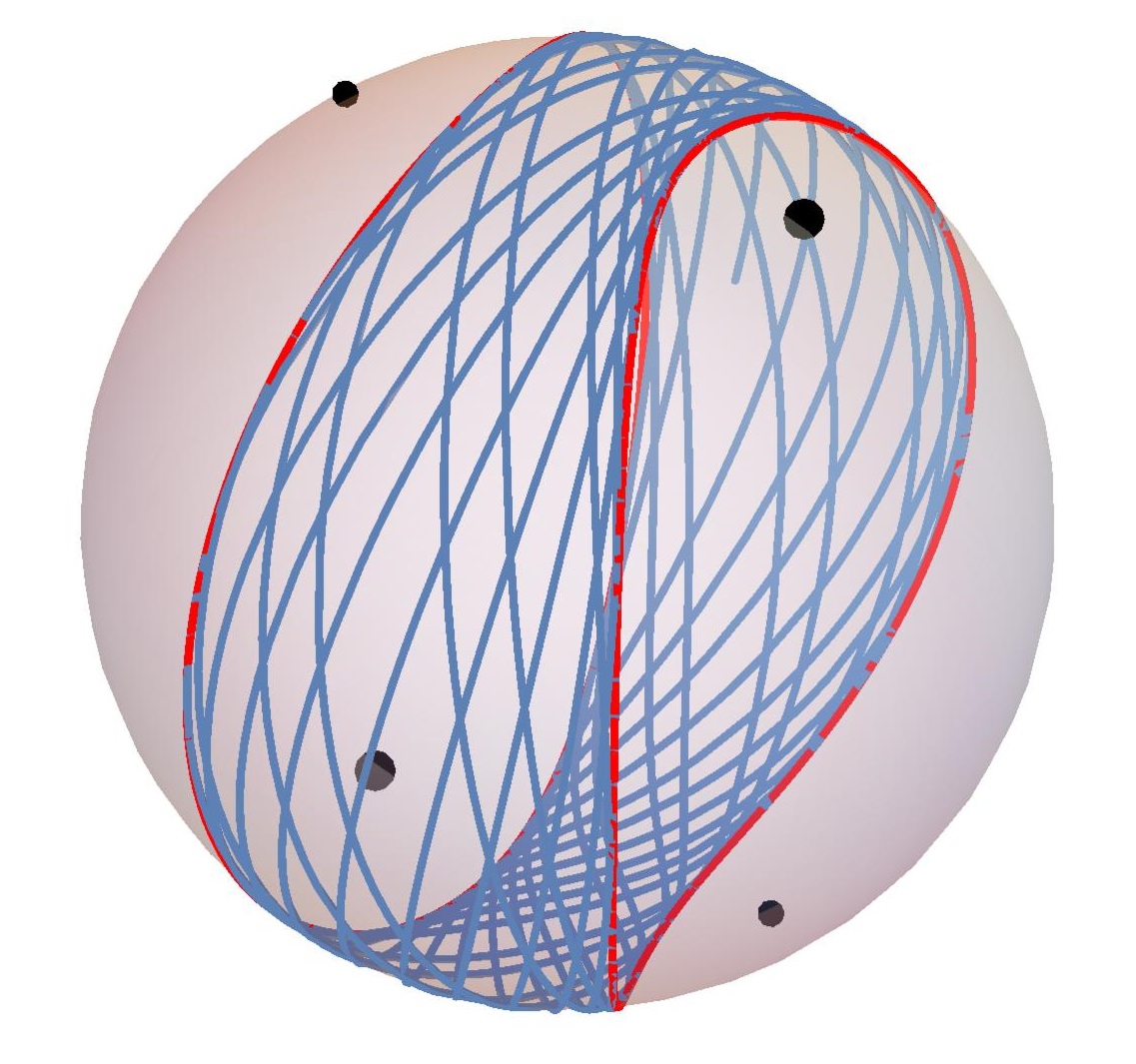}  \\
\\ (g) {\small $t_s: \frac{\bar{\sigma}}{\sigma}\Omega = -0.5 , \ \frac{\sigma}{\bar{\sigma}}G = 0 $} & (h) {\small $t_{ds}: \frac{\bar{\sigma}}{\sigma}\Omega = 0.8, \ \frac{\sigma}{\bar{\sigma}}G = 0.2$} & (i) {\small $t_{mp}: \frac{\bar{\sigma}}{\sigma}\Omega = 1.5, \ \frac{\sigma}{\bar{\sigma}}G =0.2$} \\ {\small $s_{u_0}=1$, $\, s_{v_0}=0$} & {\small $\, s_{u_0}=1$, $\, s_{v_0}=2$} & {\small $s_{u_0}=1$, $\, s_{v_0}=2$}\\
\end{tabular}
\end{center}
\begin{figure}[h]
\caption{Orbits in $S^2$. In all cases: $\gamma=\frac{1}{3}$, $\sigma=\cos \frac{\pi}{6}$, $\bar{\sigma}= \sin \frac{\pi}{6}$.}
\label{graficas1}
\end{figure}

\clearpage

\clearpage

\begin{center}
\begin{tabular}{ccc}
\includegraphics[height=3.8cm]{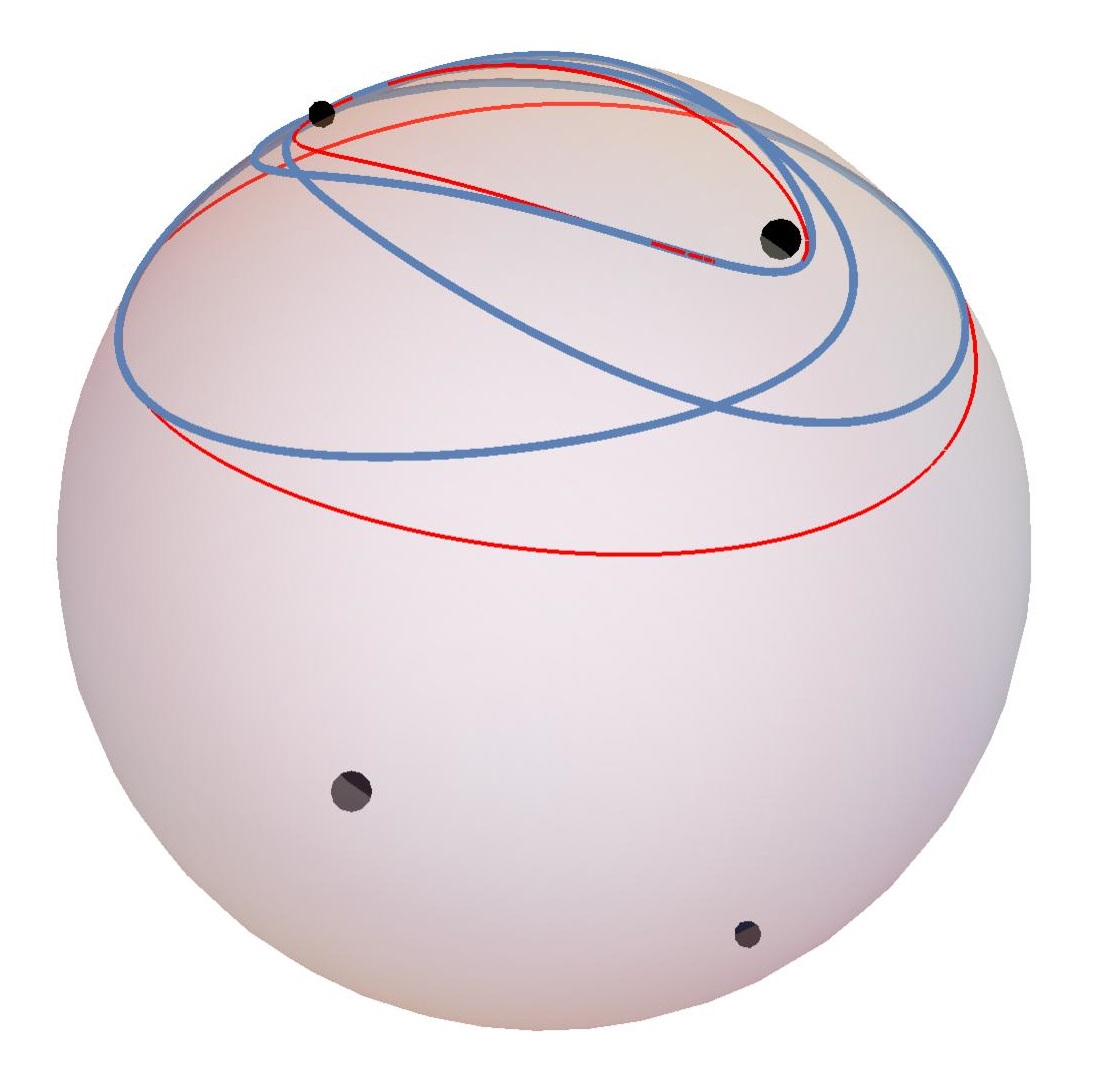}  &  \includegraphics[height=3.8cm]{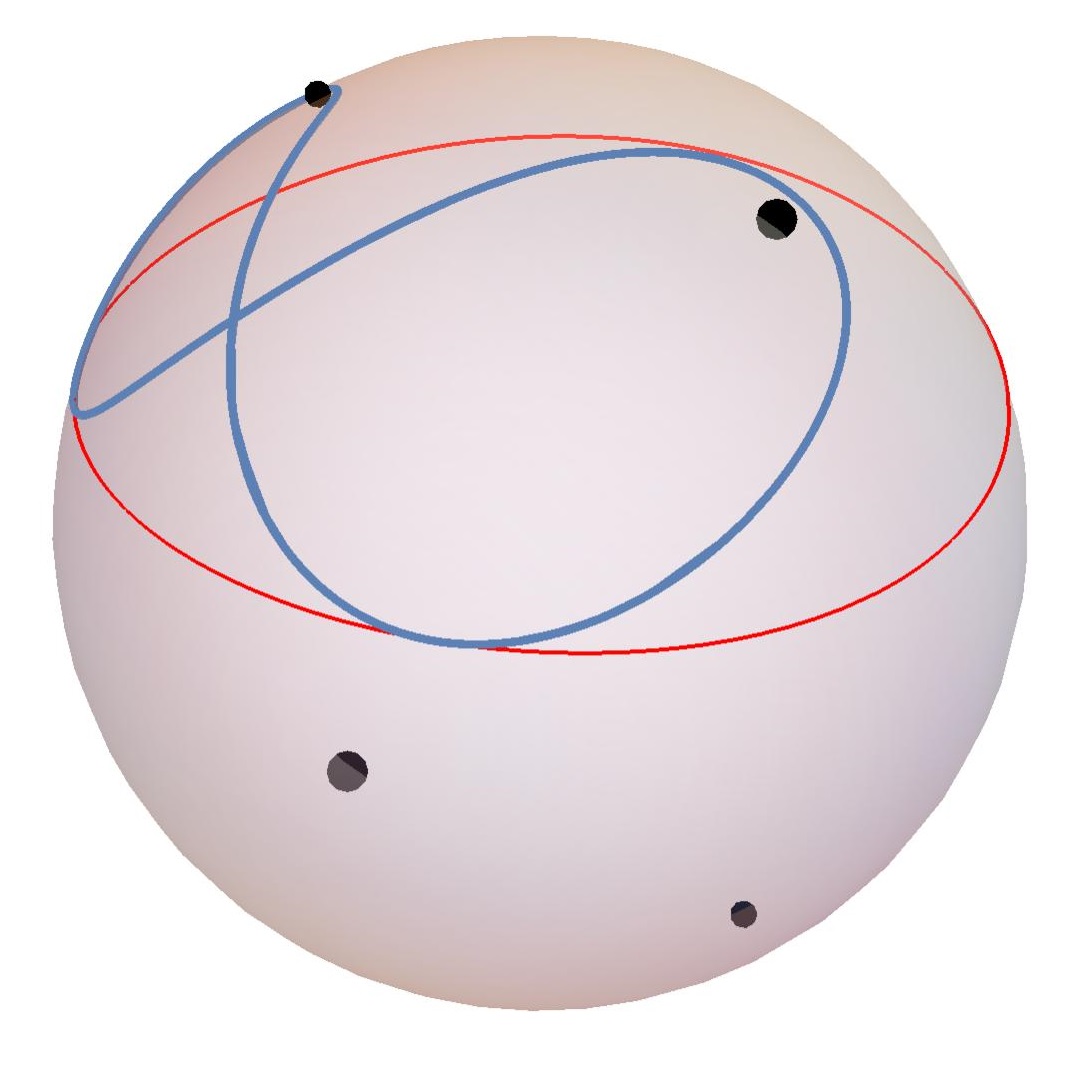} & \includegraphics[height=3.75cm]{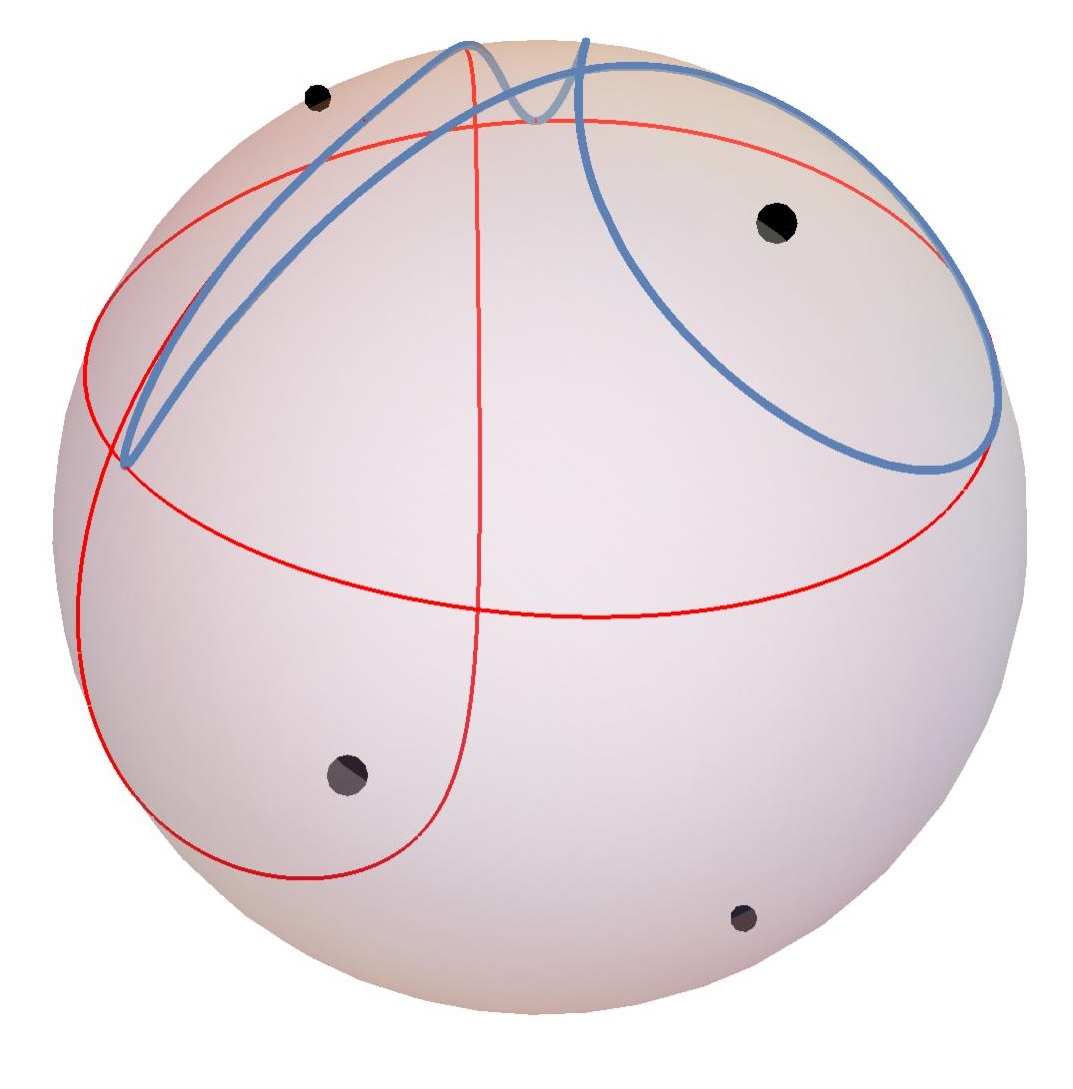} \\
 \\ (a) {\small $t_p: \frac{\bar{\sigma}}{\sigma}\Omega = -0.25, \ \frac{\sigma}{\bar{\sigma}}G \cong 0.80727$} & (b) {\small $t_l: \frac{\bar{\sigma}}{\sigma}\Omega = -0.2, \ \frac{\sigma}{\bar{\sigma}}G \cong 0.29835$} & (c) {\small $t_{s'}: \frac{\bar{\sigma}}{\sigma}\Omega = -0.25, \ \frac{\sigma}{\bar{\sigma}}G \cong  0.10725$} \\ {\small $2T_u=3T_v\, ,\, s_{u_0}=0$, $\, s_{v_0}=0$} & {\small $\, T_u=T_v\, ,\, s_{u_0}=3$, $\, s_{v_0}=0$} & {\small $3 T_u=T_v\, ,\ s_{u_0}=3$, $\, s_{v_0}=-1$}\\
\\ \includegraphics[height=3.8cm]{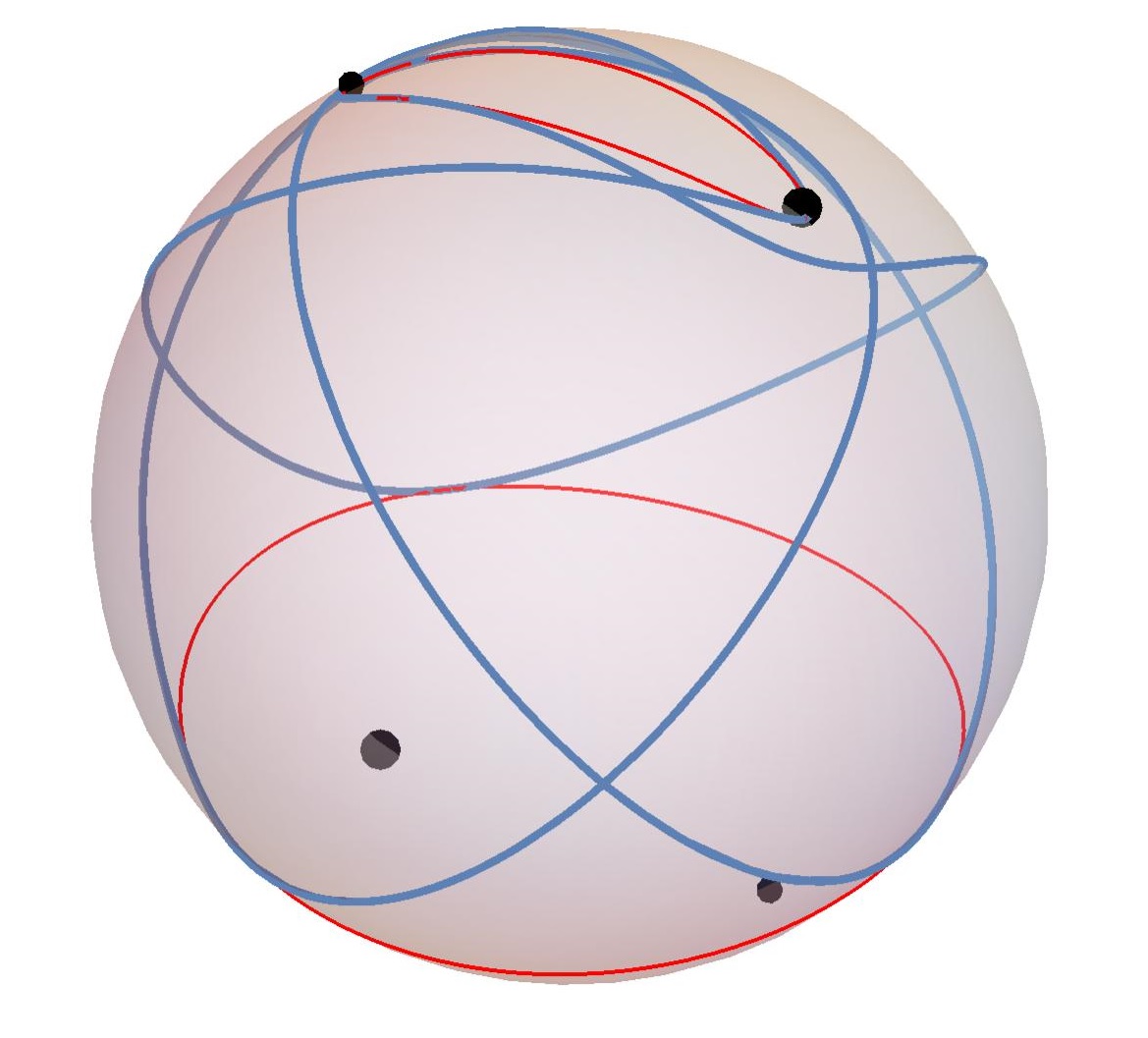}  & \includegraphics[height=3.8cm]{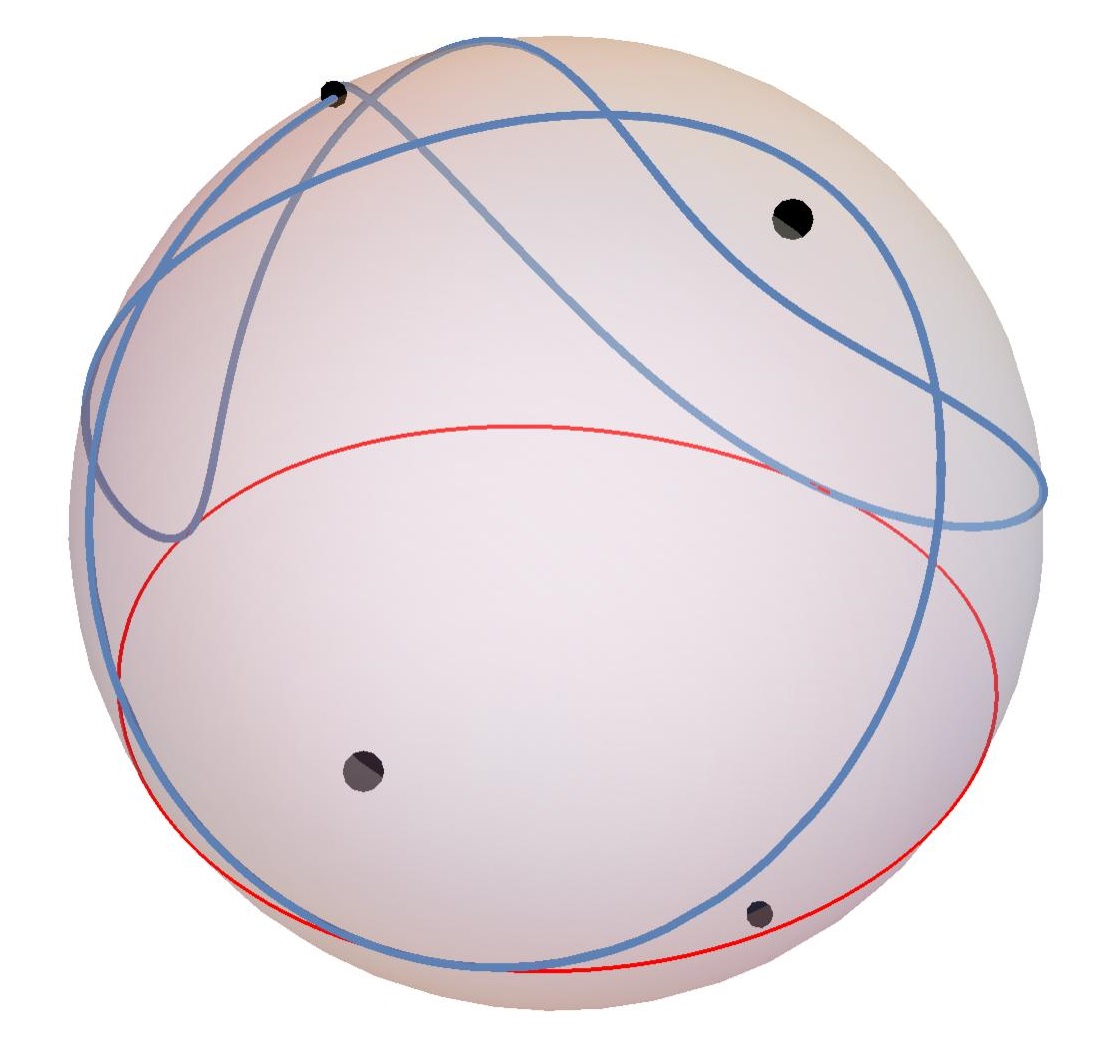}  & \includegraphics[height=3.8cm]{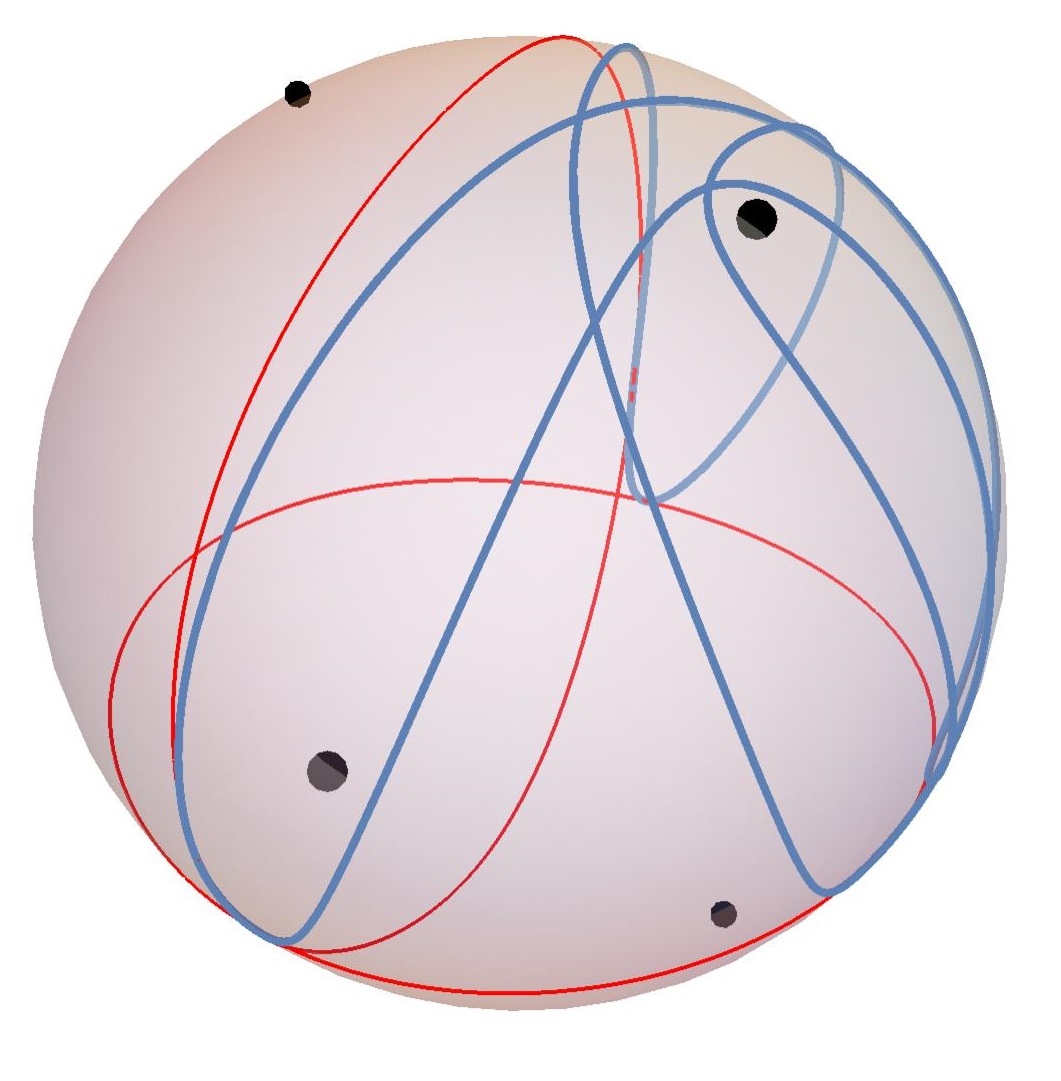}  \\
\\ (d) {\small $t_p: \frac{\bar{\sigma}}{\sigma}\Omega = 0.5, \ \frac{\sigma}{\bar{\sigma}}G \cong 1.56826$} & (e) {\small $t_l: \frac{\bar{\sigma}}{\sigma}\Omega = 0.25, \ \frac{\sigma}{\bar{\sigma}}G \cong 0.72393$} & (f) {\small $t_{s'}: \frac{\bar{\sigma}}{\sigma}\Omega = 0.3, \ \frac{\sigma}{\bar{\sigma}}G \cong 0.07292$} \\ {\small $3 T_u=4 T_v\, , \, s_{u_0}=1$, $\, s_{v_0}=0$} & {\small $\, 3 T_u=4 T_v\, , \, s_{u_0}=0$, $\, s_{v_0}=0$} & {\small $2 T_u =  T_v\, , \,s_{u_0}=3$, $\, s_{v_0}=1$}\\ \\
\includegraphics[height=3.8cm]{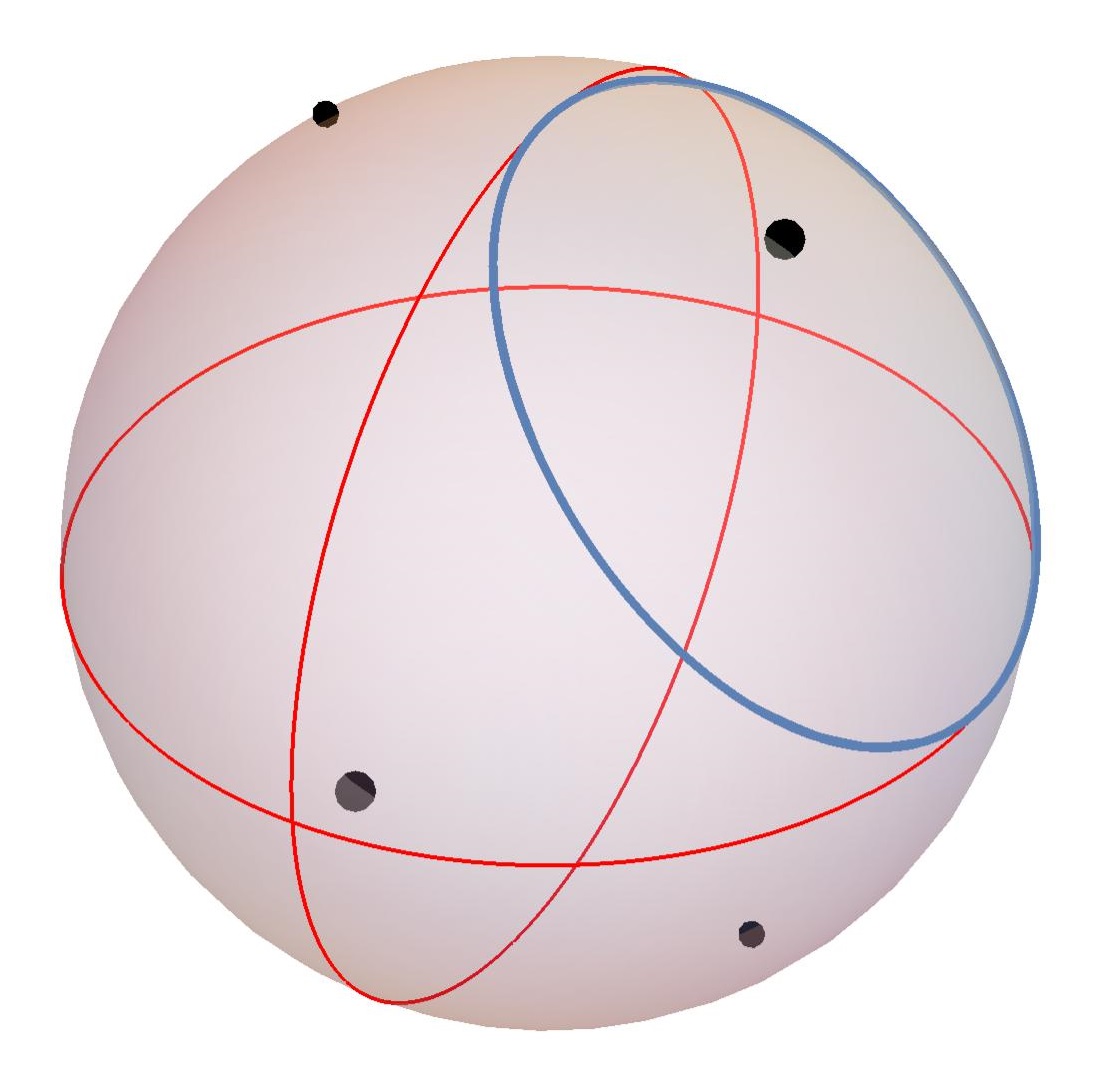}  &  \includegraphics[height=3.8cm]{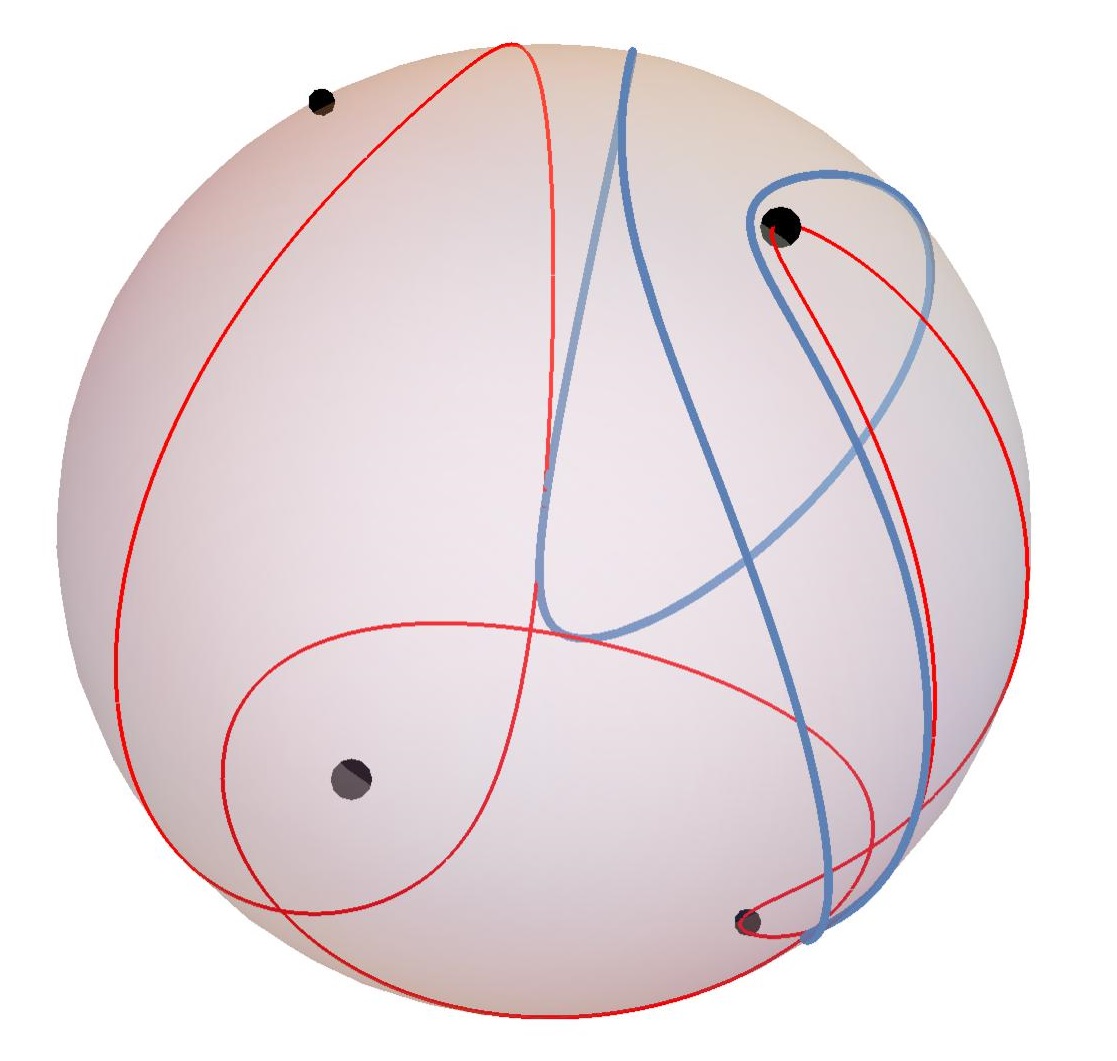}  & \includegraphics[height=3.9cm]{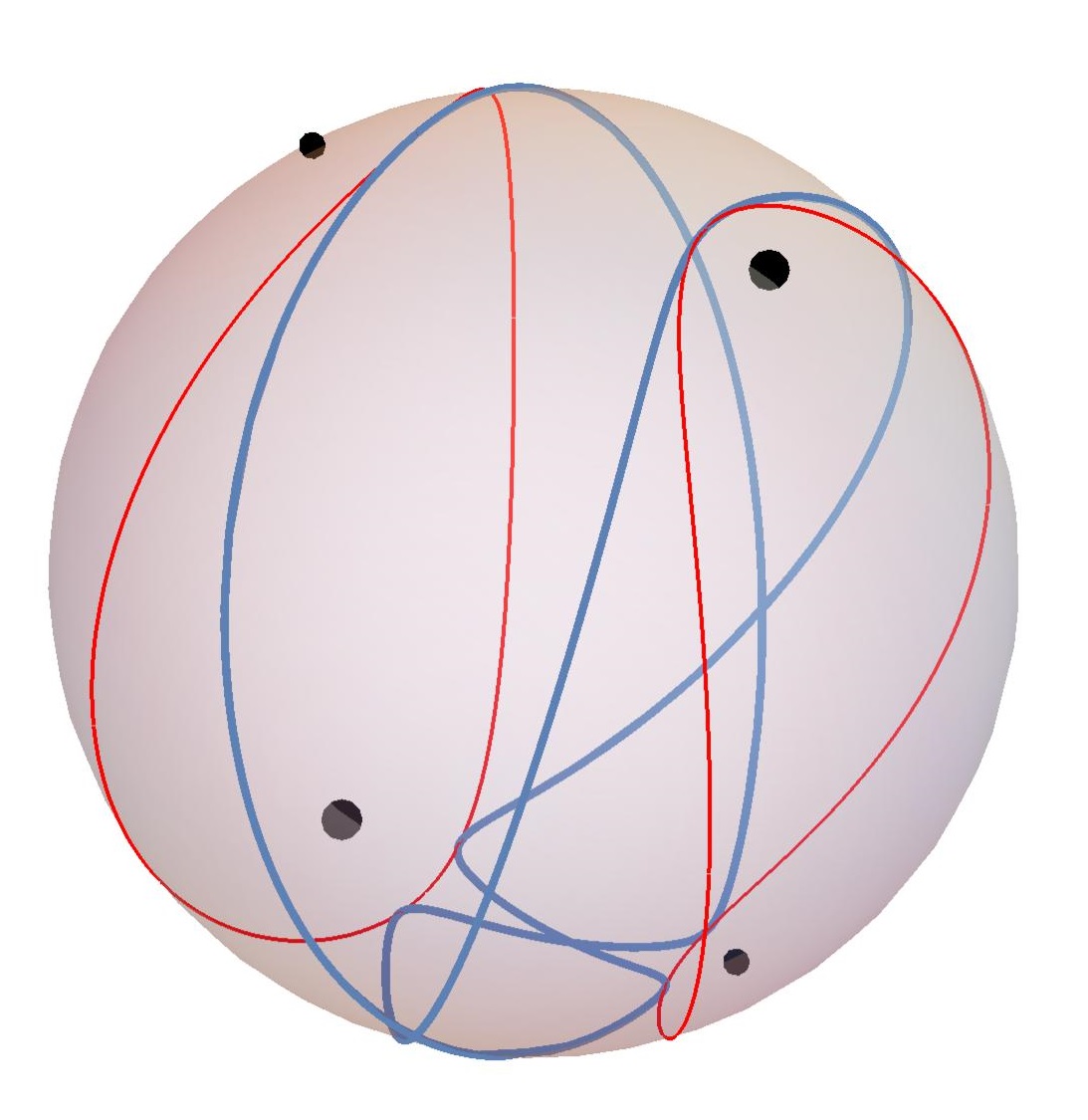}  \\
\\ (g) {\small $t_{s'}: \frac{\bar{\sigma}}{\sigma}\Omega = 0 , \ \frac{\sigma}{\bar{\sigma}}G = 0 $} & (h) {\small $t_{ds}: \frac{\bar{\sigma}}{\sigma}\Omega = 0.6, \ \frac{\sigma}{\bar{\sigma}}G \cong 0.23559$} & (i) {\small $t_{mp}: \frac{\bar{\sigma}}{\sigma}\Omega = 1.5,  \frac{\sigma}{\bar{\sigma}}G \cong 0.47580$} \\ {\small $\, T_u=T_v\,  , s_{u_0}=0$, $\, s_{v_0}=0$} & {\small $\, T_u=T_v\, , s_{u_0}=3$, $\, s_{v_0}=0$} & {\small \, $2T_u=3T_v$ \, , $s_{u_0}=0$, $\, s_{v_0}=0$}\\
\end{tabular}
\end{center}
\begin{figure}[h]
\caption{Closed orbits in $S^2$. In all cases: $\gamma=\frac{1}{3}$, $\sigma=\cos \frac{\pi}{6}$, $\bar{\sigma}= \sin \frac{\pi}{6}$.}
\label{graficas2}
\end{figure}


\clearpage

\begin{thebibliography}{}

\bibitem{Al1} Albouy, A., The underlying geometry of the fixed centers problems, in \textit{Topological Methods, Variational Methods and Their Applications}, Brezis, H., Chang, K.C., Li, S.J., Rabinowitz, P. (Eds.), Singapore: World Scientific, 2003, pp. 11-21.

\bibitem{Al2} Albouy, A. and Stuchi, T., Generalizing the classical fixed-centres problem in a non-Hamiltonian way, \textit{J. Phys. A}, 2004, vol. \textbf{37}, pp. 9109--9123.


\bibitem{Al3} Albouy, A., Projective Dynamics and Classical Gravitation, \textsl{Regul. Chaotic Dyn.}, 2008, vol. \textbf{13}(6), pp. 525--542.


\bibitem{Albouy2013} Albouy, A., There is a Projective Dynamics, \textsl{Eur. Math. Soc. Newsl.}, 2013, no. \textbf{89}, pp. 37--43.

\bibitem{Al4} Albouy, A., Projective Dynamics and First Integrals, \textsl{Regul. Chaotic Dyn.}, 2015, vol. \textbf{20}(3), pp. 247--276.

\bibitem{Alexeev} Alexeev, V.M., The generalized spatial problem of two fixed centers. Classification of motions. \textsl{Bull. of the Institute of Theoretical Astronomy}, 1965, vol. \textbf{10}(4), pp. 241--271.

\bibitem{nosPRL} Alonso Izquierdo, A., Gonzalez Leon, M.A. and Mateos Guilarte, J., Kinks in a non-linear massive sigma model, \textsl{Phys. Rev. Lett.}, 2008, vol. \textbf{101}, 131602.

\bibitem{nosJHEP} Alonso Izquierdo, A., Gonzalez Leon, M.A., Mateos Guilarte, J. and de la Torre Mayado, M., On domain walls in a Ginzburg-Landau non-linear $S^2$-sigma model, \textsl{J. High Energ. Phys.}, 2010, vol. \textbf{2010:8}, pp. 1-29.

\bibitem{Appell1890} Appell, P., De l'homographie en m\'ecanique, \textsl{Amer. J. Math.}, 1890, vol. \textbf{12}(1), pp. 103--114.

\bibitem{Appell1891} Appell, P., Sur les lois de forces centrales faisant d\'ecrire \`a leur point d'application une conique quelles que soient les conditions initiales,  \textsl{Amer. J. Math.}, 1891, vol. \textbf{13}(2), pp. 153--158.

\bibitem{Bolsinov}  Bolsinov, A.V. and Fomenko,  A.T., \textit{Integrable Hamiltonian systems: geometry, topology, classification}. Boca Raton: Chapman \& Hall/CRC, 2004.

\bibitem{Borisov2005} Borisov, A.V. and Mamaev, I.S., Generalized problem of two and four Newtonian centers, \textsl{Celest. Mech. Dyn. Astr.}, 2005, vol. \textbf{92}, pp. 371--380.

\bibitem{Borisov2007} Borisov, A.V. and Mamaev, I.S., Relations between Integrable Systems in Plane and Curved Spaces, \textsl{Celest. Mech. Dyn. Astr.}, 2007, vol. \textbf{99}(4), pp. 253--260.

\bibitem{Borisov2016} Borisov, A.V., Mamaev, I.S. and Bizyaev, I.A., The Spatial Problem of 2 Bodies on a Sphere. Reduction and Stochasticity, \textsl{Regul. Chaotic Dyn.}, 2016, vol. \textbf{21}(5), pp. 556--580.

\bibitem{Byrd} Byrd, P.F. and Friedman, M.D., \textit{Handbook of Elliptic Integrals for Engineers and Scientists}. 2nd ed. Heildelberg:  Springer-Verlag, 1971.

\bibitem{Demin1961} Demin, V.G., Orbits in the Problem of Two Fixed Centers, \textsl{Soviet Astronomy}, 1961, vol. \textbf{4}, pp. 1005--1012.

\bibitem{Euler1} Euler, L., De motu corporis ad duo centra virium fixa attracti, \textsl{Nov. Comm. Acad. Sci. Imp. Petrop.}, (i) vol. \textbf{10}, 1766, pp. 207--242. (ii) vol. \textbf{11}, 1767, pp. 152--184.


\bibitem{Euler2} Euler, L., Un corps \'etant attir\'e en raison r\'eciproque quarr\'ee des distances vers deux points fixes donn\'es, trouver les cas o\`u la courbe d\'ecrite par ce corps sera alg\'ebrique, \textsl{M\'emoires de l'Acad\'emie des Sciences de Berlin}, vol. \textbf{XVI}, 1767, pp. 228--249.

\bibitem{Higgs1979} Higgs, P., Dynamical Symmetries in a Spherical Geometry I, \textsl{J. Phys. A}, 1979, vol. \textbf{12}(3), pp. 309--323.

\bibitem{Jacobi} Jacobi, C.G.J., \textsl{Vorlesungen \"uber Dynamik}, Berlin: A. Clebsch, 1866. English translation: New Delhi: Hindustan Book Agency, 2009.

\bibitem{Killing1885} Killing, H. W., Die Mechanik in den nicht-euklidischen Raumformen, \textsl{J. Reine Angew. Math.}, 1885, vol. \textbf{98}(1), pp. 1--48.

\bibitem{Kozlov1992} Kozlov, V.V. and Harin, A.O., Kepler's Problem in Constant Curvature Spaces, \textsl{Celest. Mech. Dyn. Astr.}, 1992, vol. \textbf{54}(4), pp. 393--399.


\bibitem{Lagrange} Lagrange, J.L.,  Recherches sur le mouvement d'un corps qui est attir\'e vers deux centres fixes, \textsl{Miscellanea Taurinensia}, 1766--1769, t. IV. \textsl{Oeuvres compl\`etes}, vol. \textbf{2}, pp. 67--121.

\bibitem{Legendre} Legendre, A.M., \textsl{Trait\'e des Fonctions Elliptiques et des Int\'egrales Eul\'eriennes}, vol. 1, Paris: Huzard-Courcier, 1825.


\bibitem{Liouville1846} Liouville, J., Sur quelques cas particuliers o\`u les \'equations du mouvement d'un point mat\'eriel peuvent s'int\'egrer, \textsl{J. Math. Pures Appl.}, 1846, vol. \textbf{11}, pp. 345--378.

\bibitem{Mamaev2003} Mamaev, I.S., Two Integrable Systems on a Two-Dimensional Sphere, \textsl{Dokl. Phys.}, 2003, vol. \textbf{48}(3), pp. 156–-158.


\bibitem{Neumann}  Neumann, C., De problemate quodam mechanica, quod ad primam integralium ultra-ellipticorum classem revocatur, \textsl{J. Reine Angew. Math.}, 1859, vol. \textbf{56}, pp. 54--66.


\bibitem{Mathuna} \'{O} Math\'una, D., \textsl{Integrable Systems in Celestial Mechanics}, Boston: Birkh\"{a}user, 2008.


\bibitem{Seri} Seri, M., The problem of two fixed centers: bifurcation diagram for positive energies, \textsl{J. Math. Phys.}, 2015, vol. \textbf{56},  012902.

\bibitem{Serret1859} Serret, P., \textsl{Th\'eorie nouvelle g\'eom\'etrique et m\'ecanique des lignes \`a double courbure}, Paris: Mallet-Bachelier, 1860.

\bibitem{Vozmischeva2000} Vozmischeva, T.G., Classification of Motions for Generalization of the two Centers Problem on a Sphere, \textsl{Celest. Mech. and Dyn. Astr.}, 2000, vol. \textbf{77}(1), pp. 37--48.

\bibitem{Vozmischeva2002} Vozmischeva, T.G. and Oshemkov, A.A., Topological analysis of the two-center problem on two-dimensional sphere, \textsl{Sbornik: Mathematics}, 2002, vol. \textbf{193}(8), pp. 1103--1138.

\bibitem{Vozmischeva} Vozmischeva, T.G., \textit{Integrable Problems of Celestial Mechanics in Spaces of Constant Curvature}. Boston: Kluwer Academ. Publ., 2003.

\bibitem{Waalkens} Waalkens, H., Dullin, H.R. and Richter, P.H., The problem of two fixed centers: bifurcations, actions, monodromy, \textsl{Physica D: Nonlinear Phenomena}, 2004, vol. \textbf{196}(3), pp. 265--310.


\bibitem{Whittaker} Whittaker, E. T. and Watson, G. N., \textit{A Course of Modern Analysis}. 4th ed. Cambridge: Cambridge University Press, 1996.




\end{thebibliography}
\end{document}